\documentclass{svjour3}                   

\usepackage{amssymb}
\usepackage{graphicx}
\usepackage{natbib}
\bibpunct{(}{)}{;}{a}{}{,}

\usepackage{mathptmx}     
\journalname{SSRv}
\usepackage{epsfig}
\usepackage{psfig}
\usepackage{color}

\newcommand{\aap}{A\&A}

\newcommand{\apjl}{ApJ}
\newcommand{\apjs}{ApJS}

\newcommand{\chandra}{{\it Chandra}}
\newcommand{\asca}{{\it ASCA}}
\newcommand{\sax}{{\it BeppoSAX}}

\newcommand{\xmm}{{\it XMM-Newton}}

\newcommand{\lb}{{\left<\right.}}
\newcommand{\rb}{{\left.\right>}}
\newcommand{\mincir}{\raise
  -2.truept\hbox{\rlap{\hbox{$\sim$}}\raise5.truept \hbox{$<$}\ }}
\newcommand{\magcir}{\raise
  -2.truept\hbox{\rlap{\hbox{$\sim$}}\raise5.truept \hbox{$>$}\ }}

\newcommand{\rgzero}{r_\mathrm{g0}}
\newcommand{\heatpar}{\alpha_H}
\newcommand{\Beff}{B_\mathrm{eff}}
\def\lsim{\;\raise0.3ex\hbox{$<$\kern-0.75em\raise-1.1ex\hbox{$\sim$}}\;}
\def\gsim{\;\raise0.3ex\hbox{$>$\kern-0.75em\raise-1.1ex\hbox{$\sim$}}\;}
\def\alf{Alfv\'en~}
\def\cmc{\rm ~cm^{-3}}

\def\kms{\rm ~km~s^{-1}}

\begin{document}

\title{Large--Scale Structure Formation: from the first non-linear objects to massive galaxy clusters}

\author{S. Planelles \and D.R.G. Schleicher \and A.M. Bykov}

\authorrunning{Planelles et al.}
\titlerunning{Large-Scale Structure Formation}

\institute{
S. Planelles \at
Department of Astronomy, University of Trieste, via Tiepolo 11, I-34143 Trieste, Italy\\
INAF -- National Institute for Astrophysics, Trieste, Italy\\
\email{susana.planelles@oats.inaf.it}\\
\and
D.R.G. Schleicher \at
Institut f\"ur Astrophysik, Georg-August-Universit\"at G\"ottingen, Friedrich-Hund-Platz 1,
37077 G\"ottingen, Germany\\
\email{dschleic@astro.physik.uni-goettingen.de}\\
\and
A.M.~Bykov \at
A.F.~Ioffe Institute for Physics and
Technology, 194021, St.Petersburg, Russia \\
and St.Petersburg State
Politechnical University\\ 
\email{byk@astro.ioffe.ru}    
}

\date{Received:  ; Accepted:}

\maketitle

\begin{abstract}
The large-scale structure of the Universe formed from initially small perturbations in the cosmic density field, leading to galaxy clusters with up to $10^{15}$~M$_\odot$ at the present day. 
Here, we review the formation of structures in the Universe, considering the first primordial galaxies and the most massive galaxy clusters as extreme cases of structure formation where fundamental processes such as gravity, turbulence, cooling and feedback are particularly relevant.
The first non-linear objects in the Universe formed in dark matter halos with $10^5-10^8$~M$_\odot$ at redshifts $10-30$, leading to the first stars and massive black holes. At later stages, larger scales became non-linear, leading to the formation of galaxy clusters, the most massive objects in the Universe. We describe here their formation via gravitational processes, including the self-similar scaling relations, as well as the observed deviations from such self-similarity and the related non-gravitational physics  (cooling, stellar feedback, AGN). While on intermediate cluster scales the self-similar model is in good agreement with the observations, deviations from such self-similarity are apparent in the core regions, where numerical simulations do not reproduce the current observational results. The latter indicates that the interaction of different feedback processes may not be correctly accounted for in current simulations. 
Both in the most massive clusters of galaxies as well as during the formation of the first objects in the Universe, turbulent structures and shock waves appear to be common, suggesting them to be ubiquitous in the non-linear regime.

\end{abstract}

\section{Introduction}
\label{intro}
    
The current hierarchical paradigm of structure formation
is set within the spatially flat  {\it $\Lambda$-Cold Dark Matter}  model  
\citep[$\Lambda CDM$;][]{Blumenthal_1984} with cosmological constant, also known as the {\it concordance} model. 
Tight constraints on the parameters of the underlying cosmological model have now been 
placed thanks to the combination of different observational probes
\citep[see, e.g.][for recent reviews]{Voit_2005, Allen2011, Hamilton_2013}. 
In the resulting scenario \citep[see][and references therein]{Planck_2013_parameters}  
the Universe, whose age is estimated to be $\sim 13.8$~Gyr, 
is composed  of dark energy ($\Omega_\Lambda\approx0.7$), dark matter  ($\Omega_{{DM}}\approx0.25$)  
and baryonic matter  ($\Omega_{b}\approx0.05$), with a Hubble constant given by $H_0\approx67$~km/s/Mpc.  
In addition, the primordial matter power spectrum seems to be characterized by a power-law index 
$n\approx0.96$  with an amplitude $\sigma_8\approx0.83$.  
  
Within this paradigm, the formation of the first structures in the Universe is driven by the gravitational collapse
of small inflation--induced matter density perturbations existing in the primordial matter density field.
Predictions  from  N-body simulations \citep[e.g.][]{Klypin_1983} have  confirmed that the growth of these perturbations
gives rise to the formation of a complex network of cosmic structures interconnected along walls and filaments
concerning a  wide range of scales.

The first structures in the Universe are expected to form at redshifts of $10-30$ in dark matter (DM) halos 
of $10^5-10^8$~M$_\odot$ \citep{Tegmark97, Barkana01, Glover05, Bromm09}. A crucial 
condition for these DM halos to form stars or galaxies is the ability of their gas to cool in a 
Hubble time. To address this question, \citet{Tegmark97} have modeled the cooling in DM halos 
of different virial temperatures, 
showing that a virial temperature of at least $1000$~K is required so that efficient cooling via molecular 
hydrogen can occur. Such a temperature corresponds to a mass scale of 
\begin{equation}
 M_{H_2}\sim10^{6.5}\left( \frac{10}{1+z} \right)^{3/2}M_\odot.
 \end{equation} 
Halos of this or slightly higher masses are typically referred to as the so-called minihalos, which are generally assumed to harbor the first primordial stars in the Universe. Their formation has been explored through detailed numerical simulations starting from cosmological initial conditions,  following the formation of the first minihalos and their gravitational collapse, including gas chemistry and cooling, down to AU-scales or below \citep{Abel02, Bromm04, Yoshida08}. The first such simulations typically followed only the formation of the first peak during the gravitational collapse, hinting at the formation of rather massive isolated stars of $\sim100-300$~M$_\odot$ due to the rather high accretion rates of $\sim10^{-3}$~M$_\odot$~yr$^{-1}$. 
Subsequent studies have explored the formation of self-gravitating disks and their fragmentation at later stages \citep{Stacy10, Clark11, Greif11, Greif12, Latif13a}, indicating the formation of star clusters and binaries rather than isolated stars. The resulting initial mass function (IMF) of these stars is expected to be top-heavy, with characteristic masses in the range of $10-100$~M$_\odot$. The studies involving sink particles further suggest that low-mass protostars can be ejected from the center of the halo via 3-body interactions, thus implying the potential presence of primordial stars with less than a solar mass that could survive until the present day. Radiative feedback seems to imply an upper mass limit of $50-100$~M$_\odot$ \citep{Hosokawa11, Susa13}.

In DM halos with virial temperatures above $10^4$~K, an additional cooling channel is present via atomic hydrogen. In such DM  halos, also referred to as atomic cooling halos, cooling is always possible via atomic hydrogen lines, helium lines or recombination cooling, while the minihalos may not be able to cool if their molecular hydrogen content is destroyed by photodissociating backgrounds \citep{Machacek01, Johnson07, Johnson08, Schleicher10, Latif11}. Such halos are also more robust with respect to the first supernova explosions \citep{Wise08_3, Greif10}, and may thus give rise to a self-regulated mode of star formation. In the presence of a strong radiative background, for instance from a nearby galaxy, they may remain metal-free and collapse close to isothermally at $\sim8000$~K \citep{Omukai01, Spaans06, Schleicher10, Shang10, Latif11, Latif13c, Prieto13}. While the initial studies followed on the collapse of the first peak \citep{Wise08_2, Regan09a, Shang10}, \citet{Regan09b} aimed at following the longer-term evolution confirming the formation of a self-gravitating disk. \citet{Latif13b} recently pursued the first high-resolution investigation on the fragmentation of such halos on AU scales, finding that fragmentation may occur, but does not inhibit the growth of the resulting central objects. For the high accretion rates of $\sim1$~M$_\odot$~yr$^{-1}$ measured in their simulations, radiative feedback is expected to be negligible \citep{Hosokawa12} and the formation of very massive objects with up to $10^5$~M$_\odot$ seems feasible \citep{Schleicher13}.  Such supermassive stars are expected to collapse via the post-Newtonian instability and form the progenitors of supermassive black holes \citep[SMBHs;][]{Shapiro86}.

Depending on previous metal enrichment and the ambient radiation field, atomic cooling halos may also gather the proper conditions  for the  formation of the first galaxies. 
The formation and evolution of these galaxies, directly connected to the formation of the first stars and their associated radiative or supernova feedback, 
represent a crucial and complicated aspect of the whole cosmic history.  
In this sense, the main focus of this review will be on the formation of the first stars and SMBHs, and the reader is referred to the  reviews by \citet{Bromm09} and \citet{Bromm11} concerning the formation and properties of the first galaxies.

In the hierarchical paradigm of structure formation, the first objects are the building blocks of subsequent structure formation, leading to larger galaxies and galaxy clusters through accretion and mergers \citep[e.g.][]{Somerville12}.   
As a consequence of this connection, regardless  of the wide range of  involved scales,  a number of physical processes, such as the generation of turbulence during collapse and
the relevance of cooling and  feedback processes,  seem to be common in the formation of the different cosmic structures. 
Roughly speaking, the cosmic hierarchy is delimited, in terms of mass and formation time, by the first galaxies in the early Universe and the most massive galaxy clusters at the present day, whereas  the bulk of galaxies generally lie in-between these extreme cases. 
However,
a full understanding of galaxy evolution represents a complex and fundamental topic in cosmology that is being currently investigated by a considerable number of authors \citep[see][for a recent review on the current status of galaxy formation]{Silk_2012}. 
Given the complexity of this topic and the limited space available for this review, we avoid any description of galaxy evolution. Instead, 
since we are mostly interested in the role  that the physics of plasma plays on the formation of cosmic structures, we will focus both  on the formation of the first objects, i.e. the first stars and massive black holes, as well as on the large galaxy clusters at the present day.
These extreme scenarios will allow us to illustrate the importance of cooling, turbulence and feedback during structure
formation independently of the considered scales.

Galaxy clusters are  the largest nonlinear objects in the Universe today  and  thus a central  part of the large--scale structure (LSS). 
Clusters of galaxies, whose total masses vary from $10^{13}$ up to $10^{15}M_\odot$, 
are characterized by very deep gravitational potential wells containing a large number of galaxies ($\sim 10^2-10^3$)
over a region of a few Mpc \citep[see, e.g.][for an early review on galaxy clusters]{Sarazin_1988}. Although most of the mass in clusters is in the form of DM, 
a very hot and diffuse plasma, the intra--cluster medium (ICM), resides within the space between galaxies in clusters. 
The ICM, where the thermal plasma coexists with magnetic  fields and relativistic particles, holds the major part of the baryonic matter in clusters.
This cluster environment affects the  
evolution of the hosted galaxies by means of a number of dynamical processes such as harassment, ram--pressure stripping or galaxy mergers \citep[e.g. see][for a textbook on galaxy formation and evolution]{Mo_2010}.
The intra--cluster plasma, with typical temperatures of $T\sim 10^{7}-10^{8}$ K,  strongly emits X-ray radiation, 
causing clusters of galaxies to have  high X-ray luminosities, $L_X\sim 10^{43}-10^{45}$~erg/s. 
In addition, the ICM is quite tenuous, with electron number densities of $n_e \sim 10^{-4}-10^{-2}$ $cm^{-3}$ and, 
although it is formed  
mainly of hydrogen and helium, it also holds 
a mean abundance of heavier elements of about $\sim1/3$ of the solar abundance.
 
Given their typical extensions and their deep  gravitational potential wells, 
clusters of galaxies are fundamental for our comprehension  of the Universe, 
marking the transition between cosmological and galactic scales. 
Whereas on cosmological scales  the growth of perturbations is mainly driven by the effects of gravity on the DM component, 
on galactic scales  gravity operates in connection  with a number of gas dynamical and astrophysical phenomena.  
Given such an scenario, galaxy clusters and, in particular, the hot intra--cluster plasma represent a fascinating and complex environment harboring  a wide range of astrophysical and dynamical  processes related to both the gravitational  collapse and the  baryonic physics: 
gravitational shock waves, gas radiative cooling, star formation (SF), gas accretion onto SMBHs
hosted by massive cluster galaxies, feedback from supernovae 
(SNe) or active galactic nuclei (AGN), shock acceleration, magnetohydrodynamical (MHD) processes, gas turbulence, ram--pressure stripping of galaxies, 
thermal conduction processes, energetics associated to the populations of cosmic ray (CR) electrons and protons, etc.

All these processes are manifested by a number of cluster observables such as the thermal X-ray emission, the Sunyaev-Zel'dovich effect \citep[SZ;][]{SZ_1972}, the spectra of galaxies, or the radio synchrotron and gamma-ray emissions associated  to the population of non-thermal particles.
As a consequence, galaxy clusters reside in an  incomparable position within astrophysics and
cosmology: while the number and distribution of clusters can be used to place  constraints on 
the current model of cosmic structure formation, a thorough understanding of the complicated processes 
determining the properties of the hot intra--cluster plasma seems to be crucial to fully understand galaxy cluster observations.

In this review, we describe the formation of the large-scale structure of the Universe in the framework of the $\Lambda$CDM model. A particular focus  is both on the formation of the first objects, i.e. the first stars and massive black holes, as well as on the large galaxy clusters at the present day. In both applications, we emphasize the role of gravitational as well as non-gravitational plasma physics such as turbulence, cooling, magnetic fields or feedback processes. The overall structure of this review is as follows:
in \S\ref{sec:theory_sf} we start by reviewing the basic concepts of cosmic structure formation, 
from the early linear evolution of small density perturbations out to the complex collapse of real overdensities; 
in \S\ref{sec:mf} we overview the relevance for cosmology of a proper calibration of the halo mass function; 
in \S\ref{sec:early_universe} we describe the formation of the first halos in the early Universe;
a brief description of the self-similar model of the intra--cluster plasma is done in \S\ref{sec:self_similar_evolution}, 
whereas in  \S\ref{sec:thermo},  
the role played by non-gravitational heating and cooling processes in altering the 
predictions of such a model is discussed;  
finally, we summarize the results presented in \S\ref{sec:conclusions}.
 
Given the limited space available for this review, 
we refer the reader  to recent 
reviews about  early structure formation in the Universe \citep[e.g.][]{Bromm11}
and cosmology with  clusters of galaxies
\citep[e.g.][]{Allen2011, kravtsov_borgani12}  for a more extensive
discussion of these topics.

\section{Theory of structure formation}
\label{sec:theory_sf}

In this Section we outline the main theory of cosmic structure formation 
through the process of gravitational instability of small initial density perturbations.
We refer the reader to previous reviews or cosmology textbooks for a more detailed analysis 
(e.g. \citealt{Peebles_1993, Coles_2002}; see as well \citealt{Borgani_2008b}).

\subsection{Linear evolution of density perturbations}

The gravitational instability of a uniform and non-evolving medium versus 
small perturbations was first  addressed  by  \citet{Jeans_1902}.
Applying this theory to an expanding Universe in the linear regime, while density  perturbations are small, 
provides a general picture of cosmic structure formation.   

Let us consider an initial density perturbation field characterized by its dimensionless density contrast:
\begin{equation} 
\delta({\bf x})={\rho({\bf x})-\bar\rho \over \bar\rho} \, ,
\label{eq:del}
\end{equation}
where $\rho({\bf x})$  is  the matter density field  at the position ${\bf x}$, and
$\bar\rho=\lb \rho\rb$ is the mean mass density of the background universe. 
The primordial properties of this field are determined during the inflationary epoch.
In general, inflationary models predict  a homogeneous and  isotropic Gaussian random fluctuation field 
\citep[e.g.][]{Guth_Pi_1982}, which appears to be confirmed by observed fluctuations in the 
Cosmic Microwave Background  \citep[CMB; e.g.][]{Planck_CMB_2013}.  

To resolve the evolution  of the initial density perturbations in an expanding Universe, 
the perturbed Friedmann's equations need to be solved.
However, during the linear evolution the problem can be simplified. 
Consider that a self--gravitating and pressureless fluid dominates the 
matter content of  an expanding Universe.  
In principle, these assumptions are valid  if the perturbation is unstable, that is, if its scale
is larger than the characteristic Jeans 
scale\footnote{The Jeans length, the characteristic length scale for the self-gravity of the gas, 
is defined as $\lambda_J=\sqrt{\frac{15k_B T}{4\pi G\mu\rho_{gas}}}$,
with $k_B$ the Boltzmann constant, $T$ the gas temperature, $G$ the Newton's constant, 
$\mu$ the mean molecular weight and $\rho_{gas}$ the mass density of the gas.},
and if we deal with the evolution of DM perturbations. 
If the fluid is also assumed to be non--relativistic,
the Newtonian treatment can be applied. In this case, the evolution of density perturbations
is described by  the continuity, the Euler, and the Poisson  equations:
\begin{eqnarray}
{\partial \delta\over \partial t}+\nabla \cdot [(1+\delta){\bf u}]=0\\
\label{eq:continuity}
{\partial {\bf u}\over \partial t} + 2H(t){\bf u} + ({\bf u}\cdot\nabla){\bf u} = -{\nabla \phi\over a^2}\\
\label{eq:euler}
{\nabla^2\phi = 4\pi G\bar\rho a^2\delta}\,,
\label{eq:poisson}
\end{eqnarray}
where spatial derivatives are with respect to the comoving coordinate ${\bf x}$,
$a(t)$ is the cosmic expansion factor such that ${\bf r} = a(t){\bf x}$ is the proper coordinate,  
${\bf v} =\dot{{\bf r}}=\dot a{\bf x}+{\bf u}$ is the total velocity of a fluid element 
(with $\dot a{\bf x}$ and ${\bf u}=a(t)\dot{{\bf x}}$ giving  the Hubble flow and the peculiar velocities, respectively),
$\phi({\bf x})$ is the gravitational potential and $H(t)=\dot a/a = E(t)H_0$ is the time-dependent Hubble parameter.
In the case of a $\Lambda$CDM cosmology, when relativistic species are neglected, $E(z)$ is given by  
\begin{equation}
E(z)\equiv \frac{H(t)}{H_0}=[(1+z)^3\Omega_m+(1+z)^2(1-\Omega_m-\Omega_\Lambda)+\Omega_{\Lambda}]^{1/2} \, .
\label{eq:ez}
\end{equation}

When small density fluctuations ($\delta << 1$) are considered, 
all the non-linear terms with respect to $\delta$ and ${\bf u}$ can be ignored and, therefore, 
the above equations  can be written as 
\begin{equation}
{\partial^2 \delta\over \partial t^2}+2H(t){\partial
  \delta\over \partial t}=4\pi G\bar\rho\delta\,.
\label{eq:linear}
\end{equation}
This relation represents one of the most fundamental equations 
within the linear theory of gravitational collapse: 
it delineates the Jeans instability of a  fluid with no pressure
under the counter--effect of the cosmic expansion (represented by the 
$2H(t) {\partial \delta /\partial t}$ term). 
Since Eq.~\ref{eq:linear} is a second order differential equation in time $t$, 
its solution can be written as
\begin{equation}
\delta({\bf x},t) = \delta_+({\bf x},t_i)D_+(t)+\delta_-({\bf x},t_i)D_-(t)\,,
\label{eq:linear_solut}
\end{equation}
where $D_+(t)$ and $D_-(t)$ are, respectively,  the growing and decaying modes of 
$\delta({\bf x},t)$, and $\delta_+({\bf x},t_i)$ and $\delta_-({\bf x},t_i)$ the corresponding 
spatial distribution of the primordial matter field. 
Given that  the density growing modes only depend on time, 
the density fluctuations will evolve at the same pace throughout the cosmic volume. 
However, these density growing modes depend on the particular underlaying cosmology in such a 
way that, in different Friedmann--Lemaitre--Robertson--Walker (FLRW)
universes structures will grow in a different manner. 
 
As an example, in the case of a flat matter-dominated
Einstein--de--Sitter universe (EdS, $\Omega_m=1$, $\Omega_{\Lambda}=0$), given that
 $H(t)=2/(3t)$, $D_+(t)=(t/t_i)^{2/3}\propto a(t)$ and $D_-(t)=(t/t_i)^{-1}$. 
Therefore, in this particular case, cosmic expansion and gravitational instability proceed at the same rate.
Contrarily, it can be shown that,
in the case of a cosmological model with $\Omega_m<1$, such as a flat one with $\Omega_m=0.3$,  
there is an epoch, when the cosmological constant begins to be significant,  
at which the  characteristic time--scale of expansion 
turns out to be shorter than in the EdS case. As a consequence, after that epoch,  
cosmic expansion proceeds faster than gravitational collapse, generating 
a minor evolution in the number of collapse objects between $z\sim0.6$ and $z=0$ \citep[e.g.][]{Borgani_nature_2001}.
These results indicate that the observational determination
of the level of evolution of collapse regions (such as galaxy clusters) 
provides important constraints on cosmological parameters.
 
We can define the two--point correlation function of  $\delta({\bf x})$  
as $\xi(r)=\lb \delta({\bf x}_1)\delta({\bf x}_2)\rb$, which depends  only on the 
distance between the considered points, $r=|{\bf x}_1-{\bf x}_2|$. 
$\xi(r)$ describes whether the density field is more ($\xi(r)>0$) or 
less ($\xi(r)<0$) concentrated than the mean.  
In addition, a convenient description of $\delta({\bf x})$ is given by its Fourier representation
$\delta({\bf k})={(2\pi)^{-3/2}}\int d{\bf x} \,\delta({\bf x})e^{i{\bf k}\cdot {\bf x}}$.
If we also express $\xi(r)$ in Fourier space, it is easily demonstrable that  its 
Fourier  transform corresponds to the power spectrum of density fluctuations: 
\begin{equation}
P(k)=\lb |\delta({\bf k})|^2\rb ={1\over 2\pi^2}\int dr \,r^2 \xi(r) {\sin
kr \over kr}\,,
\end{equation}
which does not depend on the orientation of the wave-vector ${\bf k}$ but 
on its modulus.  

$P(k)$ is a fundamental quantity that provides a full  
statistical description of a uniform and isotropic Gaussian field.
Inflationary models   predict 
a perturbation power spectrum of the form $P(k)=A k^n$,
where $A$ is the normalization and $n$ the spectral index. 
More precisely, inflation
provides a nearly Gaussian density perturbation field characterized by a 
scale-invariant spectrum with $n\simeq1$ \citep[e.g.][]{Guth_Pi_1982},
which appears to be confirmed by measured CMB anisotropies 
\citep[e.g.][]{Planck_CMB_2013}.

A common practice in the analysis of cosmological density fields is that of
using filtering functions to define spatial scales. 
To analyze the collapse of primordial fluctuations on scales  $R\propto (M/\bar\rho)^{1/3}$, 
giving rise to objects of mass $M$, it is useful to  define a window function, $W_R(r)$, which filters out 
the modes on smaller scales, and the corresponding smoothed density field, 
$\delta_R({\bf x})= \delta_M({\bf x})=\int \delta({\bf y})W_R(|{\bf x}-{\bf y}|)\,d{\bf y}$.
If the Fourier transform of the window 
function\footnote{The functional form of the window function, which depends on the particular 
choice of filter, provides the 
connection between mass and smoothing scale. Two common filter functions are 
$W_R(k)={3[\sin(kR)-kR\cos(kR)]/ (kR)^3} $ and 
$W_R(k)=\exp\left(-{(kR)^2/ 2} \right)$ corresponding to 
the top--hat and the Gaussian windows, respectively. 
For each of these filters, the correspondent relation between mass and smoothing scale is
given by $M=(4\pi/3)R^3\bar\rho$ and $M=(2\pi R^2)^{3/2}\bar\rho$.} 
is $W_R(k)$, the variance of the perturbation field at the scale $R$ is given by
\begin{equation}
\sigma^2_R=\sigma^2_M = \lb \delta_R^2\rb = {1\over
  2\pi^2}\int dk \,k^2P(k)\,W_R^2(k)\,.
\label{eq:var}
\end{equation}
In principle, whereas the  functional form of $P(k)$ depends on  
$\Omega_m$, $\Omega_{\rm b}$, and  $H_0$ \citep[e.g.][]{Eisenstein_1999},
its  normalization,  which is related to $\sigma_R ^2$, 
needs to be determined by observations of the cosmic LSS  or of the CMB anisotropies. 
The most widely used parameter for this normalization is
$\sigma_8$, which is  the variance estimated within comoving spheres of   
radius $R=8\,h^{-1}{\rm Mpc}$, 
roughly matching the typical scale of massive clusters\footnote{
Early redshift surveys showed that
$\sigma \sim 1$ for spheres of $R=8\,h^{-1}{\rm Mpc}$ \citep[e.g.][]{Davis_1983}.}.
As we discuss in \S\ref{sec:mf}, an estimate of $\sigma_8$ is  
given by the halo mass function.

This linear approximation applies after recombination, while $\delta<<1$, to describe the evolution of 
density fluctuations  on an initial mass scale $M\magcir$ $M_J(z_{rec})$ $\sim 10^5 M_\odot$. 
However, this linear theory can not be used to study the growth of structures in the  strongly 
non-linear regime, where typical fluctuations reach amplitudes of about unity and
overdensities with $\delta>>1$ are plausible  (as an example,
a cluster of galaxies corresponds to $\delta  \magcir 100$).
In this case,  non-linear models or numerical simulations are required to solve the evolution.

\subsection{Non-linear evolution of spherical perturbations: the Spherical Top-Hat Collapse}
\label{sec:top_hat}

The only situation in which the non-linear evolution can be 
precisely calculated is the one addressed by the 
simple spherically symmetric collapse model \citep[e.g.][]{Gunn_Gott_1972, Bert_1985}.
This model resolves the evolution of a spherical density perturbation of radius $R$ into a virialized halo.
Initially, the spherical perturbation is assumed to have constant  
overdensity and, since it is expanding  with the background universe, 
null velocity at its border. 
The symmetry of this configuration allows us to treat the spherical perturbation as an 
isolated FLRW universe, meaning that we can describe the growth of the overdensity 
using the same Friedmann's equations as for cosmology.

For simplicity we consider that the background universe is described by
a close EdS model, in which the perturbation radius $R(t)$  behaves in the same way as the expansion factor.
Within such a model, after a short period of time
the growing mode will dominate the evolution of the  perturbation.
At the initial time $t_i$, by imposing the condition of null velocities at the edge of the spherical region,
the linear growing mode is given by $D_+(t_i)=(3/5)\delta(t_i)$ and, thus,  the corresponding density parameter is 
$\Omega_p(t_i)=\Omega(t_i)(1+\delta_i)$, where the suffix $p$ stands for the perturbation itself, and the other quantities 
for the unperturbed background universe. 

The perturbation will grow until reaching its maximum expansion at  a given turn-around time, 
$t_{ta}$. If at this moment the spherical perturbation  detaches from the expansion of the background 
and, instead, initiates to collapse, the structure will be formed.
By solving the Friedmann's equations it can be shown that,
if $\Omega_p(t_i)>1$, the perturbation will recollapse.
At $t_{ta}$, the corresponding perturbation overdensity  is given 
by\footnote{On the contrary, the linear--theory extrapolation 
to $t_{ta}$ yields  a smaller value: $\delta_+(t_{ta})\simeq 1.07$.}
$\delta_+(t_{ta})\simeq 4.6$.

After $t_{ta}$, the perturbation  decouples from the underlying cosmic expansion 
and recollapses, reaching an equilibrium  state at the time $t_{vir}$ at which the
virial condition between the kinetic $K$ and the potential $U$
energy of the perturbation is satisfied, that is, $U=-2K$.
Assuming energy conservation during the evolution into this equilibrium state, the virial condition
can be used to derived that
 $R_{ta}=2R_{\rm vir}$ and that the density at $t_{vir}$ is $\rho_p(t_{vir})=8\rho_p(t_{ta})$. 
Hence, the non-linear overdensity of the perturbation at the virialization is given by
\begin{equation}
\Delta_{vir}={\rho_p(t_{vir})\over \rho(t_{vir})}=18\pi^2\simeq 178\,.
\label{eq:vir}
\end{equation}
Despite the simplicity of the spherical collapse model, this result is quite encouraging since 
N-body simulations  find that
a density contrast of $\sim 100-200$ is quite successful in defining DM halos.
Indeed, in these simulations a common definition of halo mass is given by $M_{200}$,
defined as the mass enclosed by a sphere with an overdensity equal to $200$.

On the contrary, the linear--theory extrapolation predicts a smaller value 
for the required overdensity at the time of  collapse:
\begin{equation}
\delta_c=\delta_+(t_{vir})=\delta_+(t_{ta}) \left({t_{vir} \over t_{ta}} \right)^{2/3}\simeq 1.69\,.
\label{eq:virlin}
\end{equation}
As we will see later, this is a  key discriminant that determines  the 
halo mass function. 

The above equations, valid for an EdS cosmology, can be 
extended to any other cosmological model.
In this sense, the overdensity at virialization can be defined as $\Delta_c$ or as $\Delta_{vir}$ depending on
whether it is referred to the critical, $\rho_c(z)$, or to  the mean background matter density, $\rho_m(z)$.
These overdensities  relate to each other as $\Delta_{vir}=\Delta_c/\Omega_m(z)$
(see \citealt{Bryan1998} for an estimation of $\Delta_{vir}$ in open 
and flat $\Lambda CDM$ universes).  

\subsection{Non-linear evolution of real overdensities}
\label{sec:non_linear}

Given the  simplicity of the spherical collapse model, it is not adequate
to properly describe the non--spherical collapse of actual overdense regions. 
To this end, cosmological simulations \citep[see, e.g.][for details on the numerical techniques]{Dolag_2008_b, Borgani_2011}
are essential to deepen into  the main properties of the real gravitational collapse. 

\begin{figure}[!h!t!b!p]
\begin{center}
\includegraphics[height=13.8cm]{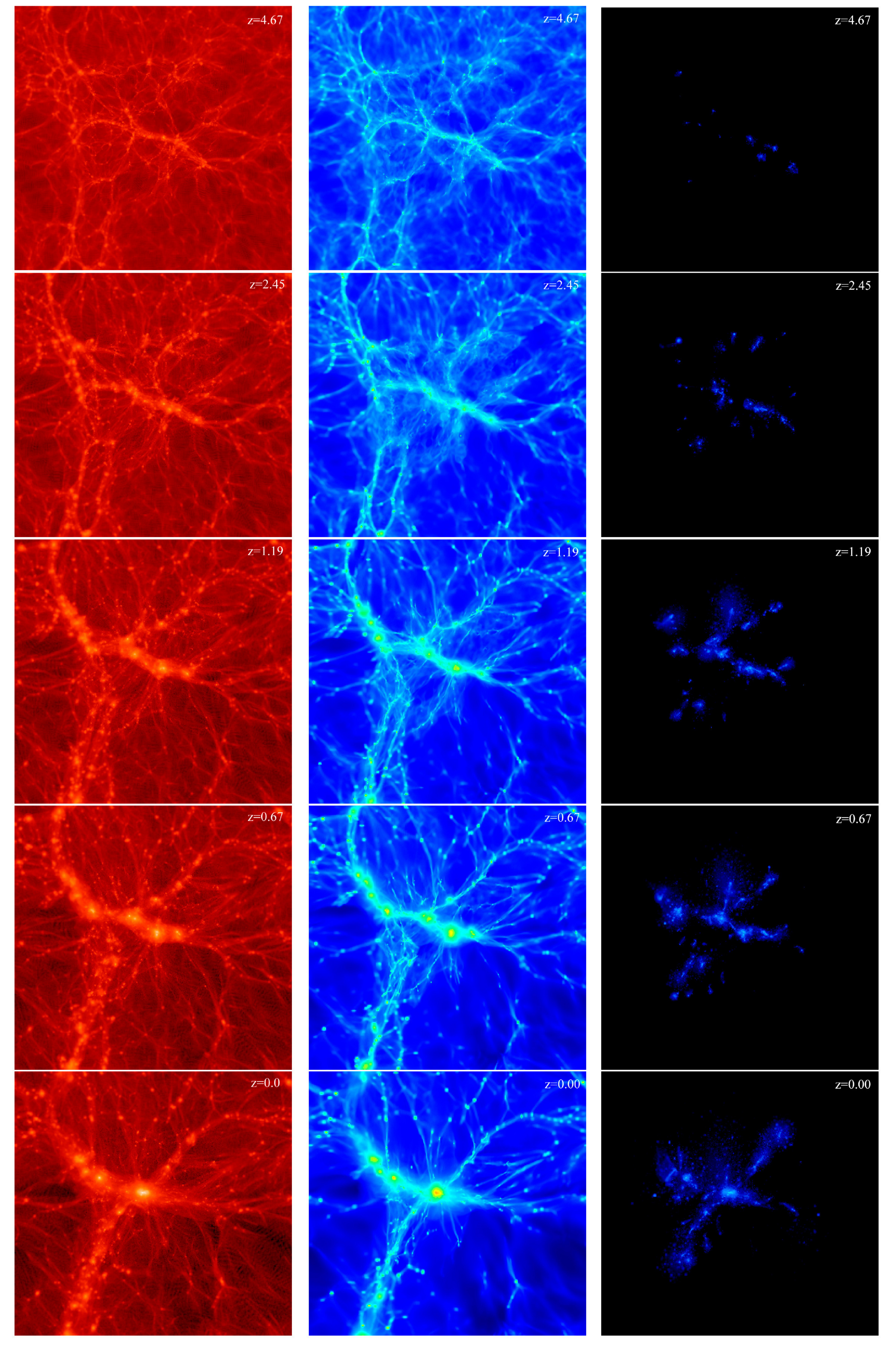}
\end{center}
\caption[Cluster formation in a cosmological context]
{Formation and evolution of galaxy clusters
  as described by a hydrodynamical simulation. Left,
  central and right columns show, respectively,  the evolution of the dark matter,
  gas and stellar densities from  $z\simeq 4$ (top panels) until $z=0$ (bottom panels). 
  At $z=0$, the biggest cluster formed has a virial mass 
  of $\sim 10^{15} M_{\odot}$ and a radius of $\sim 3$ Mpc. 
  The simulation was performed with the 
  Eulerian-AMR cosmological code MASCLET \citep{Quilis_2004}.
  Each
  panel is $64$ comoving Mpc length per edge and $5$ comoving Mpc depth.}
\label{fig:mapevol}
\end{figure}

As an example, the left panel of Fig.~\ref{fig:mapevol} 
displays the evolution of the DM density field from $z\simeq 4$ until $z=0$
as obtained from a cosmological hydrodynamical simulation. 
At early epochs, collapsed objects with low masses populate the proto--cluster region.
As the evolution proceeds, these objects merge into 
larger structures at later times. 
As can be inferred from this figure,  the actual collapse of overdense regions shows a number of  complexities
in comparison to the  top-hat collapse model: severe departures 
from spherical symmetry and constant density edges, 
filamentary matter accretion, and the existence of smaller  
overdensities within larger overdense already collapsing regions.
Given that different overdense regions have different spatial extensions 
and their time evolution proceeds differently,
the actual collapse of a cluster--scale overdensity  is a process prolonged 
in time \citep[e.g.][]{Diemand_2007}.
Besides, the non--linear  nature of this picture gives rise to merging 
events and interactions between overdensities at different scales, leading 
to an important matter redistribution within the considered regions.

Despite the complexities associated with the process of non-linear  collapse,
as the evolution proceeds, the resulting collapsing regions tend to reach an equilibrium state.
This state is described differently depending on the 
collisional  or collisionless nature of the considered component.
Indeed, the equilibrium state of the collisional gas component can 
be approximated by the condition of hydrostatic equilibrium 
(HE from now on), $\nabla\phi({\bf x}) = -\nabla p({\bf x})/\rho_{\rm gas}({\bf x})$,
under which pressure gradients and gravitational forces compensate each other. 
On the contrary, the equilibrium configuration for
the collisionless dark matter component  is provided by  the Jeans equation \citep{Binney_2008}.
With the additional assumptions of spherical symmetry and an ideal gas equation of state for the ICM gas, the resulting 
equations \cite[see, e.g.][for a complete description]{kravtsov_borgani12}
are commonly used to derive cluster masses \citep[e.g.][]{Ettori_2013}.
However, a number of processes, such as continuous matter accretion or merging events,  
can keep clusters away from equilibrium, introducing systematic uncertainties when applying the above assumptions.  

Given the complexities inherent to the non-linear process of halo formation,
the definition of a DM halo is not trivial and, as a consequence, 
there is no single definition that is agreed upon in the literature.
This incongruity has resulted in a number of different halo finding 
algorithms based on different halo boundaries and mass definitions 
\citep[see][for recent comparisons of different halo finders]{Knebe_2011, Onions2012, Knebe_2013}.
In this sense, two of the most widely 
used halo definitions are those based on the Friends-of-Friends  \citep[FoF;][]{Davis_1985} 
and the Spherical  Overdensity \citep[SO;][]{Lacey_1994} algorithms\footnote{While the FoF algorithm 
identifies DM halos with groups of DM particles separated by a distance shorter than a given linking  length
parameter,  the  SO algorithm  is based on the  mean overdensity  criterion.}.

\section{The halo mass function}
\label{sec:mf}

The halo mass function (HMF) is the 
number density of collapsed objects, at redshift $z$, with mass between $M$ and $M+dM$
in a given comoving volume.
While from an observational point of view  the HMF is  difficult to determine with high precision
\citep[e.g.][]{Rozo_2010}, it can approximately be  analyzed through 
analytic models \citep[e.g.][]{PS_74}, and
it is relatively simple to study by means of N-body cosmological simulations 
\citep[e.g.][for recent studies]{Cohn_2008,Tinker_2008, Crocce_2010, Courtin_2011, 
Bhattacharya_2011, Angulo_2012, Watson_2013, Murray_2013}. 

In this Section, after introducing  the HMF as originally derived by  \cite{PS_74}, we overview
how cosmological simulations are currently used to provide more precise calibrations
of this important prediction.

\subsection{The Press-Schechter approach}

\citet[][PS from now on]{PS_74}, based on the spherical collapse model, 
performed the first analytical attempt to derive the HMF.
The main idea of this formalism is that, any collapsed object with mass $\geq M$ by redshift $z$
stems from regions where $\delta_M\geq \delta_c$, being $\delta_M$ the linearly extrapolated density field 
(smoothed on a mass scale $M$), and $\delta_c$ the  critical overdensity for collapse. 
Motivated  by the spherical collapse model (see Eq.~\ref{eq:virlin}), 
$\delta_c\simeq 1.69$, being $z$-independent only in an EdS universe\footnote{In the general 
case, however, this overdensity depends weakly on redshift and cosmology
(for example, $\delta_c\simeq 1.675$ in a  $\Lambda CDM$ model at $z=0$).}.
Assuming a Gaussian distribution for the initial density fluctuations, the probability
of a given point to be within a region of scale $R$ satisfying the above conditions is: 
\begin{equation} 
F(M,z)={1\over \sqrt{2\pi}
  \sigma_M(z)}\int_{\delta_c}^\infty \exp\left(-{\delta_M^2\over
    2\sigma_M(z)^2} \right)\,d\delta_M\,,
\label{eq:frvol}
\end{equation}
where $\sigma_M(z)$ is the corresponding rms density fluctuation. 

From the above equation, the HMF is estimated as $\partial F(M,z)/\partial M$ 
(the fraction of independent regions evolving into objects  with  mass between $M$ and $M+dM$) 
divided by $M/\bar\rho$. An inherent implication of the PS approach is that, 
only overdense regions participate in the 
spherical collapse and, consequently, only half of the total mass content of the Universe is
considered. Thus, including a missing factor of 2, the PS mass function is given by:
\begin{eqnarray}
{dn(M,z) \over dM} & = &
{2\over (M/\bar\rho)}
{\partial F(M,z)\over \partial M} = \sqrt{2\over \pi}{\bar\rho \over M^2}{\delta_c\over
 \sigma_M(z)}\left| {d\log \sigma_M(z)\over d\log M}\right| \exp\left(-{\delta_c^2\over
2\sigma_M(z)^2} \right) =  \nonumber \\ 
 & = & {\bar\rho \over M}\psi(\nu)  \,,  
\label{eq:psmf}
\end{eqnarray}
which  only depends on the peak height $\nu\equiv\delta_c(z)/\sigma_M(z)$.

It has been shown that the functional form of $\psi(\nu)$ provided by the PS approach diverts 
significantly from the one derived from
cosmological simulations \citep[e.g.][]{ST_99, Jenkins_2001, Tinker_2010}.
In order to alleviate these discrepancies, a number of changes in the original 
model have been introduced \citep{Bond_1991, Lacey_1993}.
In this regard, the HMF has been analyzed accounting for the
ellipsoidal collapse \citep{Audit_1997, ST_99, ST_2001} or, 
within the excursion set theory, for non-gaussian primordial conditions \citep{Maggiore_2010}.
However, given the relatively simple assumptions on which these analytical prescriptions
rely, their accuracy to properly describe the HMF is limited.

\subsection{Using cosmological simulations to calibrate the halo mass function}

\begin{figure}
\includegraphics[width=6.1truecm]{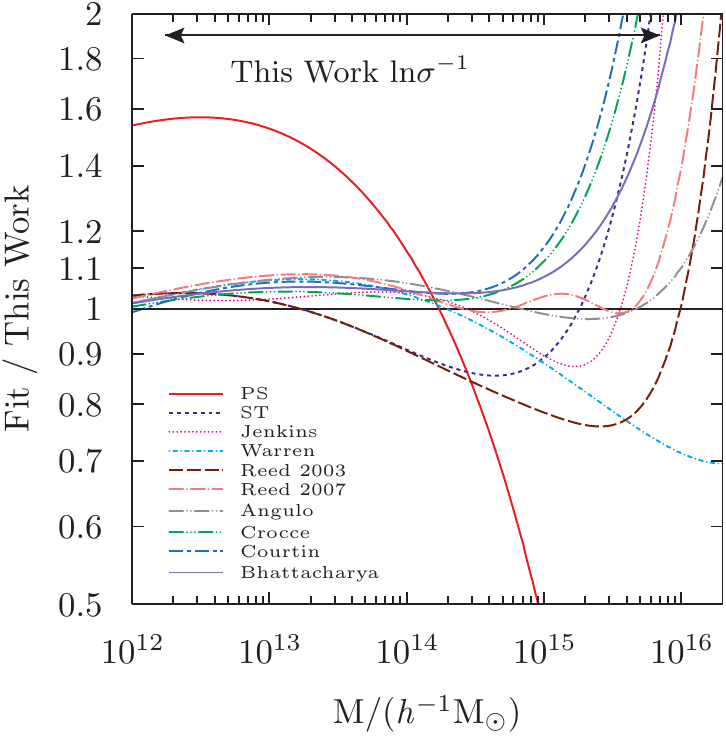}
\includegraphics[width=6.4truecm]{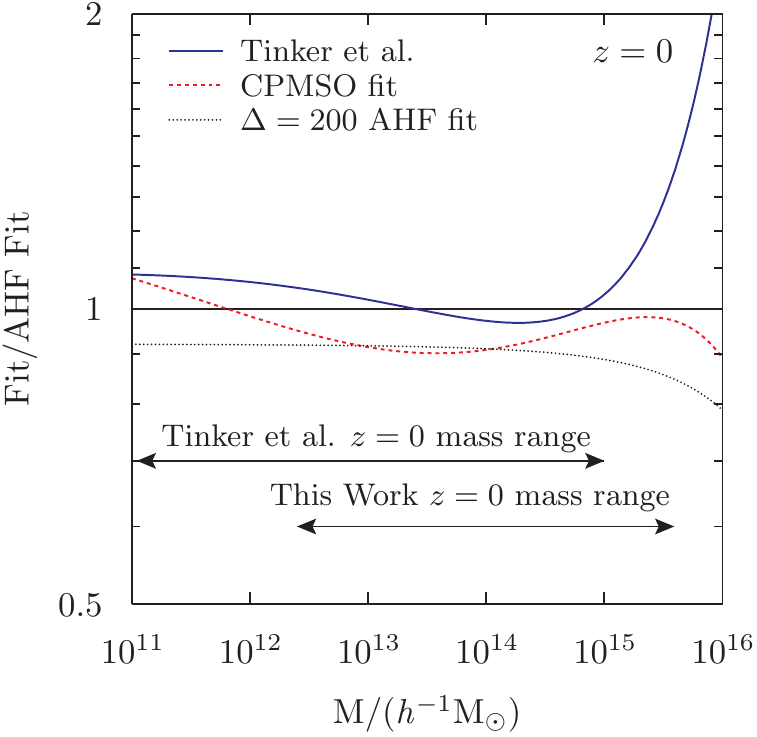}
\vspace{-0.4cm}
\caption{{\it Left panel:}  Comparison, at $z=0$, between the universal fit proposed by \citet{Watson_2013} 
for the FoF HMF and several FoF fits available in the literature.
{\it Right panel:} Comparison, at $z=0$, between the mass function obtained by \citet{Watson_2013}  with two different SO 
halo finders (AHF and CPMSO) and the fit by \cite{Tinker_2008}.
Figures from \citet{Watson_2013}. In both figures, the label `This Work'   refers to the work by \citet{Watson_2013}.}
\label{fig:mass_function}
\end{figure}

Cosmological simulations represent a powerful means to accurately calibrate the HMF
\citep[see][for a recent comparison of different HMF available in the literature]{Murray_2013}.

Given the exponential dependence of the HMF on mass and redshift, 
a precise calibration is extremely useful to place   
constraints on cosmological parameters \citep[e.g.][for recent reviews]{Allen2011, Weinberg_2012}. 
As an example, whereas at low redshift the mass function  of massive objects can be used to set  
limits on the combination of $\sigma_8$ and  $\Omega_{\rm m}$, 
the  redshift evolution of the HMF helps in breaking this degeneracy.

On the other hand, however, this exponential dependence makes 
the HMF very sensitive to the particular mass definition or to the small variations of 
$\delta_c(z)$ with redshift and cosmology. 
In this regard, many analysis of halo abundance by means of DM-only simulations 
consider that $\delta_c(z)$  is constant. 
As a consequence, they generally obtain a HMF that can be written as
an almost `universal' function of $\nu$, i.e., independent of redshift or cosmology 
\citep[e.g.][]{ST_99, Jenkins_2001, Evrard_2002, White_2002, Warren_2006, Tinker_2008, Crocce_2010,
Bhattacharya_2011, Courtin_2011}. 
However, small deviations in $\delta_c(z)$ can induce important changes in the  HMF and,  
therefore,  a precise calibration of the HMF requires to account for the correct dependencies 
of $\delta_c(z)$ \citep[e.g.][]{Courtin_2011}.

In the last years, in order to reach an accurate description of the shape of the HMF,
large N-body simulations have been used to evaluate the uncertainties induced by 
different redshifts, cosmological models and halo definitions. 
As an example, \citet{Watson_2013}, with a set of large  cosmological DM--only simulations, 
examined the redshift evolution (out to $z=30$) of the  HMF  and its dependence on  the FoF and SO halo mass definitions
\citep[see][for a comparison of the HMF provided by different halo finders]{Knebe_2011}.
In this work, they showed that the SO HMF  clearly evolves with redshift and obtained a  $z$-parameterized
fit  suitable for the whole redshift interval to within $\sim 20\%$. 
The right panel of Fig.~\ref{fig:mass_function} shows the comparison of several SO mass functions 
that they obtained at $z=0$.
On the contrary, a weaker $z$-evolution was found for the FoF HMF. 
In this case, as it is shown in the left panel of Fig.~\ref{fig:mass_function}, 
they obtained a `universal' fit  function\footnote{
The fit obtained for the HMF based on the FoF halos takes the form:
$f(\sigma)=A\left[\left(\frac{\beta}{\sigma}\right)^\alpha+1\right]e^{-\gamma/\sigma^2}$,
with $A=0.282$, $\alpha=2.163$, $\beta=1.406$, and $\gamma=1.210$. 
This fit holds for $-0.55 \le \mathrm{ln}\sigma^{-1} < 1.31$,  
corresponding to masses within $[1.8\times10^{12}$, $7.0\times10^{15}]h^{-1}\mathrm{M}_{\odot}$ at $z=0$.}
that agrees to within $\sim 10\%$ with a number of fits available in the literature at $z=0$,
a degree of deviation in accordance with the values reported in previous works
\citep[e.g.][]{Reed_2007, Lukic_2007, Tinker_2008, Courtin_2011}.

More recently, hydrodynamical simulations have demonstrated that 
baryonic cooling and heating processes can also affect the HMF
\citep[e.g.][]{Rudd_2008, Stanek_2009, Cui_2012, Cusworth_2013, Cui_2014}. 
In this regard, \citet{Cui_2014} recently found  that, in comparison to DM-only simulations,
hydrodynamical simulations accounting for cooling, star formation and SN feedback
produce an increase of the HMF, while simulations including as well AGN feedback tend to
reduce it by an overdensity-dependent  amount. 
This reduction is a consequence  of the changes that AGN feedback induces in gas density profiles and, therefore, in halo masses.
However, given our limited understanding of the physics of baryons, and  
in view of the large galaxy cluster surveys coming in the near future, 
the importance of this effect needs to be further and accurately investigated.

\section{Structure formation in the early universe}
\label{sec:early_universe}

The first objects in the Universe were expected to form in halos with $10^5-10^8$~M$_\odot$ at redshifts $10-30$. As discussed in the introduction, one may distinguish between the so-called minihalos, with virial temperatures of a few $\sim1000$~K and cooling via molecular hydrogen, and the so-called atomic cooling halos with virial temperatures of $10^4$~K. Depending on previous metal enrichment and the ambient radiation field, the latter may form the first galaxies or directly collapse into supermassive black holes. The main focus of this review will be on the latter scenario, and the reader is referred to the  reviews by \citet{Bromm09} and \citet{Bromm11} concerning the formation and properties of the first galaxies.

\subsection{Primordial star formation in the first minihalos}

A central ingredient for star formation in the first minihalos is their ability to cool via molecular hydrogen.  The latter 
is possible since H$_2$ may form even in primordial gas via the so-called H$^-$ channel
\begin{eqnarray}
\mathrm{H}+e&\rightarrow&\mathrm{H}^-+\gamma,\\
\mathrm{H}+\mathrm{H}^-&\rightarrow& \mathrm{H}_2+e
\end{eqnarray}
and the H$_2^+$ channel
\begin{eqnarray}
\mathrm{H}^++H&\rightarrow&\mathrm{H}_2^++\gamma,\\
\mathrm{H}_2^++H&\rightarrow& \mathrm{H}_2+H^+.
\end{eqnarray}
In both channels, electrons need to act as catalysts in order to promote H$_2$ formation, which is possible due to the 
electron freeze-out during cosmic recombination, leading to a ionization fraction of $\sim10^{-4}$ in the primordial Universe \citep{Peebles68}. As shown by \citet{Saslaw67}, the latter may promote molecular hydrogen formation in molecular clouds, allowing them to cool and collapse within a Hubble time \citep{Tegmark97}. The fact that the chemistry is out of equilibrium is thus crucial for the formation of the first structures, and detailed non-equilibrium calculations for the homogeneous Universe have been performed \citep[e.g.][]{Galli98, Stancil98, Puy07, Schleicher08a, Coppola12} to determine the chemical initial conditions for subsequent structure formation.

\begin{figure}
\centering
\includegraphics[scale=0.6]{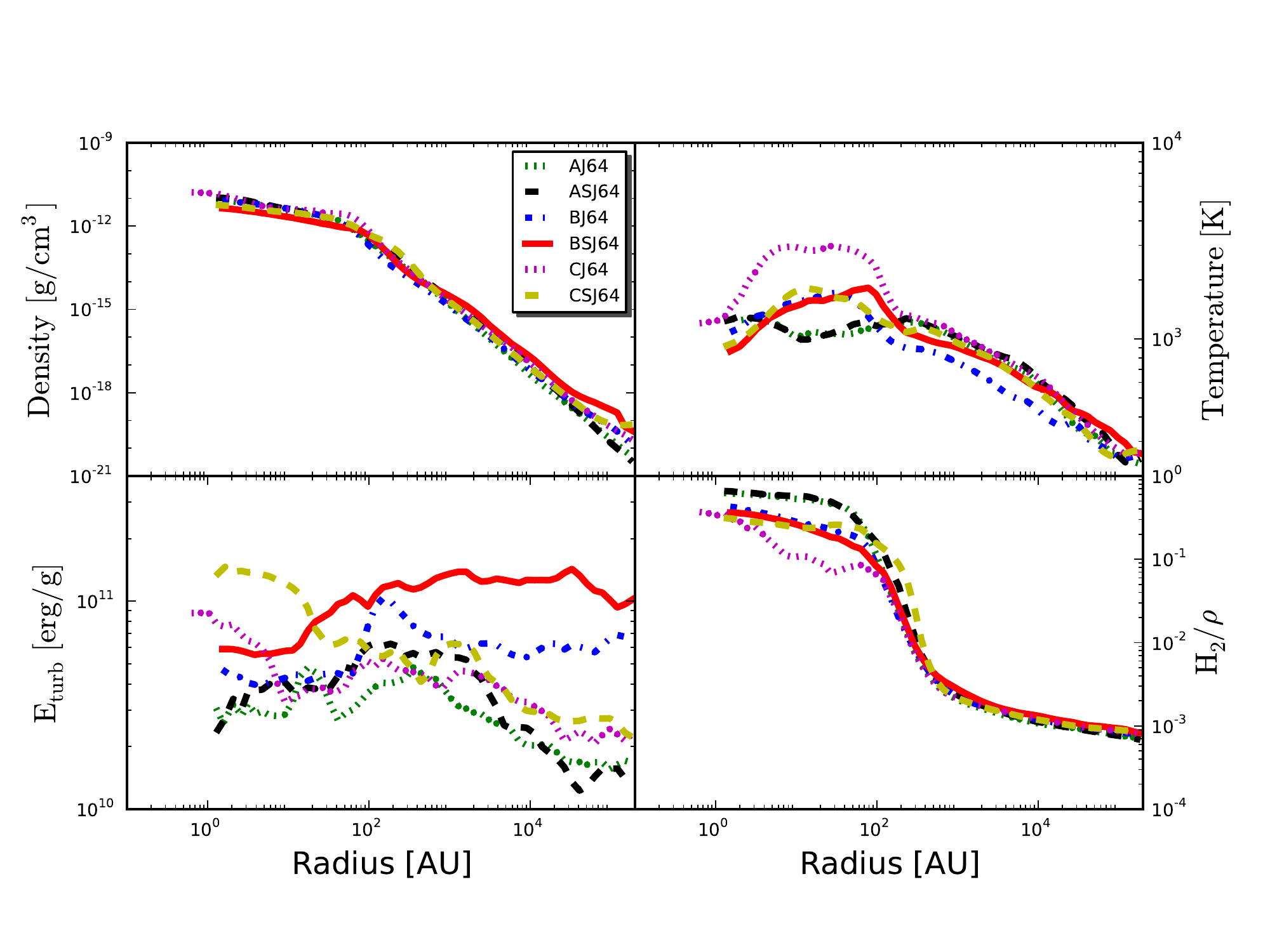}
\caption{Gravitational collapse in the first minihalos. Shown are radial profiles of the baryon density (upper left), gas temperature (upper right), specific turbulent energy density (lower left) and the H$_2$ abundance (lower right). The figure shows results for three different halos A-C with and without the turbulence subgrid-scale model. Figure by \citet{Latif13a}. } 
\label{fig:profiles_minihalo}
\end{figure}

The first cosmological simulations following the formation and subsequent collapse of the first minihalos, including detailed models for the primordial chemistry and cooling, have been performed by \citet{Abel02} and \citet{Bromm02}. These calculations were exploring the initial collapse phase where no fragmentation occurred. \citet{Yoshida06} were the first to incorporate a more detailed treatment of the microphysics at high densities in such simulations. In particular, at number densities of $\sim10^8$~cm$^{-3}$, three-body H$_2$ formation rates become relevant, which turn the hydrogen gas from a predominantly atomic into a fully molecular state. The dominant three-body reactions are given as
\begin{eqnarray}
\mathrm{H}+\mathrm{H}+\mathrm{H}&\rightarrow&\mathrm{H}_2+H,\\
\mathrm{H}_2+\mathrm{H}+\mathrm{H}&\rightarrow&2\mathrm{H}_2.
\end{eqnarray}

In this review, we illustrate the  collapse dynamics based on the recent simulation by \citet{Latif13a}. They employed the cosmological hydrodynamics code Enzo \citep{Shea04, Enzo13} including a detailed network for primordial chemistry  \citep{Abel97} with updated rates and cooling functions, as well as a subgrid-scale model for hydrodynamical turbulence \citep{Schmidt06}. Their simulation box had a total size of $300$~$h^{-1}$kpc with a root grid resolution of $128^3$, and two additional nested grids of the same resolution centered on the most massive halo. 
The simulation further employed $27$ adaptive refinement levels, ensuring a minimum resolution of $64$ cells 
per Jeans length.
While \citet{Truelove97} argued for a minimum resolution of 4 cells per Jeans length to avoid artificial fragmentation, more recent studies indicate a minimum resolution of 32 to 64 cells to capture the main properties of turbulence \citep{Federrath11, Turk12, Latif13c}.

\begin{figure}
\centering
\includegraphics[scale=0.25]{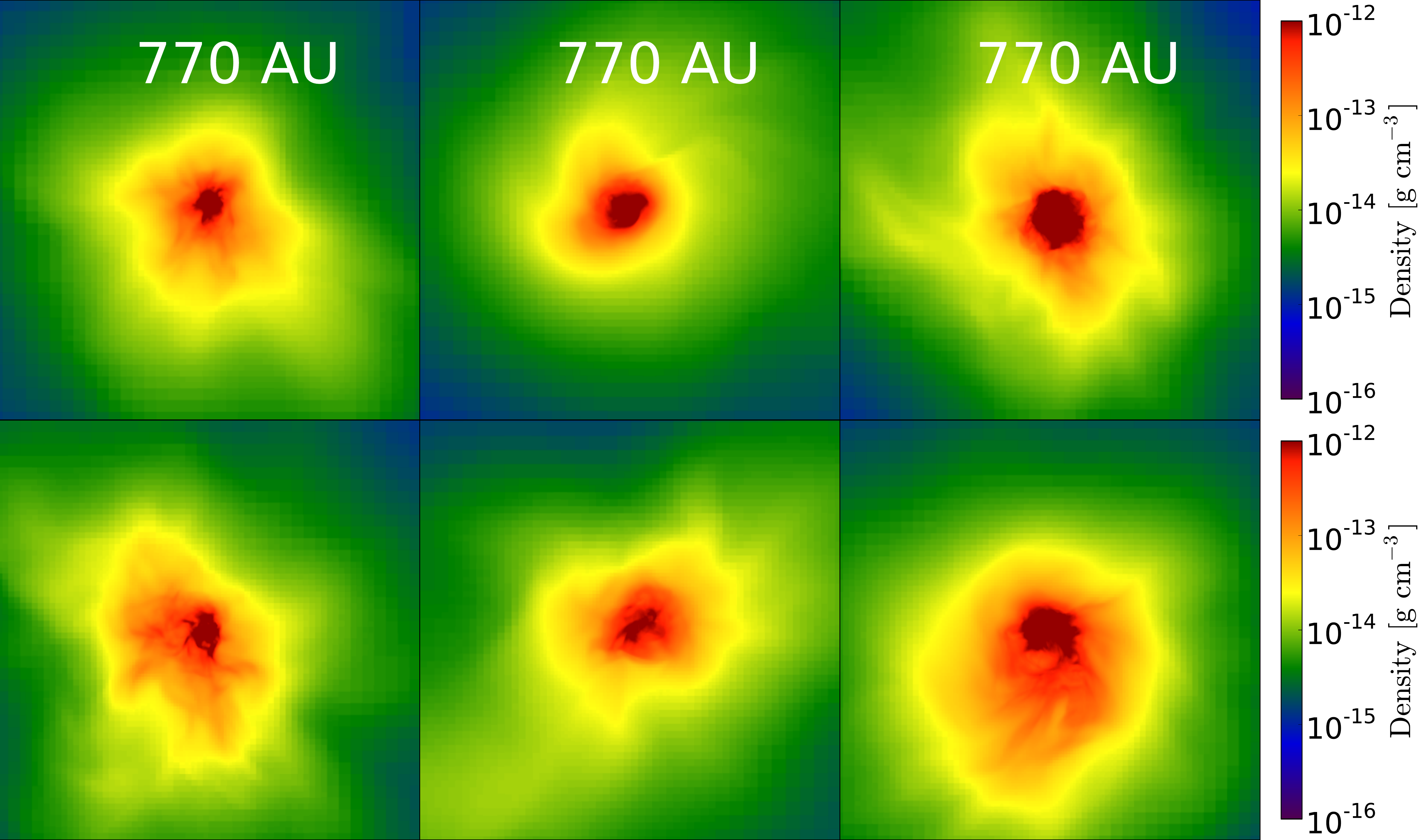}
\caption{Projections of the gas density in the central $770$~AU for three different halos. The top panel shows purely hydrodynamical simulations, while the simulations in the bottom panel include the subgrid-scale turbulence model by \citet{Schmidt06}. Figure by \citet{Latif13a}. } 
\label{fig:minihalo_density1}
\end{figure}

In Fig.~\ref{fig:profiles_minihalo}, we show the central region of this halo when the highest refinement level is reached. The gas density follows approximately an isothermal profile with $\rho\propto r^{-2}$, and flattens on scales comparable to the Jeans length at the density peak. The temperature increases towards smaller radii due to gravitational compression and increasing optical depth, and reaches a temperature of $\sim1000$~K at densities of $\sim10^9$~cm$^{-3}$ when three-body H$_2$ formation becomes relevant. The specific turbulent energy appears almost indepedent of scale, implying that turbulence is continuously re-generated during gravitational collapse via shocks and shear flows. The H$_2$ abundance is of the order $10^{-3}$ on larger spatial scales, and increases rather steeply at $\sim100$~AU when the gas becomes fully molecular as a result of the three-body reactions. The corresponding density projections in the central $770$~AU are given in Fig.~\ref{fig:minihalo_density1} for  three different halos in simulations with and without the turbulence subgrid-scale model. At this stage of the simulation, the central regions are approximately spherical with turbulent fluctuations in the density field. As also reported by \citet{Turk12}, there is no disk during this stage of the evolution.

\begin{figure}[h]
\centering
\includegraphics[scale=0.25]{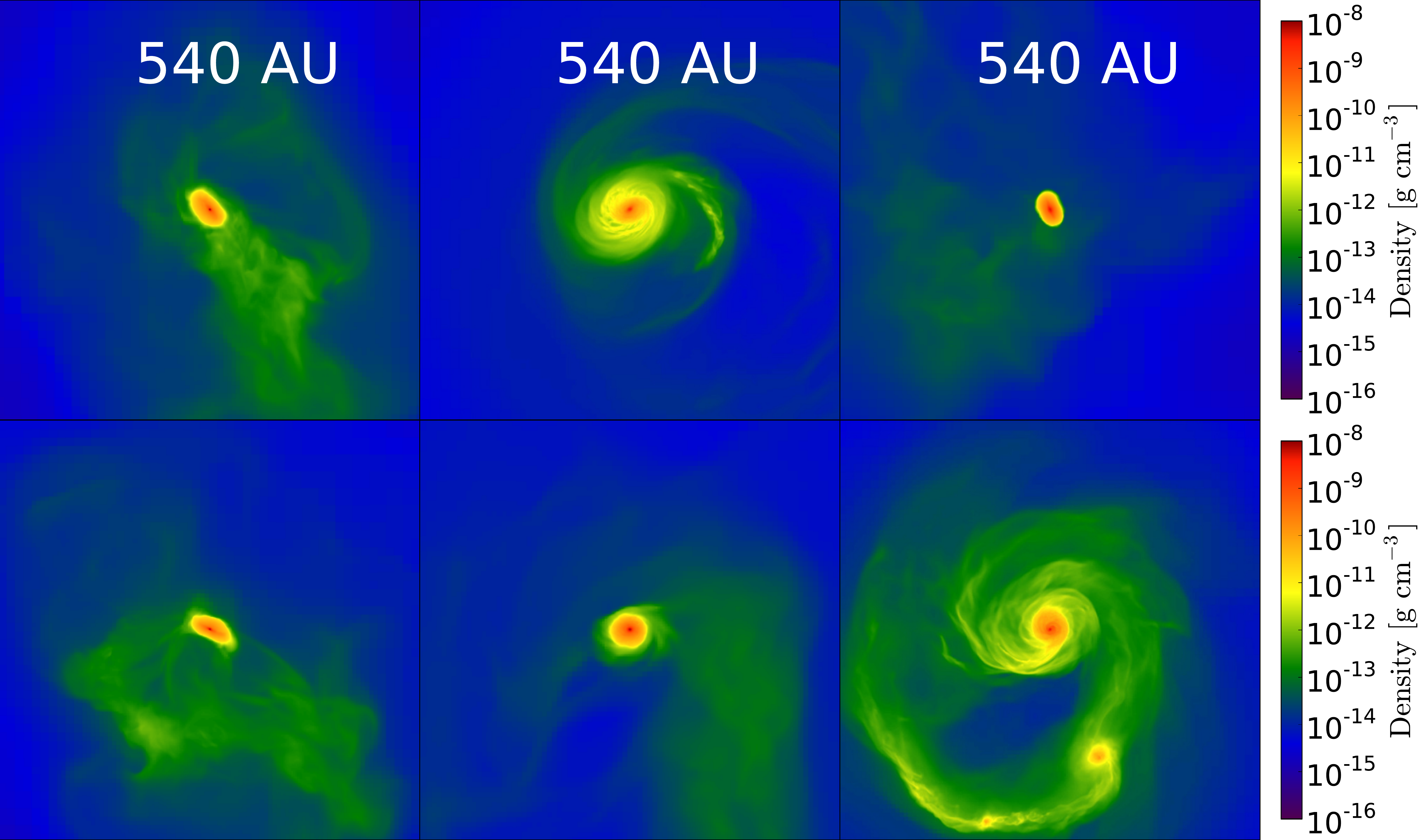}
\caption{Projections of the gas density in the central $540$~AU for three different halos approximately 40 years (5 free-fall times) after reaching the highest refinement level. The top panel shows purely hydrodynamical simulations, while the simulations in the bottom panel include the subgrid-scale turbulence model by \citet{Schmidt06}. Figure by \citet{Latif13a}. } 
\label{fig:minihalo_density}
\end{figure}

Following the evolution beyond the first peak is not straightforward, as the Jeans length always needs to be resolved with at least $4$ cells to avoid artificial fragmentation \citep{Truelove97}, thus requiring more and more refinement levels as collapse proceeds. The latter can be avoided by replacing the high-density gas with sink particles, or by turning on an adiabatic equation of state at high densities. \citet{Latif13a} followed the latter approach to pursue the subsequent evolution for five free-fall times. As shown in Fig.~\ref{fig:minihalo_density}, disk structures can be clearly recognized at the end of the simulation both in the simulations with and without the turbulence subgrid-scale model. In one of the simulations, fragmentation has already occured, while most of the runs show a central massive object with approximately $10$~M$_\odot$. Previous simulations employing a lower resolution per Jeans length indeed followed the evolution for longer times, indicating that fragmentation is expected to occur in the majority of these systems \citep{Stacy10, Clark11, Greif11}.

As for the properties of these disks at the final stage of the simulation, we note that high accretion rates of $10^{-3}-10^{-2}$~M$_\odot$~yr$^{-1}$ are found in the central $100$~AU, with considerable spatial fluctuations as expected in a real system. On the other hand, the rotational velocities range from a few up to $10$~km/s, while the radial infall velocities are of order $1$~km/s.

While most of the evolution depends only on the initial conditions and the subsequent dynamics, the three-body H$_2$ formation rate has been a major uncertainty until recently, as the rates derived by different groups 
\citep[e.g.][]{Palla83, Abel02, Flower07} showed differences by about three orders of magnitude, implying significant uncertainties for the densities at which the transition to the fully molecular state is expected to occur. A detailed description of these uncertainties was provided by \citet{Glover08}, and their implications in 3D simulations have been explored by \citet{Turk11}. Since recently, a new accurate determination of this rate is however available from quantum-molecular calculations by \citet{Forrey13} which considerably reduces the uncertainties discussed here. 

\begin{figure}
\centering
\vspace{0.7cm}
\includegraphics[scale=0.5]{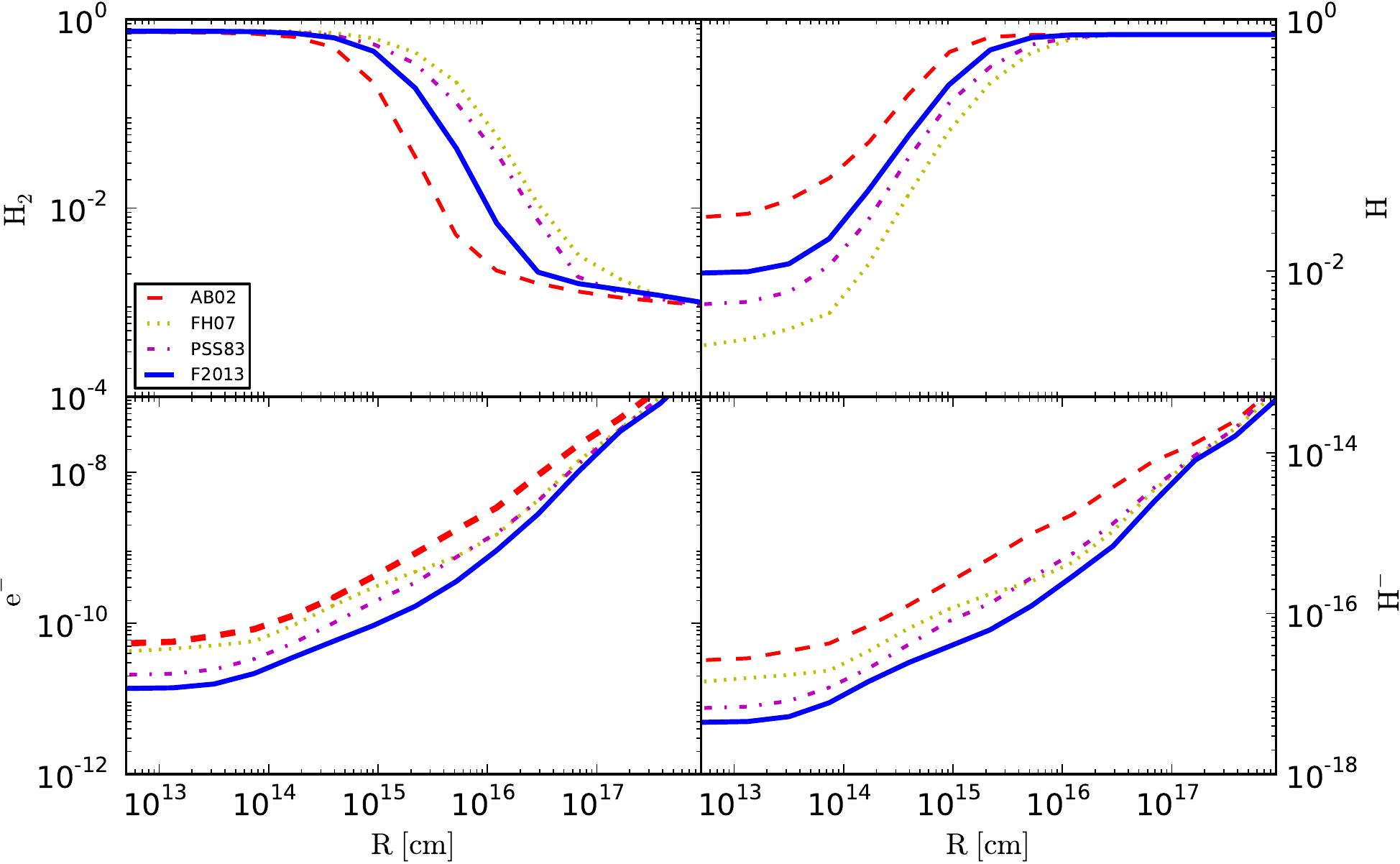}\\
\vspace{1.65cm}
\includegraphics[scale=0.5]{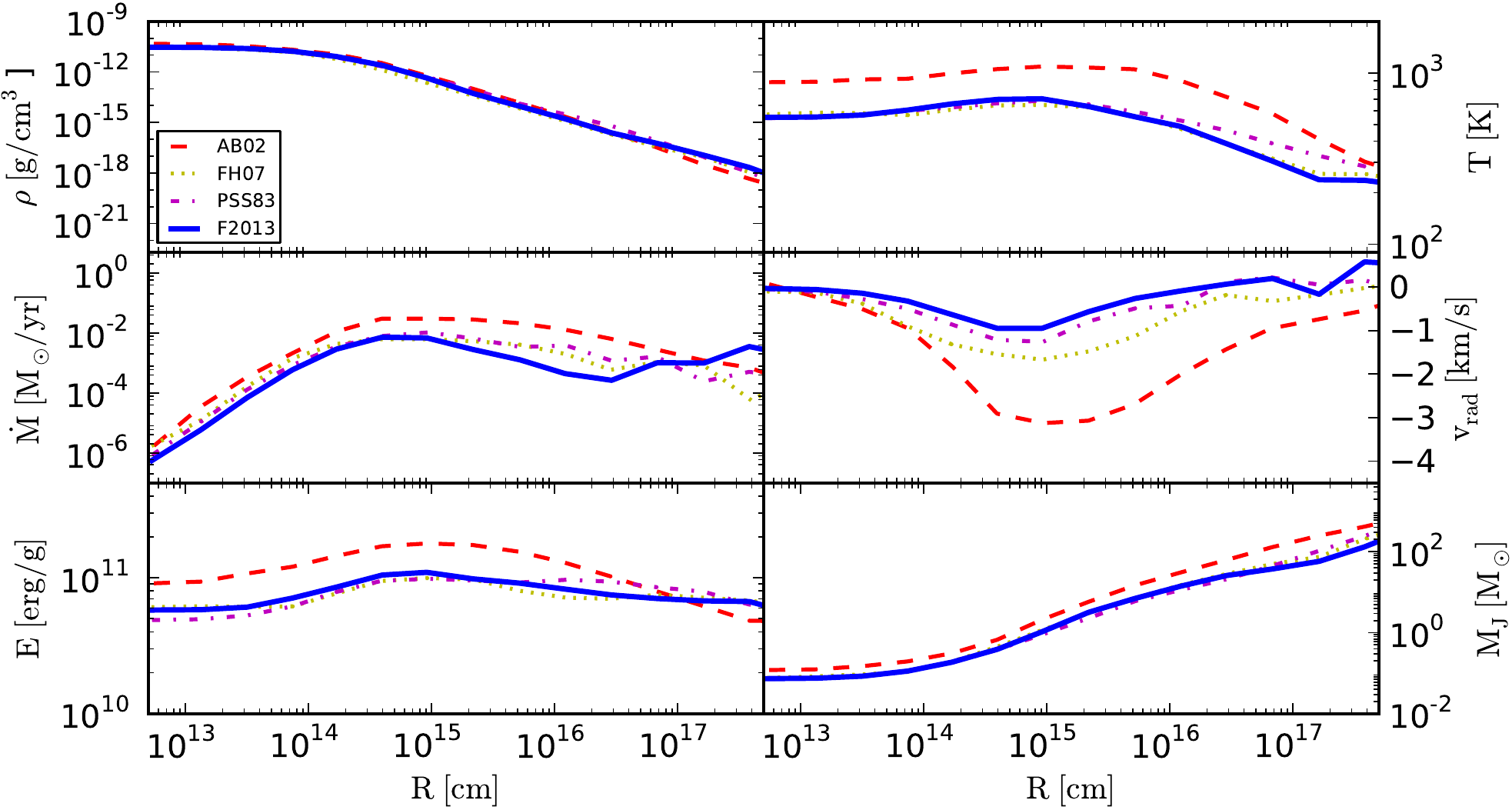}
\caption{A comparison of different three-body H$_2$ formation rates based on \citet{Palla83}, \citet{Abel02}, \citet{Flower07} and \citet{Forrey13}.  Figures by 
\citet{Bovino13b}.
{\it Top panel:}
Shown is the H$_2$ abundance (top left), the electron abundance (bottom left), the atomic hydrogen abundance (top right) and the abundance of H$^-$ (bottom right). 
{\it Bottom panel:} 
Shown is the gas density (top left), accretion rate (middle left), specific energy (bottom left), gas temperature (top right), radial velocity (middle right) and the Jeans mass (bottom right).} 
\label{fig:newrate}
\end{figure}

The impact of this rate compared to the previous rates employed in the literature has been explored by \citet{Bovino13b} in three cosmological halos 
using the chemistry package KROME\footnote{http://kromepackage.org/} \citep{Grassi_2013}. A representative example is given here in the top panel of Fig.~\ref{fig:newrate}. For the halo shown here, the resulting H$_2$ fraction lies inbetween the simulations based on \citet{Abel02} and \citet{Palla83}. We however note that in some cases, the result can be closer to one of these two. The same is also true for the abundance of atomic hydrogen. The abundances of the electrons and H$^-$ are lower compared to the other simulations, which is likely a result of the different dynamical evolution as a result of the new rates. We note that this behavior varies strongly from halo to halo without a clear trend.

The impact of the chemical evolution on the dynamics is illustrated in 
the bottom panel of Fig.~\ref{fig:newrate}. 
We note that central quantities like the gas density have a very small dependence on the rates considered here. The gas temperature shows only minor differences for most of the rates we explored, but the simulation based on \citet{Abel02} appears to have a clearly enhanced temperature compared to the other cases. The radial velocity depends rather sensitively on the adopted three-body rate and is again larger for \citet{Abel02}. The same is true for the resulting accretion rates. The Jeans mass, on the other hand, is very similar in the cases considered here, with a small enhancement for the rate of \citet{Abel02}\footnote{Based on the \citet{Forrey13} calculation, the chemical uncertainties are thus strongly reduced.}. We however note that relevant fluctuations in these quantities are still expected due to locally varying initial conditions.

We further note that the impact of additional physical processes can be expected to be relevant during primordial star formation. Magnetic fields are expected to form rapidly via the small-scale dynamo during the initial collapse \citep{Schleicher10b, Sur10, Schober12, Sur12, Turk12}, and subsequent ordering may occur as a result of large-scale dynamos in the accretion disk \citep{Pudritz89, Tan04, Silk06}. It thus needs to be determined whether jet formation can occur under realistic conditions \citep{Machida06}. Recent studies further indicate that radiative feedback can potentially set an upper limit on the stellar masses of order $50-100$~M$_\odot$ \citep{Hosokawa11, Susa13}, but more realistic 3D investigations are required for a final conclusion.

\subsection{Black hole formation in massive primordial halos}
The potential pathways to the formation of SMBHs were already sketched by \citet{Rees84}, and a detailed discussion would be largely beyond the scope of this review \citep[see][for a more general discussion]{Volonteri12}. Here, we consider the formation of massive black holes as one potential outcome of the gravitational collapse in massive primordial halos in the presence of strong photodissociating backgrounds that destroy the molecular hydrogen leading to an almost isothermal collapse regulated by atomic hydrogen.

Such scenarios were put forward by \citet{Koushiappas04}, \citet{Begelman06}, \citet{Spaans06}, \citet{Lodato07} and \citet{Schleicher10}
\citep[see as well][for an earlier work]{Bromm_2003_2}. The first numerical simulations following the gravitational collapse of these halos were reported by \citet{Wise08_2}, finding that gravitational instabilities may transport angular momentum and allow the formation of massive central objects. The thresholds required to fully dissociate the H$_2$ were carefully investigated by \citet{Shang10}, while \citet{Regan09b} pursued the first numerical simulation exploring the evolution beyond the formation of the first peak and confirming the formation of a disk at late times. While these simulations typically employed only a moderate resolution per Jeans length, \citet{Latif13c} demonstrated the appearance of extended turbulent structures once a high resolution per Jeans length is employed. Such turbulence can further aid the amplification of magnetic fields via the small-scale dynamo \citep{Schleicher10b, Latif13d}.

A central question is indeed how often the right ambient conditions exist in order to allow for such a close-to-isothermal collapse as required here. While a first investigation by \citet{Dijkstra08}  indicated that their abundance is likely sufficient to explain the observed population of SMBHs, more recent results \citep{Agarwal12, Johnson13_2}  suggest that appropriate conditions typically occur for at least one halo in a box of $1$~Mpc$^{-3}$. It is thus highly relevant to explore if such conditions lead to the formation of massive central objects.

\begin{figure}
\centering
\includegraphics[scale=0.62]{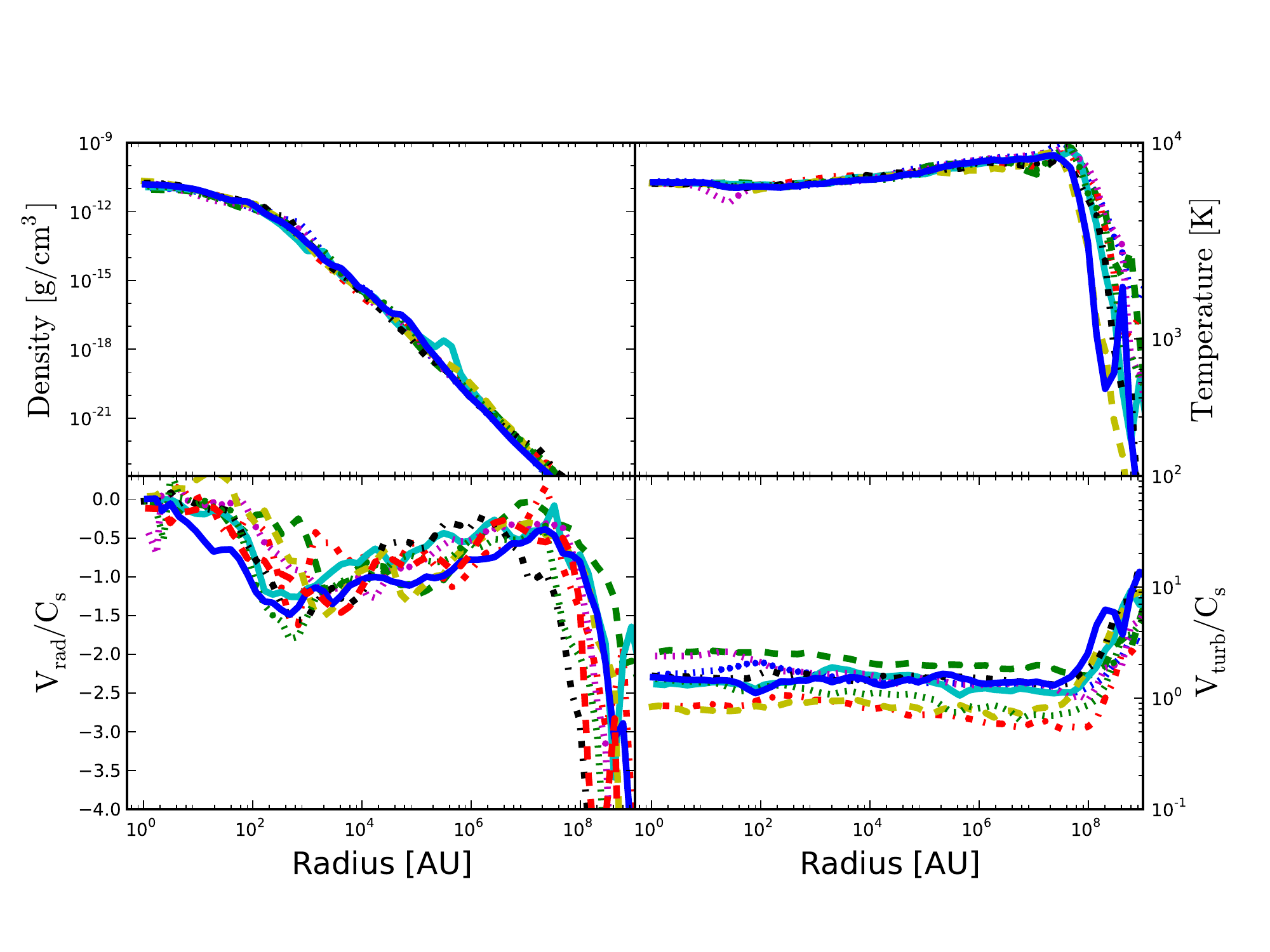}
\caption{The initial collapse in primordial massive halos. Shown are radial profiles of the gas density (top left), the radial velocity relative to the sound speed (bottom left), the gas temperature (top right) and the turbulent velocity normalized relative to the sound speed (bottom right). Figure by \citet{Latif13b}. } 
\label{fig:bhprofiles}
\end{figure}

The latter question was explored by \citet{Latif13b} with the cosmological hydrodynamics code ENZO \citep{Shea04, Enzo13}. Their simulation setup started from cosmological initial conditions at $z=100$ in a  box of $1$~$h^{-1}$Mpc, with a root grid of $128^3$, two initial nested grids of the same resolution centered on the most massive halo of $\sim10^7$~M$_\odot$, as well as $27$ additional refinement levels. The refinement level adopted here ensured a minimum resolution of at least $64$ cells per Jeans length. The simulations included primordial chemistry in the presence of a strong photodissociating background, parametrized via $J=J_{21}\cdot10^{-21}\ \mathrm{erg}~\mathrm{cm}^{-3} \mathrm{s}^{-1} \mathrm{Hz}^{-1} \mathrm{sr}^{-1}$, with $J_{21}=1000$. The simulations further included the subgrid-scale turbulence model by \citet{Schmidt06}. In order to follow the evolution beyond the first peak, the equation of state was adiabatic for densities higher than $10^{14}$~cm$^{-3}$. A total of nine such simulations has been pursued employing a different seed for the initial conditions. 

The initial state of the simulation when reaching the highest refinement level is shown in Fig.~\ref{fig:bhprofiles}. The gas density follows the expected isothermal profile with $\rho\propto r^{-2}$, which flattens on the highest refinement level on scales comparable to the Jeans scale. The gas temperature is shock heated to $\sim10^4$~K when falling onto the halo and then remains approximately constant during the collapse, as the atomic hydrogen cooling acts as a thermostat. The radial velocity normalized in terms of the sound speed increases during the initial infall, and subsequently becomes approximately constant. The infall appears enhanced on scales of $100$~AU, potentially due to the central mass, and decreases on smaller scales where the gas is not self-gravitating. Also the turbulent velocity appears approximately constant throughout the collapse, implying that turbulence is continuously regenerated via accretion shocks and shear flows.  

\begin{figure}
\centering
\includegraphics[scale=0.17]{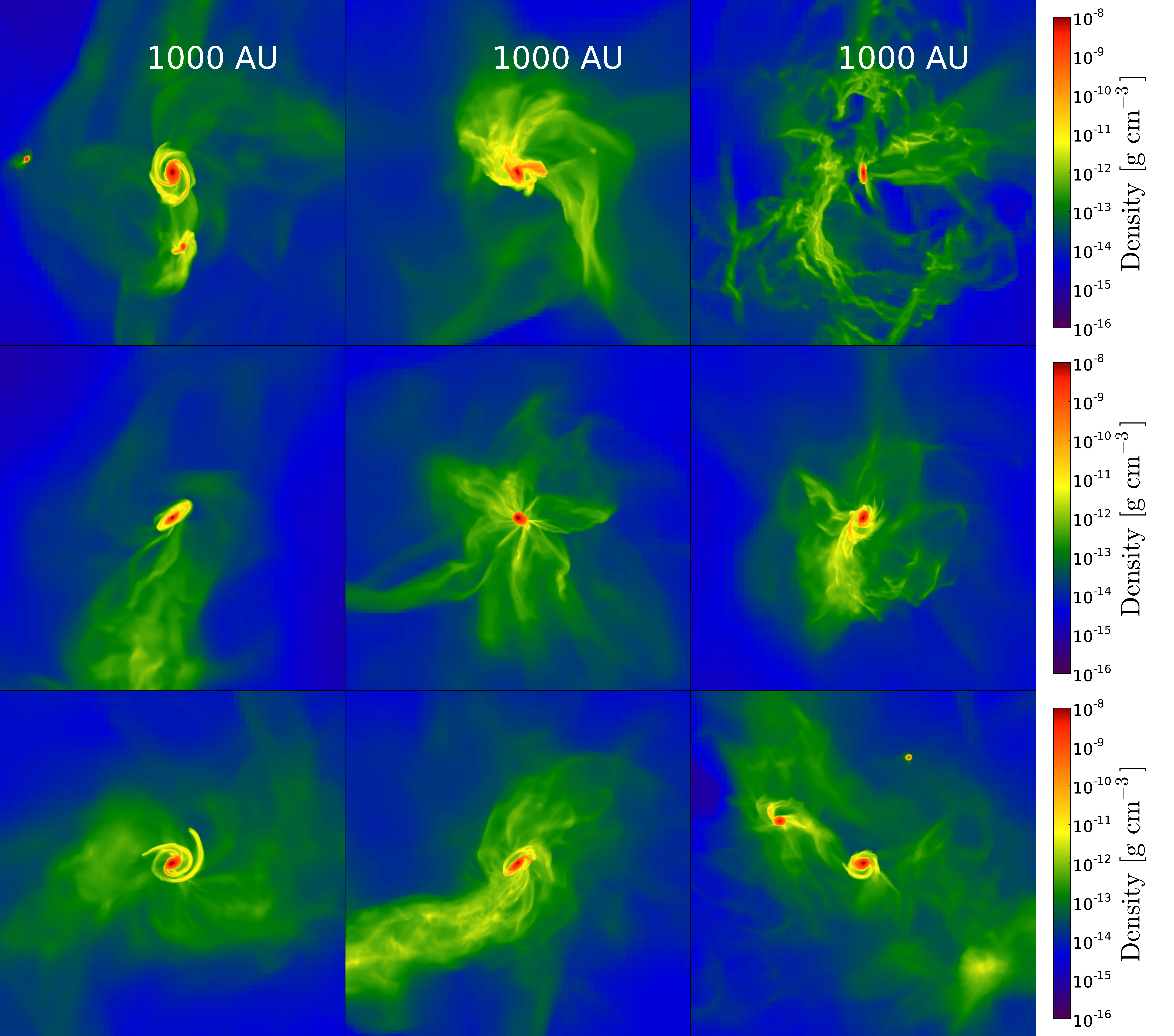}
\caption{Fragmentation in the central $1000$~AU for 9 different halos. Only 3 out of 9 halos fragment (one is not seen in this projection). Shown are density projections 4 free-fall times after the initial collapse. Figure from  \citet{Latif13b}. } 
\label{fig:fragmentation}
\end{figure}

\begin{figure}
\centering
\includegraphics[scale=0.62]{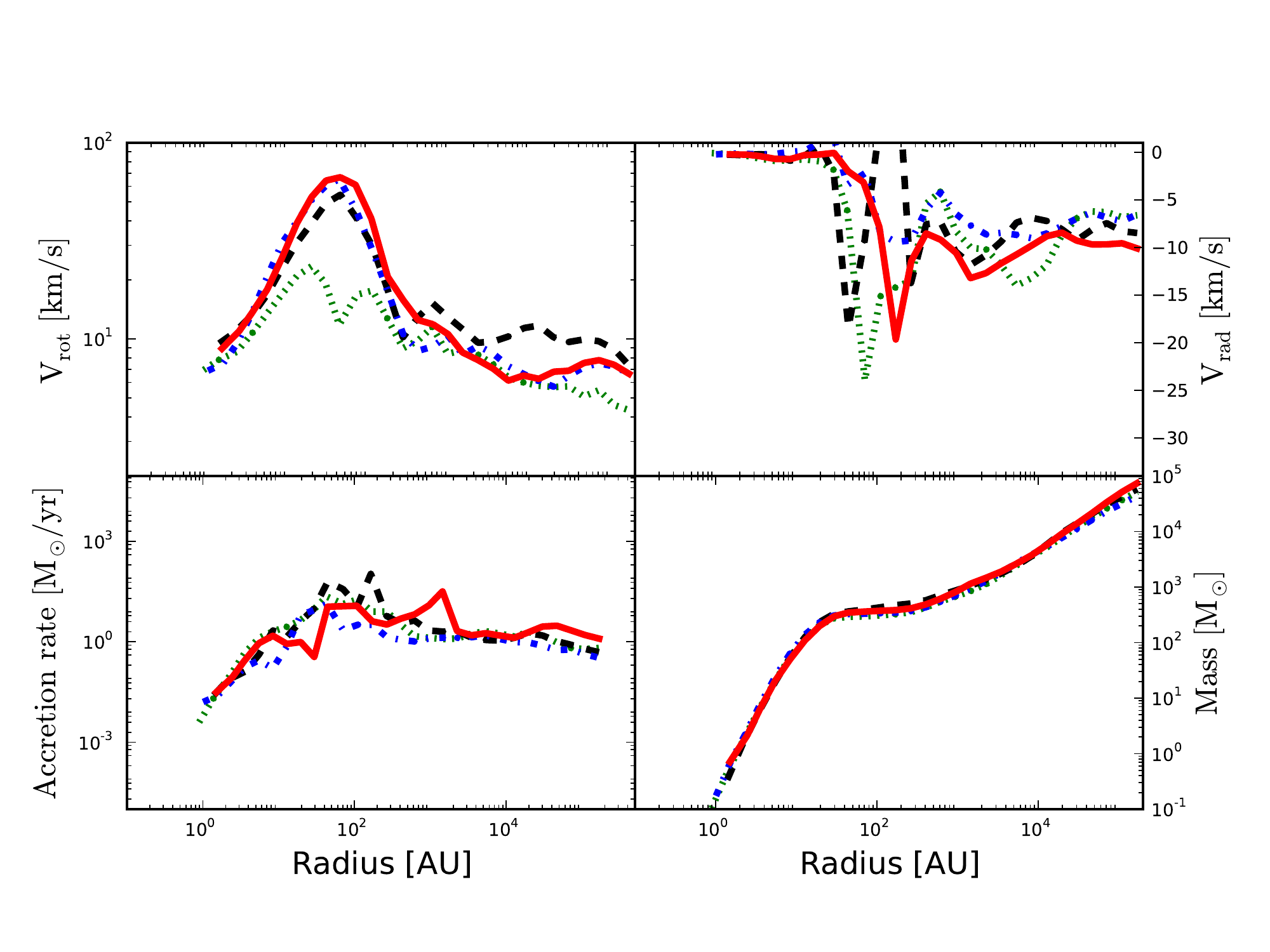}
\caption{Properties of the central region, 4 free-fall times after the initial collapse. Shown are radial profiles of the rotational velocity (top left), the accretion rate (bottom left), the radial velocity (top right) and the enclosed mass (bottom right). Figure from \citet{Latif13b}. } 
\label{fig:disks}
\end{figure}

In order to study fragmentation and the accretion onto the central object, we have followed the simulations for four free-fall times beyond the formation of the first peak. The resulting density distribution is shown in Fig.~\ref{fig:fragmentation} for nine different halos. In all cases, the central object has already reached a mass of $\sim1000$~M$_\odot$ within the central $30$~AU, with still ongoing and high accretion rates of $\sim1$~M$_\odot$~yr$^{-1}$. In these simulations, we find fragmentation in three out of nine halos (one is not visible in the projection given here). It is however possible that some of the clumps will subsequently merge, and also additional clumps may still form. The accretion occurs via self-gravitating disk structures surrounding the most massive objects. A more detailed investigation by \citet{Latif13b} further revealed that the presence of such self-gravitating disks occured only in simulations employing the turbulence subgrid-scale model, which provided additional support for the stability of these disks, while a more filamentary accretion mode occurs in purely hydrodynamical runs.  

The hydrodynamical evolution can be further assessed via Fig.~\ref{fig:disks}. The figure shows  the accretion rates of more than  $\sim1$~M$_\odot$~yr$^{-1}$. The enclosed mass scales as the radius on larger scales, corresponding to an isothermal profile, and varies more rapidly as $r^{3}$ on small scales where the density is approximately constant. The rotational velocity in the disk is of the order several $10$~km/s, while the radial velocity corresponds to $\sim10$~km/s. 
Similar results are also found by \citet{Regan_2014}.

For such accretion rates, stellar evolution calculations indicate that the resulting protostar behaves as a red giant, implying a highly extended envelope and a rather cool temperature of $\sim5000$~K \citep{Hosokawa12}. Considerations of the typical contraction timescales in the star indicate that such states can be maintained as long as the accretion rate is high, implying that radiative feedback is weak and leading to typical masses of $\sim10^5$~M$_\odot$ for the resulting protostars \citep{Schleicher13}. Even assuming that the protostars contract faster and reach the main sequence, the resulting feedback is however likely not sufficient to overcome the accretion rates \citep{Omukai02, Johnson13_1}. It therefore appears that the formation of very massive objects is feasible. These may collapse via general-relativistic instabilities to become SMBHs \citep{Shapiro86}.

The black holes forming in these early stages may later become the 
supermassive black holes observed at $z\sim 6-7$ 
\citep{Fan_2001, Fan_2003, Mortlock_2011}, the supermassive black holes observed in galactic 
centers \citep[e.g.][]{Magorrian_1998, Haering_2004}. The feedback 
of such black holes may have a significant impact both on the evolution 
of galaxies \citep{Somerville_2008, Silk_2013} as well as on the 
evolution in galaxy clusters. The latter will be explained in further 
detail in \S\ref{sec:thermo} of this manuscript.

\section{Galaxy clusters: Self-similar model}
\label{sec:self_similar_evolution}

Galaxy clusters, the largest and most massive objects in our Universe,  
form from the smaller units into a sequence of mergers. 
In this Section we briefly introduce the main predictions of the self-similar model   
\citep{Kaiser_86}  which, based on relatively simple assumptions, 
is able of estimating the main ICM properties and important correlations between them.

The self-similar model relies on several basic assumptions. 
On the one hand, an EdS background cosmology and a power--law shape for  the power spectrum of 
fluctuations are assumed. These conditions imply that there is not a characteristic scale of collapse.
On the other hand, since gravity is supposed to be the unique driver of halo collapse and gas
heating, there are not  additional characteristic scales introduced in the process of cluster formation.
This model also assumes clusters to have
spherical symmetry and to be in HE  (e.g. \citealt{Borgani_2008_b};
see also \citealt{kravtsov_borgani12} for a recent review and extensions of this model).

Under these assumptions,  the mass inside a spherical region of radius $R_{\Delta_{c}}$
enclosing a mean density equal to $\Delta_{c} \rho_{c}(z)$ at redshift $z$
is given by $M_{\Delta_{c}}=(4\pi/3)  \Delta_{c} \rho_{c}(z) R_{\Delta_{c}}^3 $, where 
$\rho_{c}(z)=\rho_{c0} E^2(z)$ is the critical cosmic density at $z$, 
being $E(z)$ given in Eq.~\ref{eq:ez}
and $\rho_{c0}$ the critical density at $z=0$.
Therefore, the cluster radius scales as
$R_{\Delta_{c}}\propto M_{\Delta_{c}}^{1/3} E^{-2/3}(z)$.
In addition, if the condition of HE is valid and the gas in clusters is distributed in a similar way 
to the DM, then it is satisfied that $k_BT \propto M_{\Delta_c}/R_{\Delta_c}$, which 
can be used to include in the above relation the dependence on the ICM temperature as
\begin{equation}
M_{\Delta_{c}}\,\propto \,T^{3/2}E^{-1}(z)\,.
\label{eq:mt_ss}
\end{equation}
This equation can now be used to derive additional scaling relations 
between different  X-ray observables. 
For instance, the X-ray luminosity goes like the product of the   
emissivity and the cluster volume, that is,
$L_X \propto \rho_{gas}^2 \Lambda(T, Z) R_{\Delta_c}^3$, where $\rho_{gas}$ is the gas density
and $\Lambda(T, Z)$, which depends on the gas temperature and metallicity, is the cooling function 
associated to a particular emission process.
The X-ray emission of the ICM plasma is mainly contributed by thermal  Bremsstrahlung and line emission.
At high temperatures ($T\magcir2\ keV$),  where 
thermal Bremsstrahlung dominates, the cooling function goes like
$\Lambda(T)\propto T^{1/2}$ \citep[e.g.][]{Sarazin_1986, Peterson_2006} and, therefore, 
$L_X$ scales with temperature as  
\begin{equation}
L_X\,\propto\,M_{\Delta_{c}}\rho_{c} T^{1/2}\,\propto \, T^2 E(z)\,.
\label{eq:lt_ss}
\end{equation}
For  $T\mincir2$\ keV, however, the dependence on temperature   
is more intricate because line emission becomes more important
than the free--free radiation and,  as a consequence, 
the above relation has to be adjusted  by accounting for the metal contribution.

An additional key quantity describing the ICM thermodynamics  
is the entropy \citep{Voit_2005} which is usually defined as
$K={\rm k}_{\rm B} T n_{\rm e}^{-2/3}$, with $n_{\rm e}$ being the electron number density. 
Therefore, for pure gravitational heating the entropy  scales as
\begin{equation}
K_{\Delta_{c}}\propto T E(z)^{-4/3}\,.
\label{eq:entr_ss}
\end{equation}

It is important to stress that the shape of the above relations and their  $z$--dependence 
are a natural consequence of both the particular assumptions of the self-similar model and  
the redshift dependence of $\Delta_{c} \rho_{c}(z)$ associated to the  assumed 
SO mass  definition \citep{kravtsov_borgani12}. 
Therefore, given that the standard cosmological model only introduces minor departures from self--similarity, 
observational deviations from these predictions
can be used to determine the effects of additional physical processes other than gravity.

In general, hydrodynamical simulations including only the effects of gravity
\citep[e.g.][]{Navarro1995, Eke1998, Nagai2007}
are able to reproduce the shape of the above X-ray scaling relations. 
However, as we discuss below, a number of  observations 
show some important deviations from these predictions, 
indicating an additional contribution from non-gravitational processes.

\subsection{Observational deviations from self-similarity}
\label{sec:deviations}

A number of X--ray observations point against the simple self--similar scenario,
especially at the scale of small clusters and groups. 
Indeed,  galaxy clusters observations have confirmed that,  
despite the simplicity and the important predictions provided by the adiabatic model
just described, there are still some important issues that deserve a further investigation: \\

\begin{itemize}

\item {\it \bf The cooling flow problem.}
At high temperatures, when the free--free radiation dominates the ICM X-ray emission,
the characteristic time scale of the gas to cool down can be written as
$t_{cool}\simeq 6.9\times 10^{10}{\rm yr}\,(n_e/10^{-3}{\rm cm}^{-3})^{-1}$
$(T/10^8 K)^{1/2}$ \citep[e.g.][]{Sarazin_2008}. 
According to this expression, as the gas temperature decreases, gas cooling becomes faster. 
In addition, given its dependence on the gas density, whereas 
$t_{cool}$ in outer cluster regions is usually longer than the age of the Universe, 
it can reach much shorter values ($t_{cool}\sim 10^8-10^9 \ yr$) in   
denser, central regions of cooling core clusters.  

Some of the main features of these cooling core clusters are the following
\citep[see][for a review and references therein]{Sarazin_1986}:
(i)  they have X-ray surface brightness with very high central values; 
(ii) their gas temperature, which is very low at the center, increases with radius;
(iii) within the cooling core regions,  $t_{cool}$ is lower than the Hubble time; 
and (iv) they show an increasing iron abundance towards the interior. 
These general features  are detected in a considerable number of clusters  
which are the so-called cool core clusters (CC). 
In general, observations report that these properties are most often found in dynamically relaxed 
clusters \citep[e.g.][]{Fabian_1994, Chen_2007}.
 
These observations led to the classical 
{\it cooling flow model} \citep[e.g.][]{Sarazin_1986, Sarazin_1988}.
This model is based on the assumption that there is not a heating source to compensate the gas radiative cooling.
Therefore, the gas cools and falls inward subsonically, eventually reaching very low central temperatures. 
When these temperatures go below the characteristic X-ray emitting values, 
strong emission lines are produced.   

However, the well-known cooling flow problem stems from the observation of several facts:
(i) the expected cooling rates predict emission lines stronger than observed;
(ii) the ratio between the central and mean cluster temperatures only remains at a factor $\thicksim$1/3 \citep[e.g.][]{Peterson_2003};
(iii) the mass deposition rates,  the amounts of cool gas and the
star formation rates are generally  smaller than estimated from the expected cooling rates \citep[e.g.][]{Edge_2003}.
These discrepancies, which imply the existence of a central heating source,
gave rise  to the well--known  {\it cooling flow problem}:  
which mechanism (or mechanisms) prevents the intra--cluster gas from cooling down to low temperatures?  \\

\item {\it \bf Self-similar scaling relations?} 
\begin{figure}
\centering
\includegraphics[width=7cm]{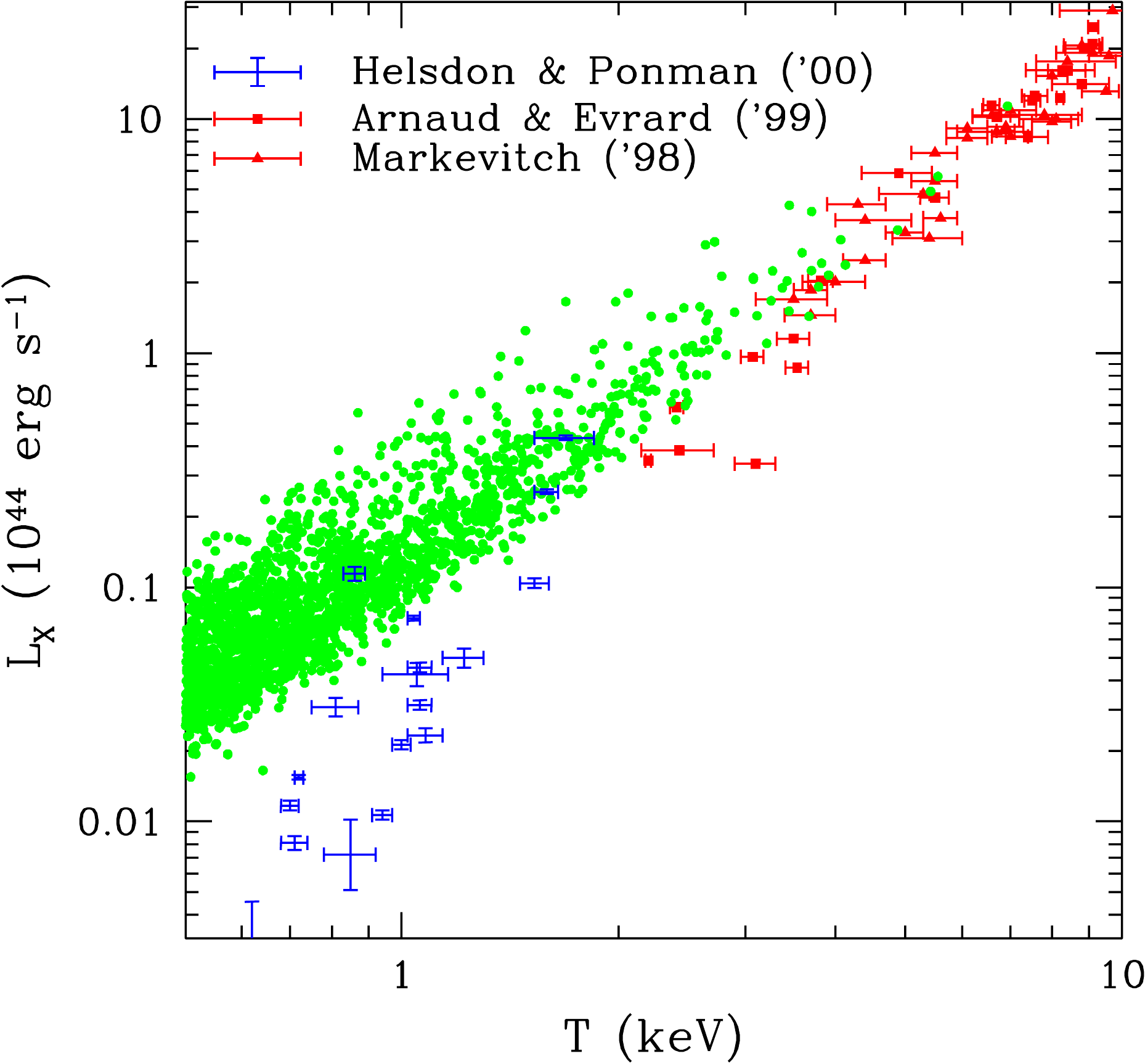}
\includegraphics[width=6cm,angle=270]{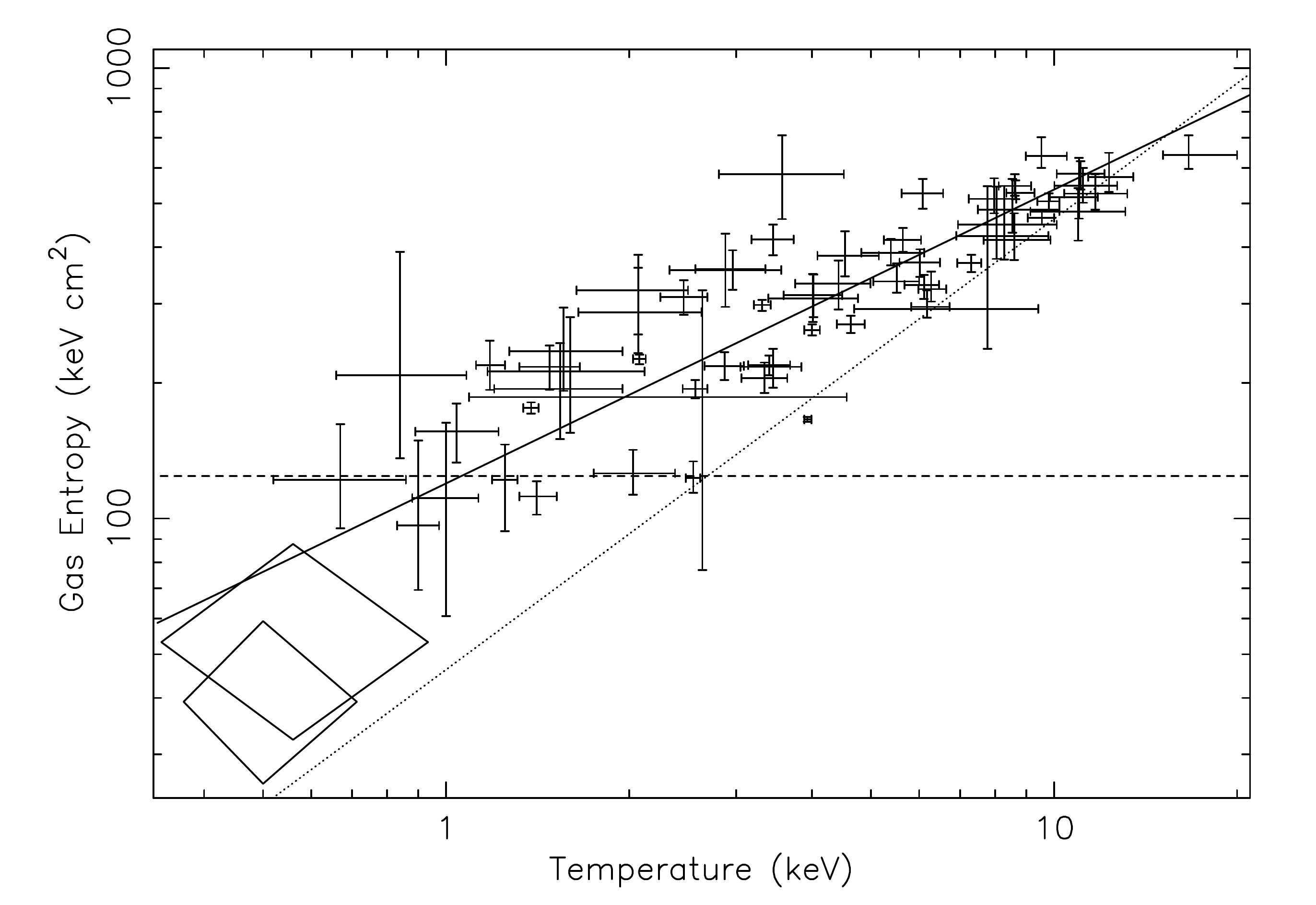}
\includegraphics[width=6cm,angle=90]{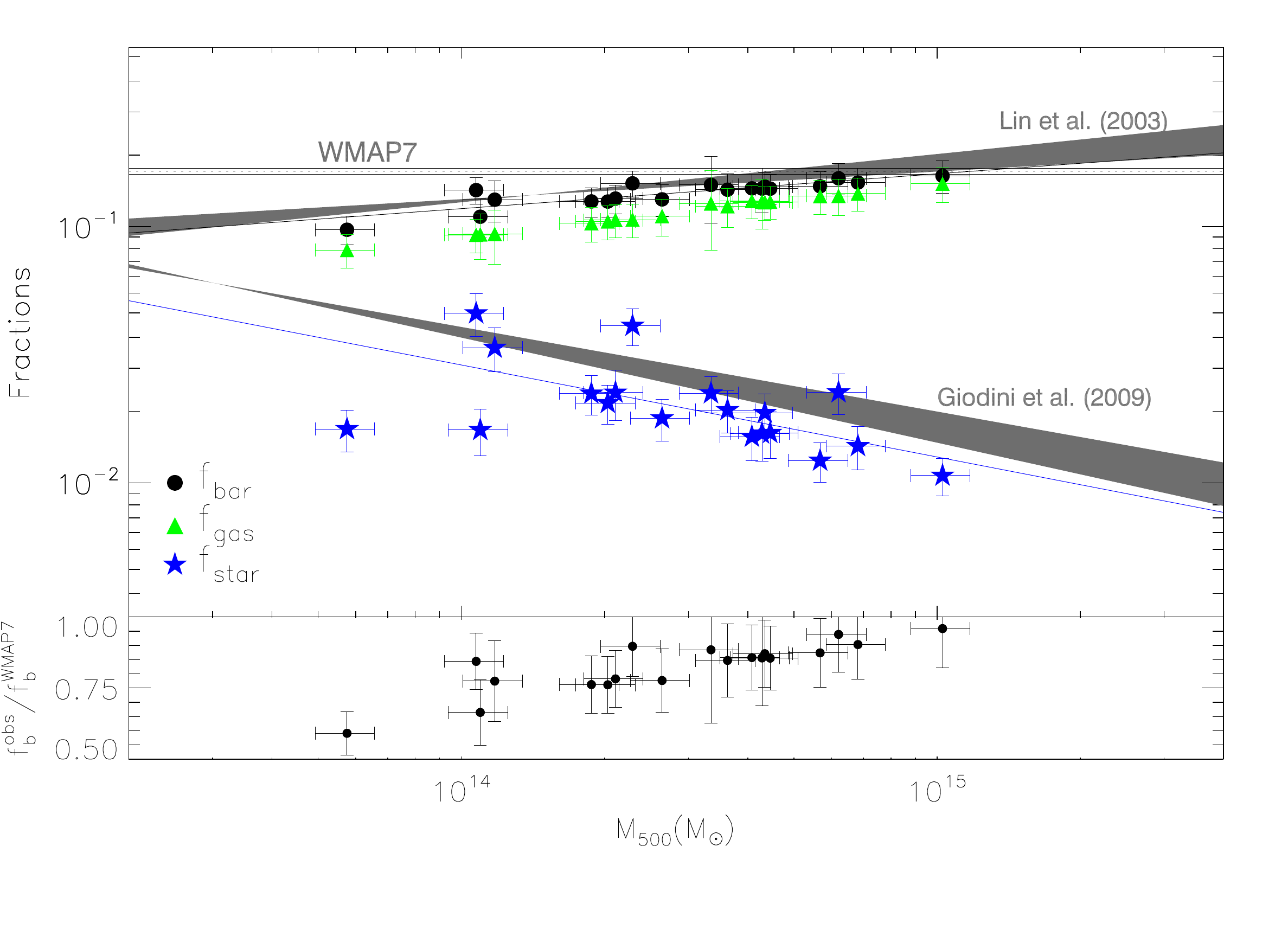}
\caption{ \textit{Top panel:} Comparison of the $L_{X}-T$ relation as obtained 
from different X-ray observational samples and from the radiative simulations
by \citet{Borgani2004} (green dots). These simulations assume null metallicity and 
include cooling,  star formation and SN feedback. Figure from \cite{Borgani2004}.
\textit {Middle panel}: Relation between gas entropy and
temperature for a sample of galaxy clusters and groups.
The solid line represents the observational relation,  
whereas the dotted line stands for the self-similar prediction. 
Figure from \citet{Ponman_2003}.
\textit{Bottom panel}: Observations of the baryonic, gas and stellar mass fractions 
as a function of total mass. 
The power--law fits obtained for the baryonic and stellar mass fractions 
are given by the black and blue solid lines, respectively.
For comparison, the corresponding  best-fits derived by \citet{Lin2003} and 
\citet{Giodini2009} are also included. 
The lower panel shows the ratio between observational and WMAP-7 baryon mass fraction. 
Figure from \citet{Lagana2011}.} 
\label{fig:obs_scaling}
\end{figure}
Contrary to what is expected from pure gravitational models,
early X-ray observations of  galaxy clusters 
demonstrated that  observational scaling laws do not scale self-similarly
\citep[see][for a recent review]{Giodini_2013}.

The first indication of self--similarity breaking was the
$L_X-T$ relation \citep[e.g.][]{Marke_1998, Allen_2001, Ettori_2004, 
Pratt2009, Maughan_2012}, for which observations report  
a steeper  slope ($\approx3$) than the expected self-similar value of 2
\citep[e.g.][]{Marke_1998, Osmond_2004}, a 
departure that  becomes even larger for systems with $k_{\rm B}T\lesssim3.5$ keV  
\citep{Ponman_1996, Balogh_1999, Maughan_2012}. 
In general, the observed $L_{X}-T$ relation shows a relatively large scatter 
\citep[e.g.][]{Pratt2009}, which is  mainly due to both the strong emission associated 
to CC clusters  and to unrelaxed systems for which the HE condition is not a 
good approximation \citep[e.g.][]{Maughan_2012}.
A common practice to reduce this scatter 
consists in excluding either cluster cores \citep[e.g.][]{Marke_1998, Pratt2009},
or CC systems \citep[e.g.][]{Arnaud_1999}.
As an example, the top panel of Fig.~\ref{fig:obs_scaling} shows a comparison
of the luminosity-temperature relation as obtained from different X-ray observational samples
\citep{Arnaud_1999, Marke_1998, Helsdon_2000} and 
from the radiative simulations with SN feedback  by \citet{Borgani2004}.  
As we can see, numerical results are consistent  with observations for 
$k_B T \magcir 2 \, keV$.  However, the agreement is not so encouraging at $k_B T \mincir 2 \, keV$,   
where simulations may include additional or more efficient feedback processes to further reduce the X-ray emission. 

Consistently with the $L_X-T$ scaling,  the relation between X-ray luminosity and mass is also steeper than expected
from self--similarity \citep[e.g.][]{Reiprich_2002, Chen_2007}.
As shown in the bottom panel of Fig.~\ref{fig:obs_scaling}, a possible reason for 
this may be related  with the observations of
an  increasing trend of the gas  mass  fraction with total mass  
\citep[e.g.][]{Balogh_2001, Lin2003, Sanderson_2003, Vikhlinin2006, Pratt2009, Dai2010, Lagana2011}.
The lower  gas content observed in low mass systems would generate a  lower X-ray emission  and,  
therefore,  a steepening of the scaling laws. 

As can be inferred from the  middle panel of  Fig.~\ref{fig:obs_scaling},
observations of central regions of small galaxy clusters and groups
report higher entropy values than expected from pure gravitational predictions, 
thereby generating a  flattening of the observed $K-T$ relation \citep{Ponman_2003}.
This increase of the gas entropy in low mass systems prevents the gas from falling 
into the center, reducing the central amounts of gas and, as a consequence, leading to the 
steepening of the observed $L_{X}-T$ relation 
\citep[e.g.][]{Evrard_1991, Tozzi_2001, Borgani_2001, Voit_2002}.

As for the total mass-temperature relation,  
observations generally report a self-similar behavior for massive galaxy clusters
\citep[e.g.][]{Arnaud_2005, Vik_2009}, whereas a slightly steeper slope 
is obtained for  smaller objects (e.g. \citealt{Arnaud_2005, Sun2009};
see also \citealt{Kettula_2013} for a recent weak lensing calibration of this relation). 
In addition, given its small intrinsic scatter, which 
is mainly due to the existence of substructures \citep[e.g.][]{Ohara_2006, Yang_2009},
the $M_{\rm tot}-T$ relation turns out to be extremely useful for
cosmological applications with galaxy clusters.\\

\item {\it \bf Temperature and entropy radial profiles.}
Independent X-ray  observations have confirmed that the ICM temperature distribution  is not  isothermal.
Indeed, as it is shown in the left panel of Fig.~\ref{fig:profiles}, the cluster temperature radial profiles are characterized by 
negative gradients at $r\gtrsim 0.1R_{180}$ \citep[e.g.][]{Piffaretti_2005, Pratt_2007}, being
nearly self--similar out to the most external regions  \citep[e.g.][]{Marke_1998, degrandi_2002, Vik_2005}.
However, the shape of these profiles in inner cluster regions 
partially depends on the dynamical state of the considered systems:
dynamically relaxed clusters generally show decreasing temperature profiles
towards the core, while unrelaxed systems show, instead, different patterns.
In principle, while gas radiative cooling might be able to generate 
low temperatures in the central regions of CC clusters, some source of energy feedback  
may avoid  an excess of cooling, thus reducing the  resulting star formation rates in these systems.
 
As for the entropy radial profiles,  pure gravitational models predict that, 
in outer cluster regions, entropy scales  with radius as $K\propto r^{1.1}$ \citep{Tozzi_2001}. 
However, recent observations \citep[e.g.][]{Simionescu2011, Walker_2012, Eckert_2013_1}
have reported  that these profiles flatten in central cluster regions
and their dependence on radius at larger radii is gentler than predicted (see right panel of Fig.~\ref{fig:profiles}). 
The particular radius at which the profiles  become flat depends on a number of factors, 
being shorter for CC than for non cool--core  (NCC) clusters \citep[e.g.][]{Sanderson_2009, Pratt_2010}.
Moreover, while high--mass clusters show a higher mean core entropy \citep[e.g.][]{Cavagnolo_2009}
and nearly self--similar profiles in outer regions, smaller systems are characterized by shallower profiles. 
This distinction between low and high mass systems, which  is a clear indication  of the
higher efficiency  of non-gravitational processes on smaller objects 
\citep[e.g.][]{Cavagnolo_2009, Pratt_2010, Maughan_2012},
highlights that entropy profiles provide an important fingerprint of the different physical 
processes breaking cluster self-similarity \citep[e.g.][]{Voit_2002, Voit_2003}.

\begin{figure}
\centering
\includegraphics[width=5.5cm]{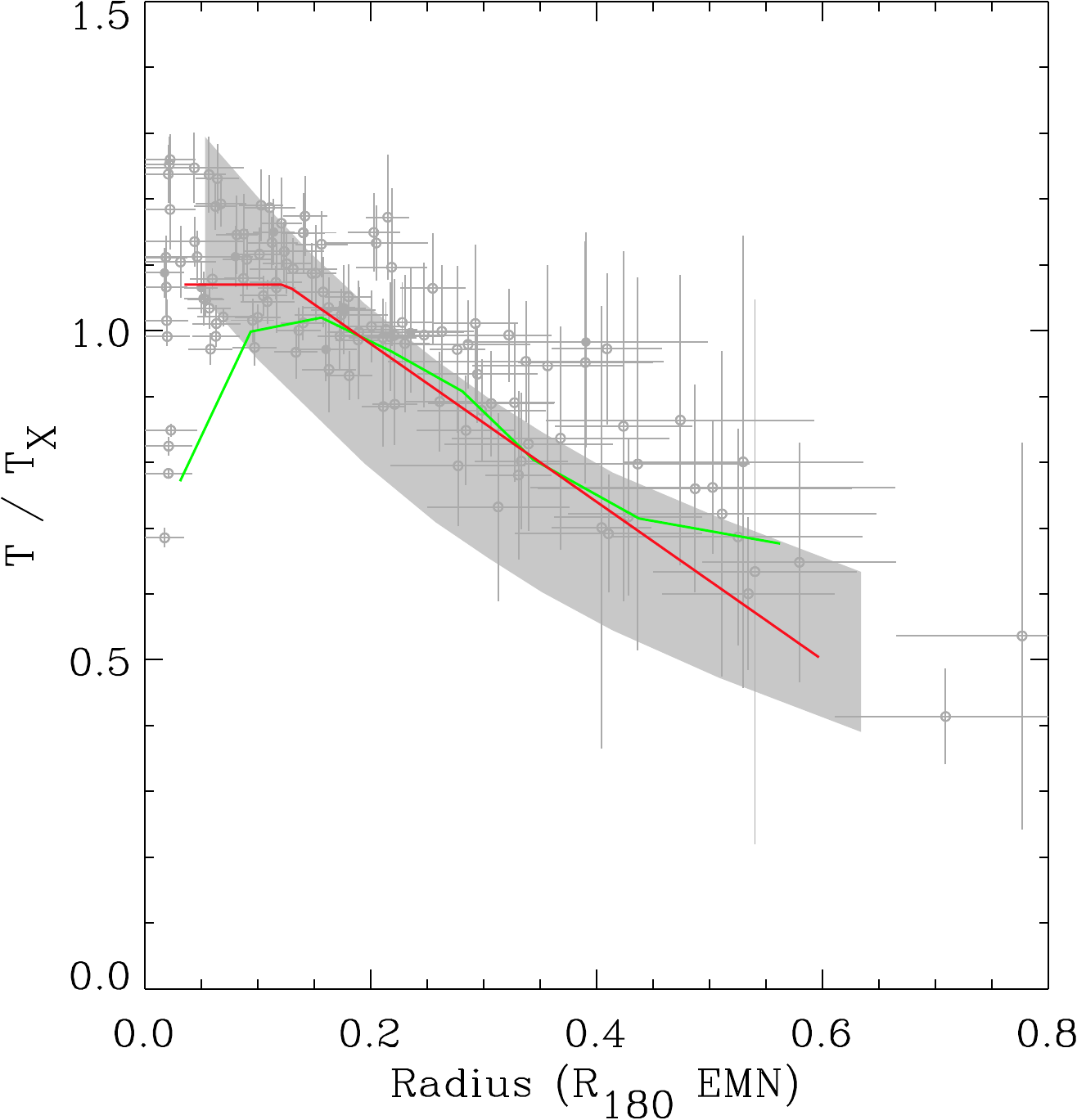}
\hspace{0.2cm}
\includegraphics[width=5.5cm]{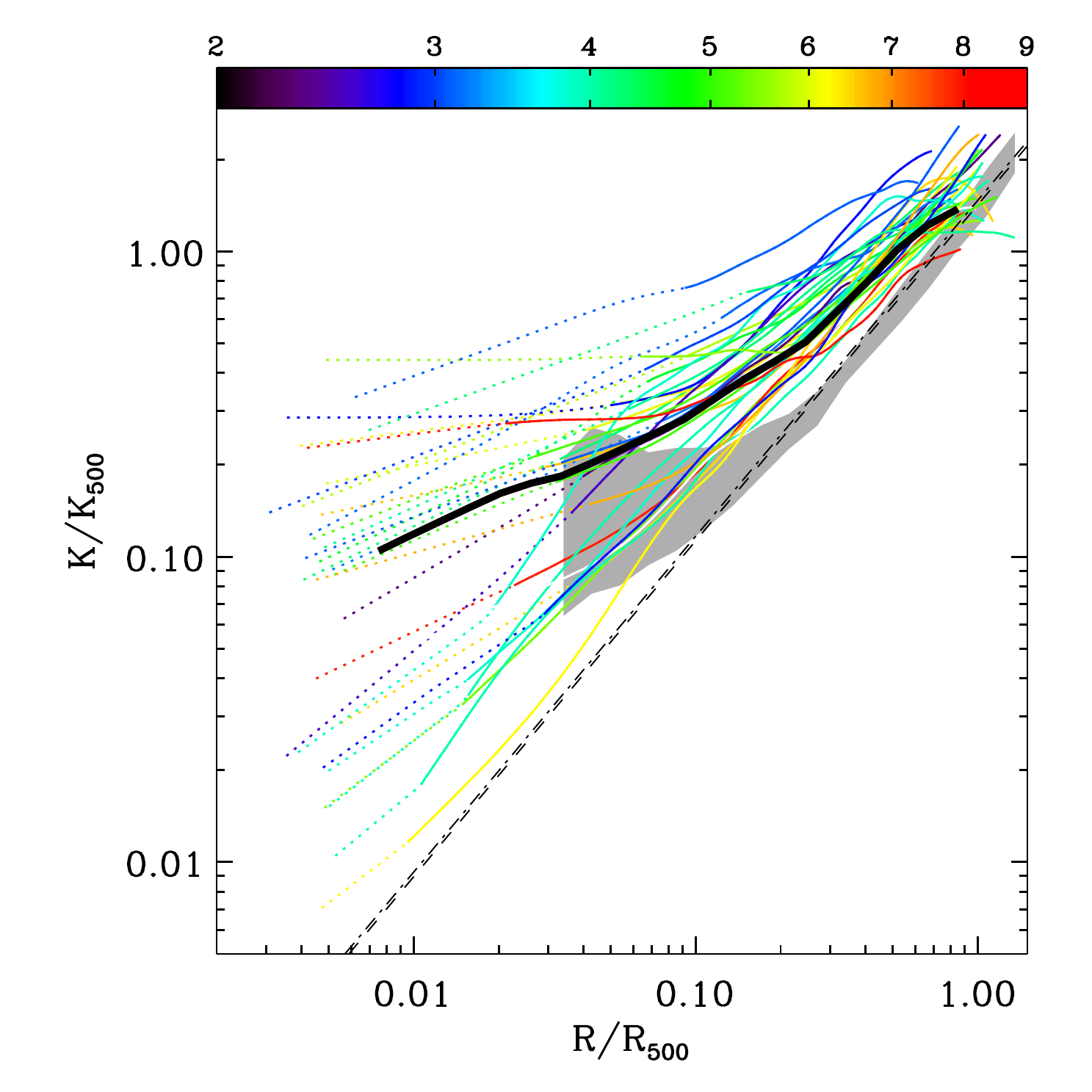}
\caption{{\it Left panel:} Comparison of the dimensionless temperature profiles 
from \xmm\ observations of 15 nearby clusters  by \cite{Pratt_2007} (dots with error bars)
with the average profiles from \asca\ \citep[grey band,][]{Marke_1998},
and the observations of cooling core clusters from  \sax\  
\citep[green line,][]{degrandi_2002} and  \chandra\ \citep[red line,][]{Vik_2005}. Figure from \citet{Pratt_2007}.
{\it Right panel:} Comparison of the scaled entropy profiles of the REXCESS sample 
\citep[colored lines according to the temperature of the clusters,][]{Pratt_2010}
with the theoretical results  from the non-radiative simulations by \citet{Voit_2005}.
The thick black line represents the median of all the observed profiles, whereas
the dashed line stands for the best power law fit to the median profile of the simulated clusters.
Figure from \citet{Pratt_2010}.}  
\label{fig:profiles}
\end{figure}

\end{itemize}

The existing disparity between the self-similar  predictions and 
the observations clearly indicates that, in addition to the effects of gravity,
additional processes related to the physics of baryons should be also taken into account.
Therefore, additional non--gravitational processes operating within galaxy clusters 
and groups, such as radiative cooling, star formation and its inferred energy feedback, 
or AGN feedback, should interact among them in order to solve the aforementioned problems.
As we discuss below,  efforts to obtain a completely consistent and satisfactory model 
able to reproduce the observations are still ongoing 
\citep[e.g.][]{Borgani2004, Voit_2005, McCarthy_2008b}.

\section{The physics of the intra-cluster plasma}
\label{sec:thermo}

Gravitational processes lead the evolution of initial density fluctuations into DM halos. 
However, as we review in this Section, the thermodynamics  of the ICM 
is determined by the combined action of gravity, acting on both DM and baryonic components,  
and  a number of non-gravitational processes  such as radiative cooling,  star formation  and AGN feedback.

\subsection{Structure formation shock waves}
\label{sec:shocks}

As explained in \S\ref{sec:non_linear}, within the hierarchical $\Lambda$CDM  paradigm
the formation of DM halos is driven by the gravitational collapse of initial matter density perturbations.
Figure \ref{fig:mapevol}, which displays the formation and evolution of galaxy clusters as obtained
from hydrodynamical cosmological simulations, indicates 
that in addition to DM (left panels), the evolution of the hot diffuse gas 
(middle panels) and the stellar (right panels) components are significantly connected to each other.
At $z=0$, galaxy clusters are identified as high--overdense DM regions at the 
intersection of large matter filaments. 
Whereas the gas distribution  follows closely the DM distribution, the pattern of the stellar component is much clumpier.

As a consequence of the hierarchical  process of structure growth, 
clusters of galaxies often undergo major merger episodes, which
are one of the most energetic phenomena  in the Universe   \citep{Sarazin_1988}.  
These mergers, associated to the collisions of 
proto--cluster structures of similar masses 
at velocities of several thousands of kilometers per second, 
generate shocks and compression waves in the ICM. 
On large  scales, these hydrodynamical shocks  are very efficient in releasing 
an important  amount  of the  energy  associated with the collision ($\sim10^{61}-10^{65}$ ergs)
as thermal energy in the final system, heating and compressing the hot intra--cluster gas and,
therefore, increasing its entropy \citep[e.g.][]{Quilis_1998, Miniati_2000}.
On smaller scales within already collapsed structures, 
additional weaker shocks are developed as a consequence of subhalo mergers,
accretion processes, or random gas flows. These internal shocks are relevant in  the 
thermalization of the intra--cluster gas \citep{McCarthy_2007b},  contributing significantly to the virialization of halos.

In addition, shocks also generate ICM turbulence and mixing
\citep{Norman_1999, Ricker_2001, Nagai_2003, Dolag2005},
redistribution  and amplification of  magnetic fields \citep{Roettiger_1999}, 
and are likely sources of non-thermal emission in galaxy clusters  \citep[e.g.][]{Bykov_2000}.
Indeed, cosmological simulations have shown that an important 
non-thermal pressure support in galaxy clusters  is provided by 
ICM turbulence \citep[e.g.][]{Dolag2005, Vazza_2011a},   
relativistic CR particles \citep[e.g.][]{Pfrommer_2007, Vazza_2012}, and magnetic fields \citep{Dolag_1999}.
In general, none of these processes alone can regulate
cooling, but they may be more efficient when AGN feedback
\citep{Voit_2011} or plasma instabilities  \citep[e.g.][]{Sharma_2012} are also considered.

Therefore, it is clear that  cosmological shocks leave their imprint on relevant ICM plasma properties
\citep[see][for a review]{Bykov_2008}. 
First, the rise of the gas entropy and thermal pressure support  induced by shocks 
represents a natural alternative way of breaking the self-similar scaling 
\citep[e.g.][for a multi-fluid accretion shock model, derived $K\propto T^{0.8}$]{Bykov_2008}. 
Second,  shocks surrounding mildly overdense, non-linear, non-collapsed structures, such as  sheets and filaments, 
heat the intergalactic medium and determine  the evolution of
the warm--hot intergalactic  medium  \citep[WHIM; e.g.][]{Cen_1999, Dave_2001}. 
Finally, the diffusive shock acceleration process
\citep[DSA; e.g.][]{Drury_1986, Blandford_1987} can  convert  a part of the thermal 
population of particles  into non-thermal  CRs, resulting therefore in both
thermal and non-thermal energy constituents \citep{Kang_2002, Kang_2005, Kang_2007}.  
Whereas the CR electrons  may be responsible of the observed radio 
halos \citep[e.g.][]{Ferrari_2008, Giovannini_2009}
and radio relics \citep[e.g.][]{Ensslin_1998} in clusters \citep[e.g.][]{Pfrommer_2008},
the CR protons may generate  $\gamma$-ray emission.   

\begin{figure*}
\centering\includegraphics[width=12 cm]{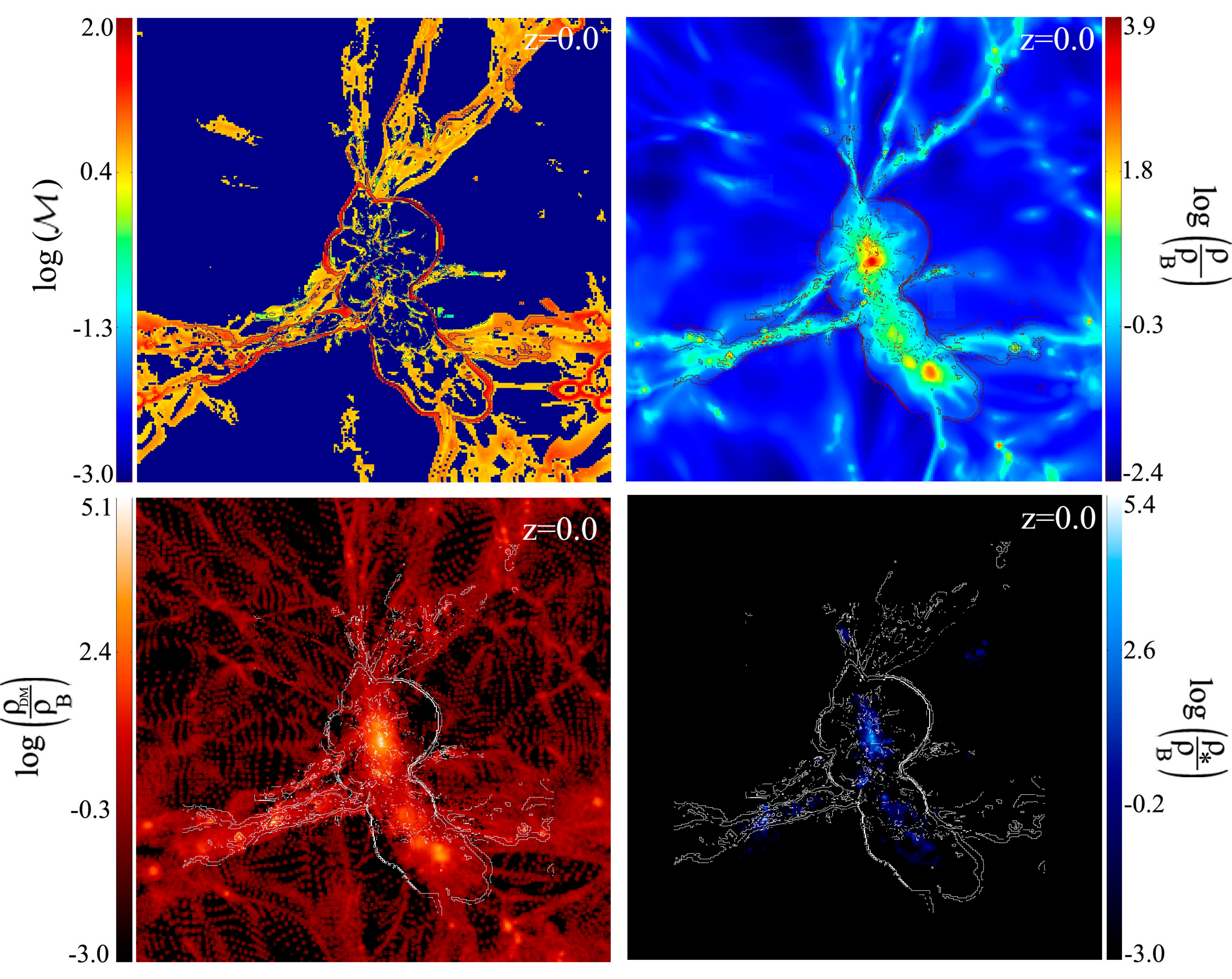}
 \caption
 {Distribution of shock Mach numbers (upper left panel) and its comparison  
 with the DM ($\rho_{DM}/\rho_{_{B}}$, lower left), gas ($\rho/\rho_{_{B}}$, upper right)  
 and stellar ($\rho_*/\rho_{_{B}}$, lower right)  overdensities at $z=0$. 
In all the panels, the contours of the shock waves with high Mach numbers are overplotted.
Each panel represents a projection of $0.2$ Mpc thickness and $64$ Mpc side length. 
Figure from \citet{Planelles_2013}. }
\label{fig:contornos}
\end{figure*}

From an  observational point  of view, detecting large-scale shocks is non-trivial because 
they take place in the periphery of  galaxy clusters where the X-ray emission,
due to the relatively low gas density, is weak. 
In spite of this, large-scale shocks have been identified in the form of  radio relics   
in more than $30$ clusters \citep[e.g.][]{Bonafede_2009, vanWeeren_2009}. 
Some merger--induced internal shocks have been also detected with Mach numbers\footnote{The  Mach number,  which
characterizes the strength of a shock, is given by $\mathcal{M}=v_{s}/c_{s}$,
where $v_{s}$ is the shock speed  and $c_{s}$ is the sound speed ahead of the shock.} 
${\mathcal M}\sim1.5-3$ \citep[e.g.][]{Marke_2007}.
From the theoretical point of view, several approaches to study shocks have been pursued 
using semi-analytical models  \citep[e.g.][]{PS_74, ST_99}
as well as numerical  simulations based on both grid--based,   
single-grid \citep[e.g.][]{Quilis_1998, Miniati_2000, Ryu_2003, Vazza_2009} 
and AMR schemes \citep[e.g.][]{Skillman_2008, Planelles_2013}, 
and particle--based codes \citep[e.g.][]{Pfrommer_2006}. 
As an example of the results obtained from these numerical analysis, 
Fig.~\ref{fig:contornos} shows the distribution of the  
shock waves detected in a cosmological AMR simulation compared 
with the distributions of DM, gas  and stellar  overdensities at $z=0$.
This figure clearly highlights that cosmological shock waves, which occupy  the whole 
simulated volume in a complex way, accurately trace the cosmic web.     
External shocks, with quasi-spherical shapes and at relatively large distances 
from the center of the structure where they appear, 
are characterized by high Mach numbers (${\mathcal M}\magcir 20$).
On the contrary, within collapsed objects,  weaker shocks with ${\mathcal M} < 2-3$ 
are present, contributing significantly  to the dissipation of  kinetic energy \citep[e.g.][]{Skillman_2008}.  

In spite of all the studies and of the long--standing debate about 
the use of different numerical techniques  \citep[e.g.][]{Agertz_2007},
the identification of cosmological shocks still represents an issue.

\subsection{Non-gravitational processes}

`Gravitational feedback', mainly in the form of shock and compression waves, 
contributes to most of the intra-cluster gas heating.
However, as discussed in \S\ref{sec:deviations}, the discrepancies with the self-similar model 
observed in small clusters and groups and in inner cluster regions, indicate the existence
of additional non-gravitationally induced  cooling and heating processes.
In the following, we explain some of the main non-gravitational processes that need to be incorporated
in hydrodynamical simulations to reproduce the self-similarity breaking.

\subsubsection{Gas radiative cooling}

Radiative cooling, which plays a major role in the ICM emissivity, 
can significantly contribute to  break self-similarity to the observed level.
To understand the role of cooling, it is useful to express the cooling time in terms of the gas 
entropy and temperature,  which in the Bremsstrahlung regime can be approximated by \citep[][]{Sarazin_2008}
\begin{equation}
t_{cool}\approx17 \left( \frac{K}{130\, keV cm^{-2}}\right)^{3/2}\left(\frac{k_{\rm B}T}{2\, keV}\right)^{-1} \ Gyrs.
\end{equation}
Therefore, for a galaxy cluster with $k_{\rm B}T\sim 2.5$ keV,  the cooling time is lower than the Hubble time for an entropy 
of  $K\mincir130$ $keV cm^{-2}$. This means that, 
the lower entropy gas within galaxy clusters will cool before, being evacuated earlier from the hot  gas 
in cluster core regions. 
This low-entropy gas will be superseded by higher entropy gas coming from outer regions, 
thus increasing the mean gas entropy \citep{Voit_Bryan_2001}.

\begin{figure}[t]  
  \vspace{0.0cm}
  \centerline{
    \includegraphics[width=2.5truein]{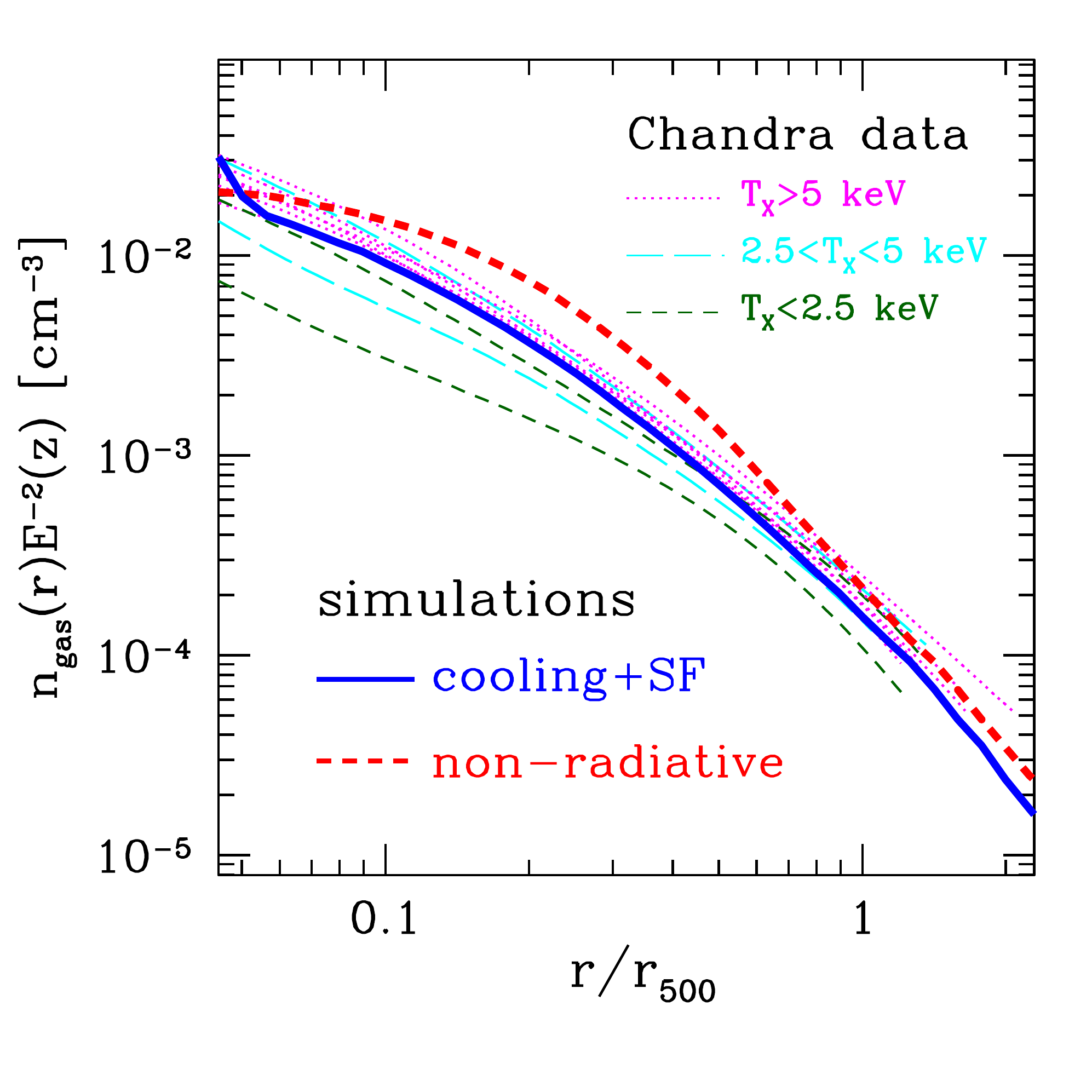}\hspace{-0.4cm}
    \includegraphics[width=2.5truein]{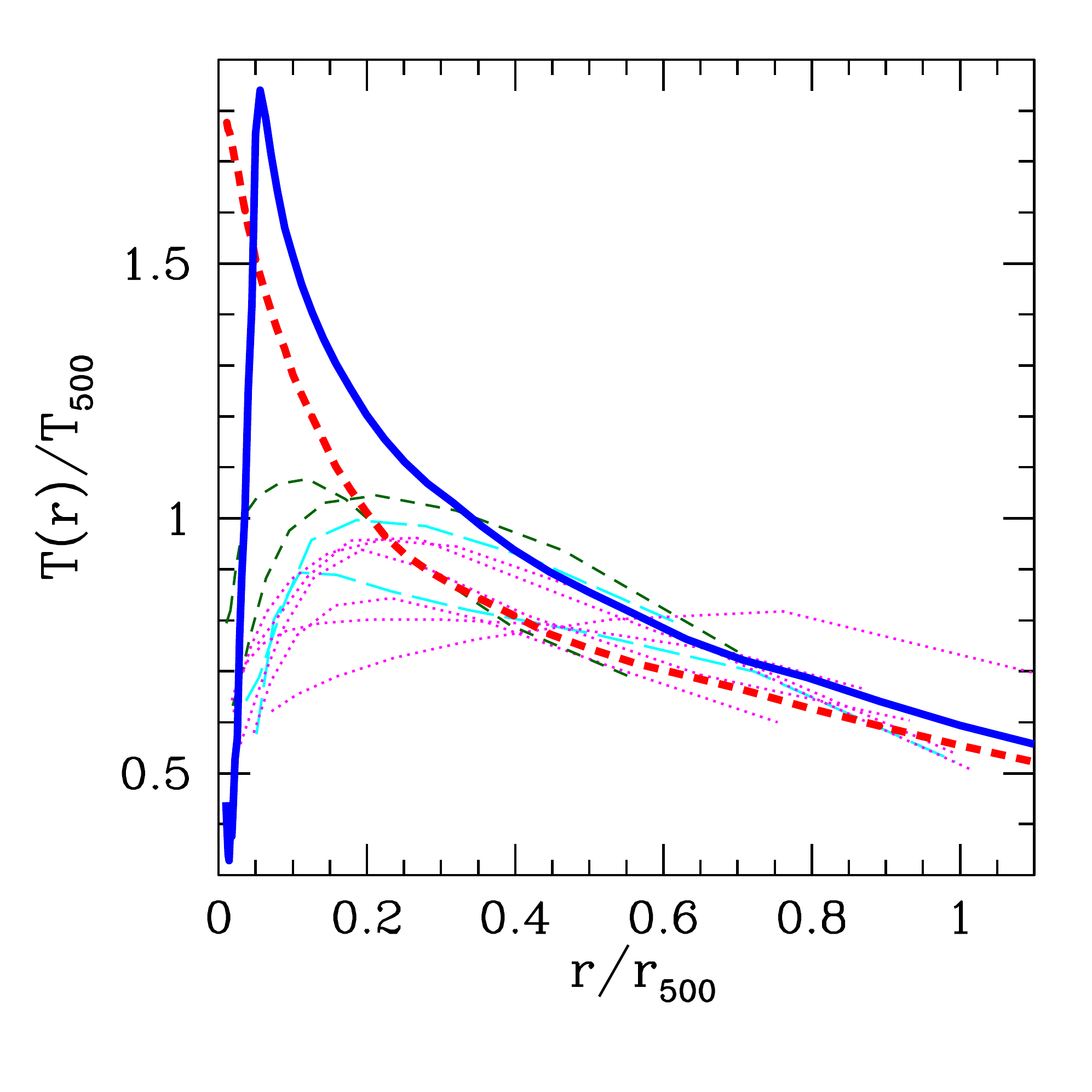} } \vspace{-0.8cm}
  \centerline{  
    \includegraphics[width=2.5truein]{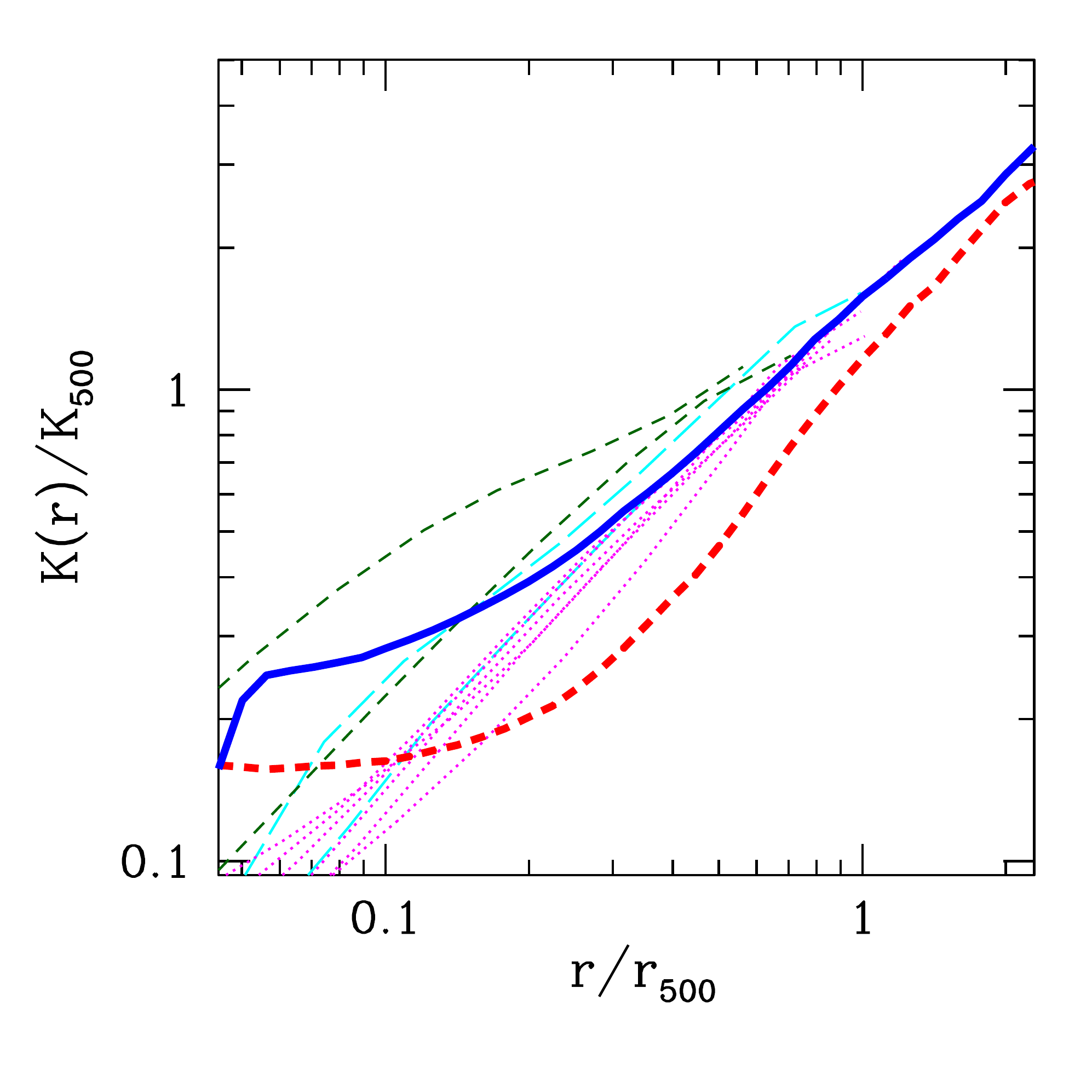}\hspace{-0.4cm}
    \includegraphics[width=2.5truein]{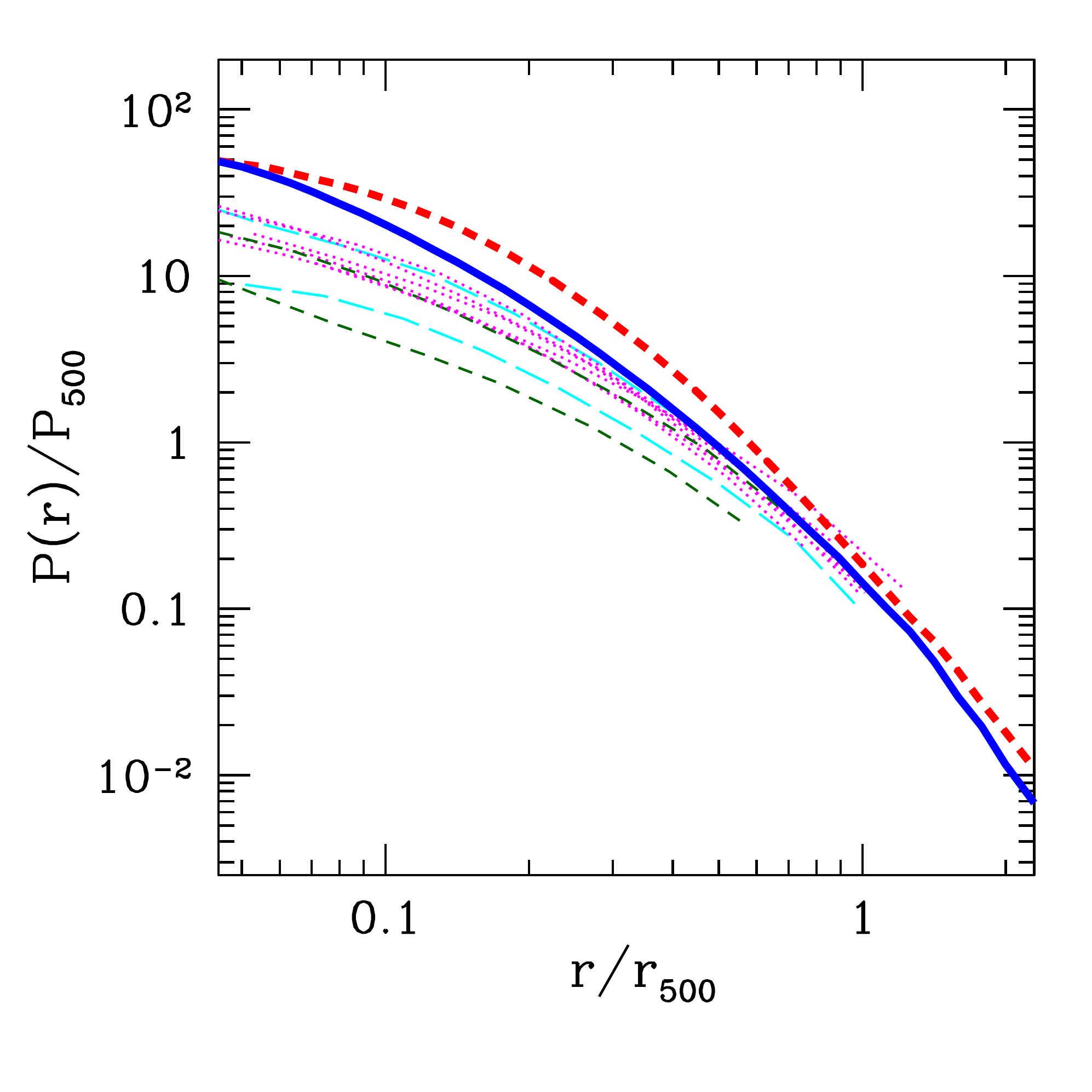} } \vspace{-0.4cm} 
  \caption{
   From top to bottom clockwise:  mean radial profiles of  gas density, temperature, pressure, and 
   entropy  of relaxed clusters at $z=0$ identified in cosmological simulations. 
    There are two sets of simulations, one non-radiative (red thick dashed lines)
    and another accounting for $cooling+SF$  (blue thick solid lines). 
    The observed \chandra\ cluster sample by \citet{Vikhlinin2006} is used for comparison 
    (thin lines in different colors indicating the temperature of the systems).
    Figure from \citet{Nagai_2007b}.}
\label{fig:nagai_comp}
\end{figure}

Hydrodynamical simulations including radiative cooling support this prediction
\citep[e.g.][]{Pearce_2000, Muan_2001, Dave_2002,Valdarnini_2002, Kay2004, Nagai_2007b}. 
As an example, Fig.~\ref{fig:nagai_comp} shows the comparison between the mean ICM profiles 
of relaxed clusters derived from cosmological simulations by \cite{Nagai_2007b}
and the observations by \citet{Vikhlinin2006}.
In particular, there are two sets of simulations, one accounting only for 
non-radiative physics ({\it{NR}}) and another taking into account radiative cooling, 
star formation and metal enrichment ({\it{CSF}}). 
From the analysis of this figure  we infer that, 
whereas the {\it{NR}} simulations produce ICM profiles in conflict with observations,
the {\it{CSF}} runs produce a better match, at  least  outside cluster core regions. 
Specifically, at $r\magcir 0.5 r_{500}$,  the {\it{CSF}} runs produce both an increase in the 
gas entropy in accordance with the observed values, and nearly self-similar  pressure profiles. 
Another effect of cooling is the reduction of the gas density in inner regions (see top-left panel of Fig.~\ref{fig:nagai_comp}).

However radiative cooling has also some undesirable effects.
As it is shown in the top-right panel of  Fig.~\ref{fig:nagai_comp},
an unexpected effect of cooling is that of increasing the ICM 
temperature towards the cluster center together with the steepening of the 
temperature profiles.
This effect results from the dearth of central pressure
support induced by cooling, which makes outer gas to fall sub-sonically  
to the center while being heated adiabatically. 
In addition,  given the runaway nature of radiative cooling, 
it suffers from overcooling, thus transforming quite large amounts of gas into stars. 
In fact, radiative simulations including different forms of stellar feedback  
obtain that around 35-40\% of the cluster baryon content  
is converted into stars  \citep[e.g.][]{Nagai_2004},  a value which is  significantly larger 
than observed \citep[$\sim 10-15\%$; e.g.][]{Balogh_2001, Lin2003, Gonzalez2007}.
In principle, these two shortcomings of cooling represent two
sides of the same problem \citep{Borgani_2011}. 
A proper source of gas heating (or most likely, a combination of several) able of 
counterbalancing cooling, keeping the gas pressurized in inner regions, and 
controlling the star formation, may provide the solution to this complicated issue.
However, as it will be discussed in the following sections, unearthing such a mechanism
is non-trivial and represents nowadays  an important  challenge in the numerical description of clusters.

\subsubsection{Star formation and its associated feedback}

SN explosions contribute to both the heating of the surrounding medium  and the distribution of  metals  
from star--forming regions into the hotter intra--cluster plasma.
Given that SN--driven winds are  a  by--product of the star formation process,  SN feedback was 
suggested to produce a realistic and self-regulated  cosmic star formation rate \citep[e.g.][]{springel_hernquist03}.  
We note that even beyond feedback models, appropriate subgrid--scale 
models for hydrodynamical turbulence may be required for the description 
of the hydrodynamical state  \citep[e.g.][]{Schmidt06, Iapichino_2008, Maier_2009}. 
Moreover, given the relatively low resolutions reached by present-day cosmological  simulations, 
the physics of the interstellar medium can not be properly  described \citep{Borgani_2011},  
a problem that persists even in relatively well--resolved
simulations of individual galaxies   
(e.g. \citealt{Tasker_2009, Dobbs_2010, Wang_2010, Bournaud_2010}; see also \citealt{Mayer_2008} for a review).  
Given the current limitations, stellar  feedback is usually incorporated in cluster simulations
via phenomenological prescriptions of star formation  and SN
heating \citep[e.g.][]{Braun_2012} at moderately low resolutions. 

\begin{figure}
\includegraphics[width=6.2cm, angle=0]{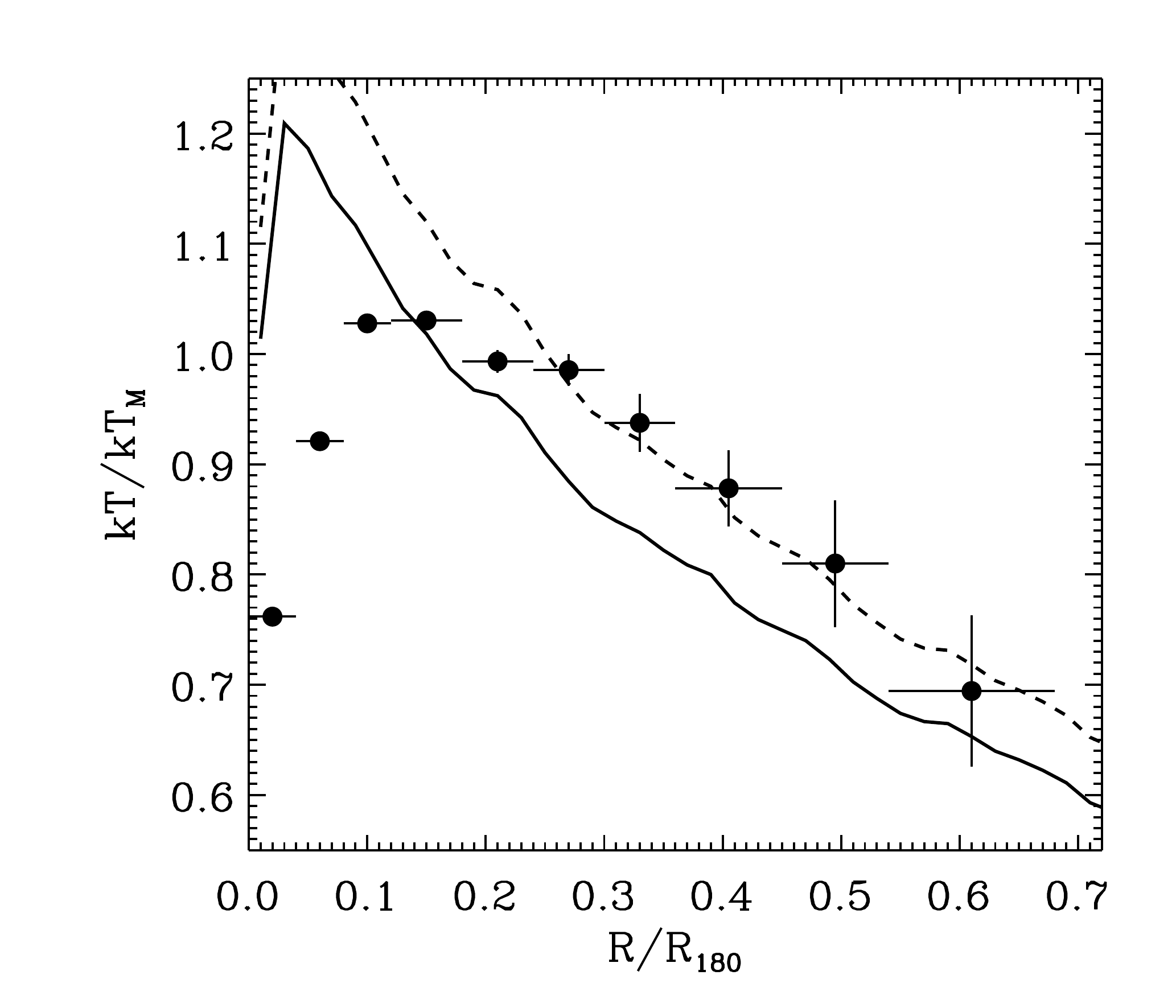}
\includegraphics[width=6.2cm, angle=0]{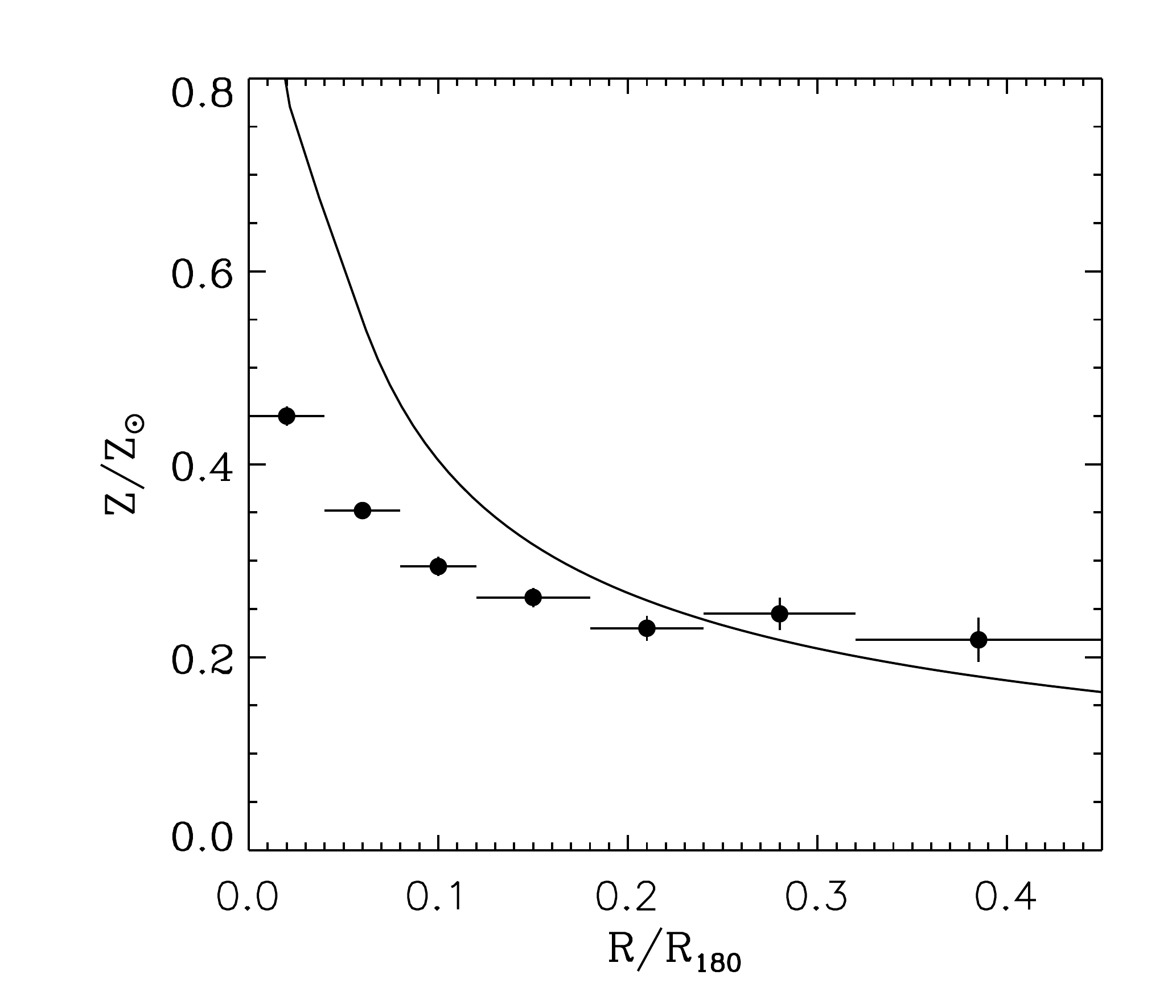}
\caption{{\it Left panel}: Comparison between the mean temperature profile 
of a sample of about 50 local ($z\mincir 0.3$)  and hot ($k_BT_X> 3$ keV)  \xmm\ 
clusters \citep[dots with error bars;][]{Leccardi_2008_T} and
the mean profile obtained from cosmological simulations including 
radiative cooling, star formation and SN feedback \citep[solid line;][]{Borgani2004}.  
The dashed line stands for the mean simulated profile rescaled  by 10\%.
Figure from \citet{Leccardi_2008_T}.
{\it Right panel}: Comparison of the mean metallicity profile
for the same sample of \xmm\ clusters  (dots with error bars)   
with the one  derived by \cite{Fabjan_2008} from 
simulations  performed with the SPH code GADGET-2 \citep{springel05} 
assuming the chemical enrichment model by \cite{tornatore_etal07}.
Figure from \citet{Leccardi_2008}.} 
\label{fig:leccardi_molendi}
\end{figure}

In general, cosmological simulations show that SN feedback
can help in partially offsetting radiative cooling, flattening the  temperature profiles,
and reducing the cluster stellar mass fractions \citep[see][and references therein]{Borgani_2011}.
However, in spite of these promising results, stellar feedback alone is not efficient enough 
to produce the observed thermal structure of CC clusters. 
As an example, the left panel of Fig.~\ref{fig:leccardi_molendi}  
shows the comparison between the temperature profiles 
of a sample of relaxed clusters as derived from observations and from simulations including SN feedback.
We see that, whereas in outer cluster regions, $r\magcir 0.2R_{180}$, simulations
recover quite well the observed profiles, 
within inner  regions the agreement  is not so satisfactory.
The low efficiency of SN feedback in compensating the cooling properly produces additional undesirable results  \citep[e.g.][]{kravtsov_borgani12}:
in general, the levels of core entropy, although reduced, remain significantly larger than reported by observations;
the BCGs have stellar masses  larger than observed; and,  
the excessive star formation is also translated into an excessive metal production in cluster central regions
(see right panel of Fig.~\ref{fig:leccardi_molendi}).
%

\subsubsection{AGN feedback}
 
Currently, the most favored mechanism to explain the ICM self--similarity breaking and the
cooling flow problem is the AGN heating resulting from gas accretion onto a central SMBH.
Indeed, many cluster observations confirm the 
effects of AGN heating on the ICM plasma \citep[e.g.][]{McNamara_2007, Chandran_2009}.
First,  the existence of SMBHs at the nuclei of galaxies 
\citep{Magorrian_1998} and the observed correlations between the BH masses 
and the halo and bulge properties of the host galaxies \citep{Ferrarese_2000} 
point to an scenario in which  galaxy formation and BH growth must proceed together.  
Second, as already reported by many observations \citep[e.g.][]{Burns_1990, Ball_1993, Sanderson_2006}, 
almost every dynamically relaxed CC cluster has an active central radio emitting source,
which has been associated by X--ray observations with the presence of 
cavities or bubbles in  the X--ray emitting gas around the central galaxy \citep[e.g.][]{Birzan_2004}. 
Third, there is a clear connection between the ICM X-ray luminosity within the core of clusters and  the 
mechanical \citep[e.g.][]{Birzan_2004} and radio luminosities \citep[e.g.][]{Eilek_2004} of  the central AGN.
Another important point in favor of this heating mechanism is that AGN feedback is a self-regulated process,
compensating in a natural way radiative cooling \citep[e.g.][]{Rosner_1989}. 
This is due to the fact that the efficiency of AGN feedback is proportional to the rate at which the central SMBH accretes intra--cluster gas
while radiative cooling takes place.
Therefore, if the feedback efficiency is too large, the ICM is naturally over--heated and the gas accretion is reduced.
On the contrary, if the gas accretion rate is too low, the intra--cluster plasma cools faster, the accretion rate onto the central BH 
increases and, correspondingly, the associated AGN feedback efficiency.  
In addition,  AGN heating is supposed to be strong enough to reduce the star formation in the BCGs.

\begin{figure}
\includegraphics[width=6.2cm, angle=0]{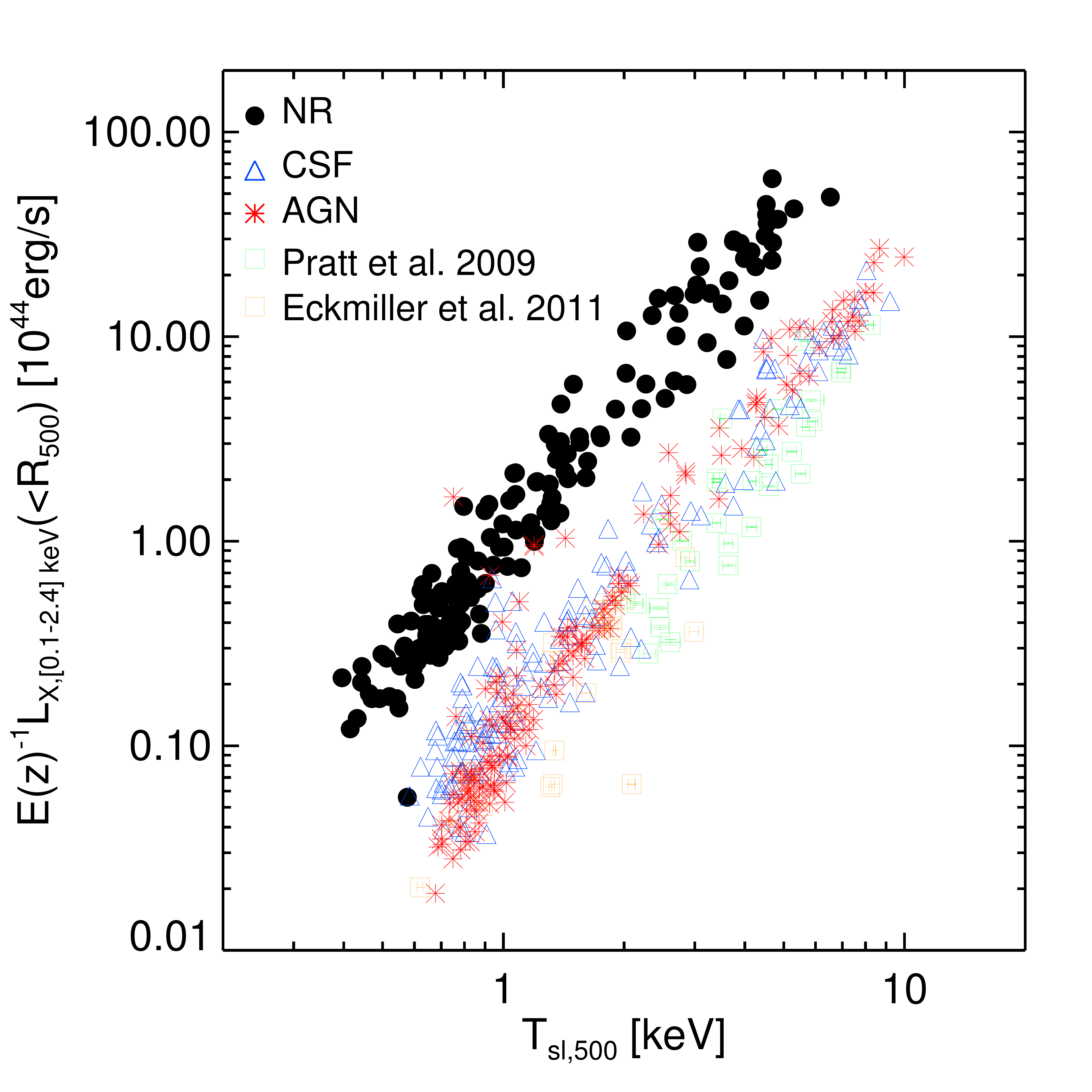}
\includegraphics[width=6.2cm, angle=0]{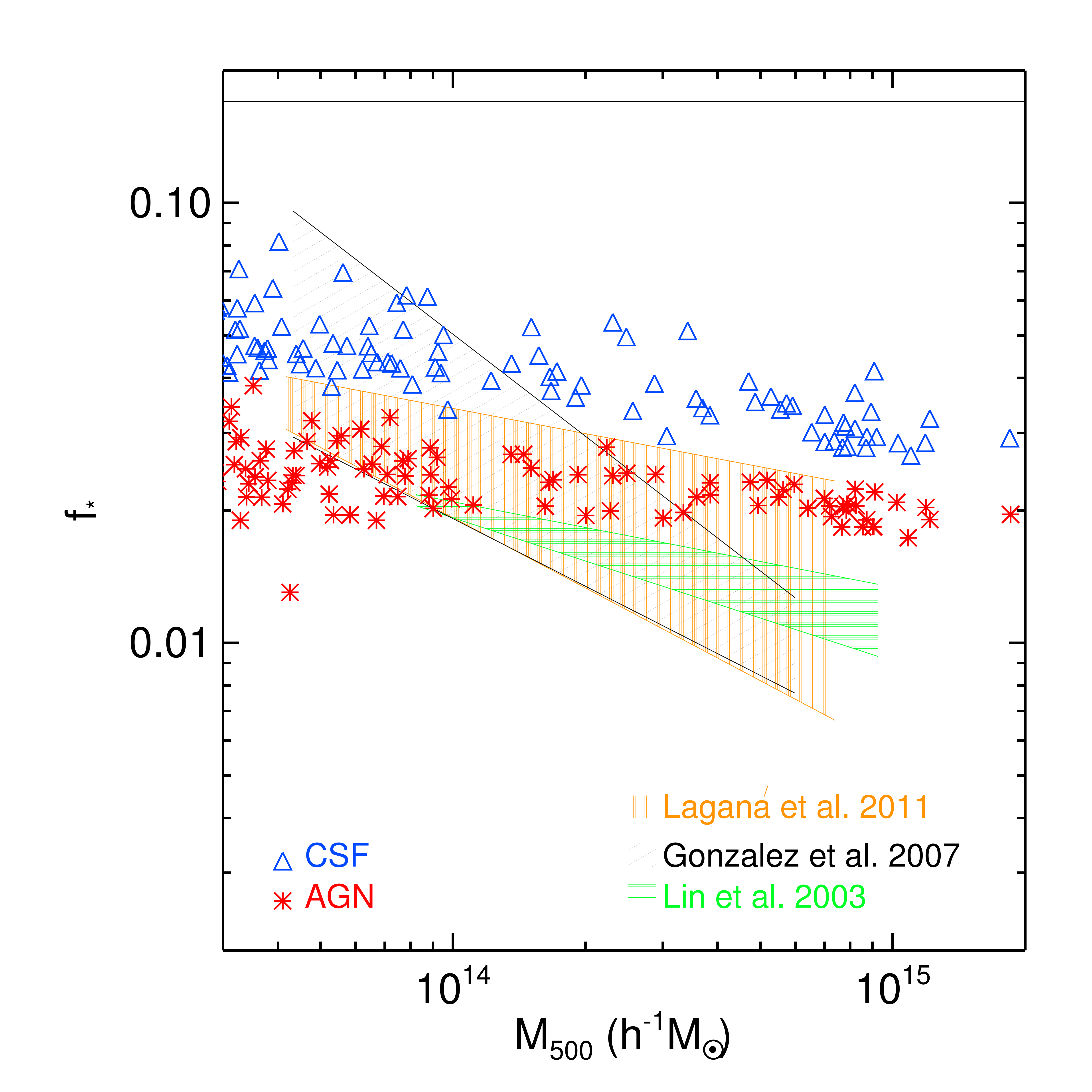}
\caption{{\it Left panel}: $L_X-T$ relation for the sample of
  groups and clusters identified in the simulations by  \cite{Planelles_2013_b}. 
  Results are shown for a set of non--radiative simulations ({\it NR}), and for two sets of radiative simulations, 
  one including cooling, star formation, SN feedback and metal enrichment ({\it CSF}), and another one
  accounting as well for the effects of AGN feedback ({\it AGN}).  Figure from  \cite{Planelles_2013_b}.
{\it Right panel}: Stellar mass fraction as a function of cluster mass 
 as obtained in the simulations by \cite{Planelles_2012}. 
  Results from radiative  simulations with ({\it AGN}) and without ({\it CSF})  AGN feedback are shown.
  The horizontal continuous line stands for the assumed baryon mass
  fraction in the simulations. Figure from  \cite{Planelles_2012}.
  In both panels, data from different observational samples is used for comparison.}
\label{fig:agn_results}
\end{figure}

However, despite the strong reasons in favor of this AGN feedback cycle,  
implementing such a self-regulated mechanism in simulations
represents a challenging task  \citep[e.g.][and references therein]{Borgani_2011}.
In this sense, the first attempts to build competent models of AGN heating consisted in theoretical
studies accounting for the effects of AGN feedback out of  cosmological context \citep[e.g.][]{Churazov_2001, Quilis_2001}.
In the last years, however, different implementations and refinements of AGN feedback models have been 
included in cosmological simulations
\cite[e.g.][]{springel_etal05, sijacki_etal07, Puchwein2008, McCarthy_2010, 
Puchwein2010, fabjan_etal10, short_2010, Battaglia2012, Martizzi_2012, RagoneFigueroa_2013}. 
Given the limited spatial and temporal resolutions achievable by current  simulations, phenomenological models
are needed to include this form of energy feedback. 
In these models, the rates of AGN energy injection are usually computed  by adopting the Bondi gas accretion 
onto the central SMBHs \citep{Bondi1952}\footnote{It is important to point out that
the Bondi approach is the simplest model of gas accretion. A number of studies \citep[e.g.][]{Hobbs_2012} have already highlighted 
the main drawbacks of this approach and the necessity of adopting alternative and more realistic schemes.}. 
In addition to thermal AGN feedback, some observations report that BHs in the center of galaxies
generate relativistic jets that shock and heat the neighboring ICM.
In the light of these observations, the effects of kinetic AGN feedback in the form of 
AGN--driven winds have been also analyzed by several authors 
\citep[e.g.][]{Omma_2004, Dubois_2011, Gaspari_2011, Barai_2013}.

Simulations including different prescriptions of these phenomenological models
have indeed reported some promising achievements 
\citep[e.g.][]{sijacki_etal07, Puchwein2008, fabjan_etal10, Martizzi_2012, Planelles_2012, Planelles_2013_b, LeBrun_2013}.
As it is shown in the right panel of Fig.~\ref{fig:agn_results}, 
AGN feedback seems to be very efficient in  attenuating   
the star formation in high--mass galaxy clusters, producing therefore stellar mass fractions
in better agreement with observational data. 
In addition, as shown in the left panel of Fig.~\ref{fig:agn_results},
AGN heating can also reduce the amount of hot gas in small clusters and groups, thus reproducing better the
observed $L_X-T$ relation and partially resolving the disagreement that otherwise existed for small systems
(see top panel of Fig.~\ref{fig:obs_scaling}).
Besides, AGN feedback has been also shown to be quite effective in dispersing heavy elements 
throughout the intra--cluster plasma, producing  a better consistency with the observed ICM metallicity profiles
\citep[e.g.][]{fabjan_etal10, McCarthy_2010, Planelles_2013_b}.

\begin{figure}
\centering\includegraphics[width=12.5cm, angle=0]{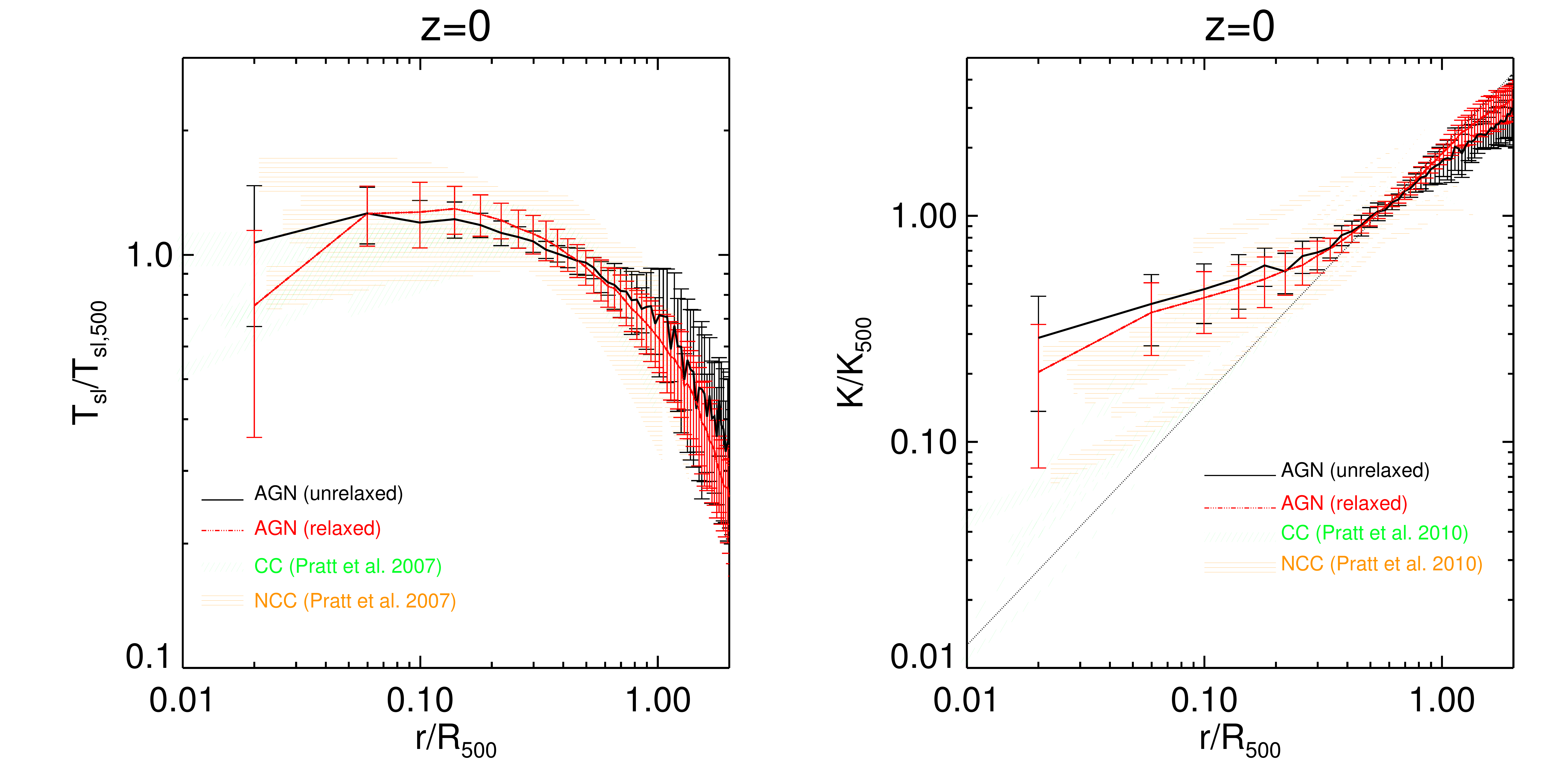}
\caption{
Mean temperature (left panel) and entropy (right panel) radial profiles
for the sample of relaxed/unrelaxed galaxy clusters identified in a set of simulations including 
AGN feedback \citep[adapted from][]{Planelles_2013_b}. 
Error bars indicate $\pm1-\sigma$ dispersion around the mean profile.
In both panels, radial profiles of CC and NCC clusters from the REXCESS sample \citep{Pratt_2007, Pratt_2010}, 
which are represented by colored shadowy areas, are used for comparison.
The dotted black line in the right panel shows the self-similar expectation for the entropy ($K\propto r^{1.1}$).}
\label{fig:planelles_profles}
\end{figure}

Despite the above successes,  a number of discrepancies between observations and  simulations still exist. 
As an example, Fig.~\ref{fig:planelles_profles} shows the mean temperature and entropy  radial profiles
for the sample of relaxed and unrelaxed massive clusters identified in the 
AGN simulations by \cite{Planelles_2013_b}. 
The lack of diversity between the simulated profiles of relaxed and unrelaxed systems is 
at odds with the observed profiles of CC and NCC clusters.
This indicates that, even including  AGN feedback and accounting for 
metal--dependent cooling rates, simulations are still not able to produce the
correct cooling/heating interplay in cluster cores.
The entropy values in inner regions are also higher  than reported by observations. 
In addition, although the stellar masses of the BCGs obtained in  these simulations are reduced, they are still
much larger than observed (\citealt{RagoneFigueroa_2013}; see as well \citealt{Kravtsov_2014} for a recent observational analysis of the 
stellar mass--halo mass relation).
Moreover, in a recent work, \citet{Gaspari_2014}  investigated the isolated effect of kinetic and thermal AGN feedback on 
the $L_X-T$ relation of galaxy clusters and groups. They showed that, even with different parameterizations of these commonly used 
AGN models,  it is not possible to break self-similarity to the desired level without actually breaking as well the cool-core structure
of the considered systems.

These results suggest that, in order to describe the observational properties of the intra--cluster plasma 
in inner core regions and beyond, a proper scheme of AGN feedback may be complemented by additional physical processes.
In this sense, a number of mechanisms,  such as  the effects of CRs in AGN--induced bubbles \citep[e.g.][]{Sijacki_2008}, 
the heating  induced by galaxy motions \citep[e.g.][]{Kim_2005}, or  thermal conduction \citep[e.g.][]{Zakamska_2003}, 
have been suggested. However, further investigation is required to find the correct interaction 
between these and additional plasma physical processes.

\subsection{Additional plasma physical processes}
\label{sec:additional}

Intergalactic gas heating at the cluster formation stage occurs by means of conversion
of the energy of a cold gravitationally accelerated baryonic matter
into the energy of hot thermal gas at cosmological shocks. The
process of relaxation of the kinetic energy of the cold plasma flow
to the quasi-equilibrium thermal distribution in rarefied cosmic
plasmas is non-trivial. This is because the Coulomb collision rate
is slow in the rarefied intergalactic medium and the relaxation
processes are collisionless which means they  are due to the
collective plasma wave-particle interactions. Therefore, the
standard textbook single fluid hydrodynamic and MHD approaches are
not {\sl a priori} valid for the description of the collisionless
plasma flows.

\subsubsection{Gas heating, magnetic field amplification and particle
acceleration in collisionless cosmological shocks} 

The Coulomb mean free path  of a proton of velocity $v_7$ (measured
in 100 $\kms$) in the WHIM of overdensity
$\delta$ can be estimated as $\lambda_{p} \approx 3.5 \times
10^{21}~ \cdot v_{7}^4 \cdot \delta^{-1} \cdot (1+z)^{-3} \cdot
(\Omega_bh^2/0.02)^{-1}$ cm, where $\Omega_b$ is the baryon density
parameter and we assume the Coulomb logarithm to be about 40. The
protons are magnetized in the flow (i.e. their gyro-frequencies are
higher than the mean frequencies of the Coulomb collisions)  if the
 magnetic field magnitude exceeds  about 10$^{-18}$ G.

The microscopic plasma scale, called the ion inertial length, is
defined as $l_i = c/\omega_{\rm pi} \approx 2.3 \times 10^7
n^{-0.5}$ cm, where $\omega_{\rm pi}$ is the ion plasma frequency
and $n$ is the ionized ambient gas number density measured in
$\cmc$. This scale determines the widths of the transition region of
the supercritical collisionless shock waves. In the strong enough
collisionless shocks (typically of a Mach number above a few)
resistivity cannot provide energy dissipation fast enough to create
a standard shock transition  on a microscopic scale
\citep[e.g.][]{Kennel_ea85}. Ion instabilities are important in
such shocks, the so--called {\it supercritical} shocks. At the microscopic
plasma scale the front of a supercritical shock wave 
is a transition region occupied by magnetic
field fluctuations of an amplitude $\delta B/B \sim 1$ and
characteristic frequencies of about the ion gyro-frequency.
Generation of the fluctuations is due to instabilities in the
interpenetrating multi-flow ion movements. The viscous transition
region width is typically a few hundreds ion inertial lengths for a
parallel shock, while it is about ten times shorter for a transverse
shock. The ion inertial length in the WHIM
can be estimated as $l_i \approx 5.1 \times 10^{10} \cdot
\delta^{-1/2} \cdot (1+z)^{-3/2} \cdot (\Omega_0h^2/0.02)^{-1/2}$
cm, providing the width of the collisionless shock transition region
is smaller by many orders of magnitude than the Coulomb mean free
path (that is in the kiloparsec range). The question whether
collisionless shocks can form in plasmas with magnetic pressure much
smaller than the plasma pressure is of fundamental importance.
Two-dimensional particle-in-cell (PIC) simulations of the structure
of non-relativistic collisionless shocks in unmagnetized
electron-ion plasmas performed by \citet{kt08,kt10} revealed that
the energy density of the magnetic field generated by the
Weibel-type instability within the shock transition region reaches
typically 1\%-2\% of the upstream bulk kinetic energy density. The
width of the shock transition region was found in their simulation
to be about 100 ion inertial lengths $l_i$, independent of the shock
velocity. A tiny fraction (much less than a percent) of the incoming
protons can be reflected from the collisionless shock transition
region and these particles are subject of Fermi acceleration if the
shock upstream flow carries magnetic fluctuations. In the case of
strong shocks with high \alf and sonic Mach numbers the accelerated
particles can get a substantial part (tens of percent) of the shock
ram pressure. The pressure of non-thermal accelerated particles may
mediate the shock flow as it was apparently observed by Voyager 2 in
the heliosphere termination shock \citep[see, e.g.][]{florinskiea09}. 
Moreover, evidences of strong non-adiabatic
amplification of fluctuating magnetic fields by anisotropic
distributions of accelerated particles observed in strong shocks of
young supernova remnants \citep[see, e.g.][]{helderea12}.

\begin{figure}
\centering
\includegraphics[width=7.2cm, angle=0]{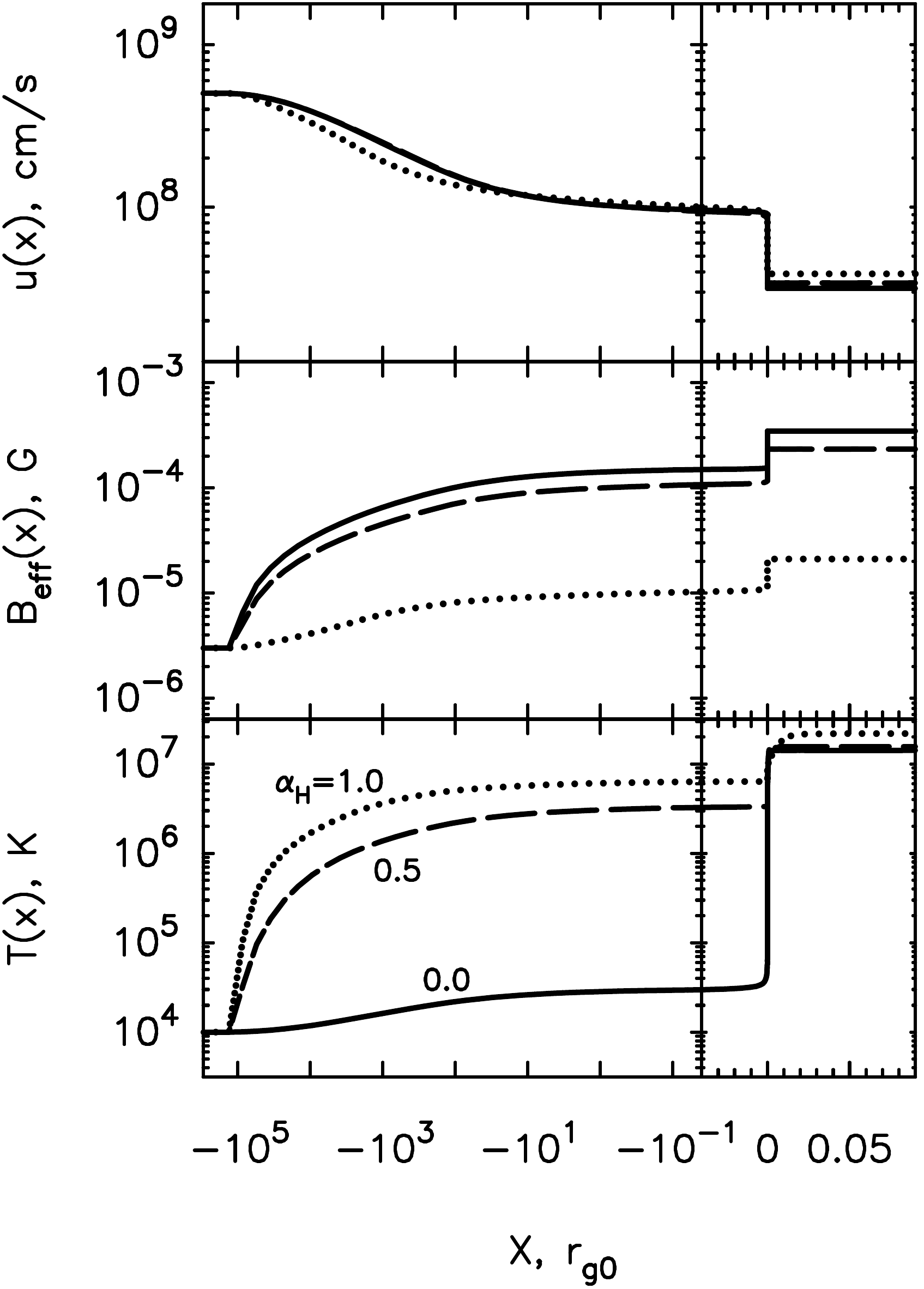}
\caption{
Profile of a strong shock in WHIM simulated by
\citet[][]{vbe08} in the non-linear Monte Carlo model with different
values of the fraction $\heatpar$ of CR driven magnetic turbulence
dissipated in the precursor. The solid, dashed and dotted lines
correspond, respectively, to $\heatpar = 0$, $0.5$ and $1.0$. The
plotted quantities are the bulk flow speed $u(x)$, the effective
amplified magnetic field $\Beff(x)$ and the thermal gas temperature
$T(x)$. The shock is located at $x=0$, and note the change from the
logarithmic to the linear scale at $x=-0.05\:\rgzero$.}
\label{fig:ubt_1e4_1e6}
\end{figure}

Hybrid plasma simulations with kinetic treatment of ions and fluid
electron description \citep[see,
e.g.][]{winskeea90,treumann09,bs13} allow us to study domains of
some thousands gyroradii of incoming proton around the
non-relativistic shocks. Recent two-dimensional hybrid simulations
by \citet{gs12}, who modelled quasi-parallel shock with the \alf
Mach number ${\mathcal M}_a=6$ 
revealed energetic power-law ion distribution of
index about -2 in the shock downstream. The energetic non-thermal
particle population contained about 15\% of the incoming upstream
flow. Limited dynamical ranges of PIC and hybrid simulations do not
allow us yet to study the formation of extended non-thermal tails of
relativistic particles accelerated by non-relativistic shocks. On
the other hand,  simulations of DSA based
on kinetic and Monte-Carlo modeling indicated the formation of the
extended tails of the non-thermal particles. The energetic particles
accelerated by a strong shock have hard spectral indexes and
therefore, the CR pressure is dominated by the high--energy end of
particle distribution. These particles may penetrate into far
upstream and modify the shock flow by the CR-pressure gradient
\citep[see, e.g.][]{Blandford_1987,je91,md01,ab06,vbe08}. Energetic particles
of the highest energy escape into the upstream region providing
energy outflow and allowing the shock compression ratio to be higher
while the post--shock gas temperature and entropy appear significantly
reduced comparing to that in the standard single fluid shock. It is
important to note that the DSA is a very
complicated highly non-equilibrium non-linear process with a strong
coupling between the thermal and non-thermal components. Fast
growing instabilities of the anisotropic distributions of energetic
particles result in efficient production of strong magnetic
turbulence in the shock upstream \citep[see
e.g.][]{McKVlk82,bell04,ber12,SchureEtal2012}.

In Fig.~\ref{fig:ubt_1e4_1e6} we illustrate the effects mentioned
above with simulated velocity, magnetic field, and gas temperature
profiles of a strong shock of velocity 5000$\kms$ with the far
upstream gas temperature of 10$^4$ K and magnetic field of about
$\mu$G. The simulation was made with the non-linear Monte-Carlo
model by \citet[][]{vbe08,vbe09} which accounts for efficient CR
acceleration, strong non-adiabatic magnetic field amplification due
to CR-driven instabilities in the shock upstream with magnetic
turbulence dissipation, and the escape of highest energy particles
to the shock upstream. The models of non-linear DSA 
predict hard spectra of accelerated relativistic
particles, which often show  concave spectral shapes instead of the
power-laws. Strong amplification of the fluctuating magnetic fields
in the upstream flow by CR-driven instabilities is expected in the
models of DSA. Important physical effects
to be learned from the strong collisionless shock modeling are: (i)
potentially sizeable (above ten percent) energy leakage from the
system with the ultra-relativistic particles accelerated at the
shock and escaping through the shock upstream back into
intergalactic medium, and (ii) a possibility of strong
super-adiabatic magnetic field amplification by CR-driven
instabilities \citep[see for a review][]{bbmo13}.

\begin{figure}[t]
\includegraphics[width=6.0cm, scale= 1.2, angle=0]{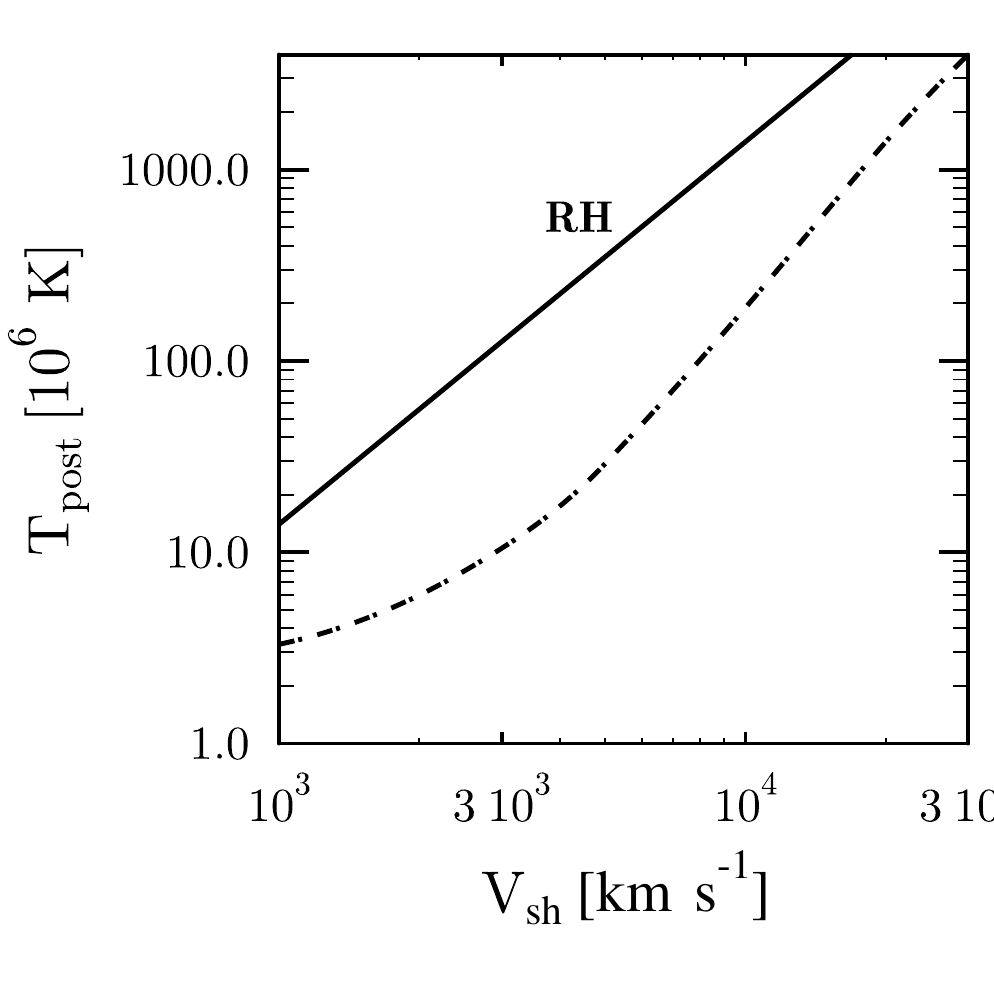}
\includegraphics[width=6.0cm, scale = 1.2, angle=0]{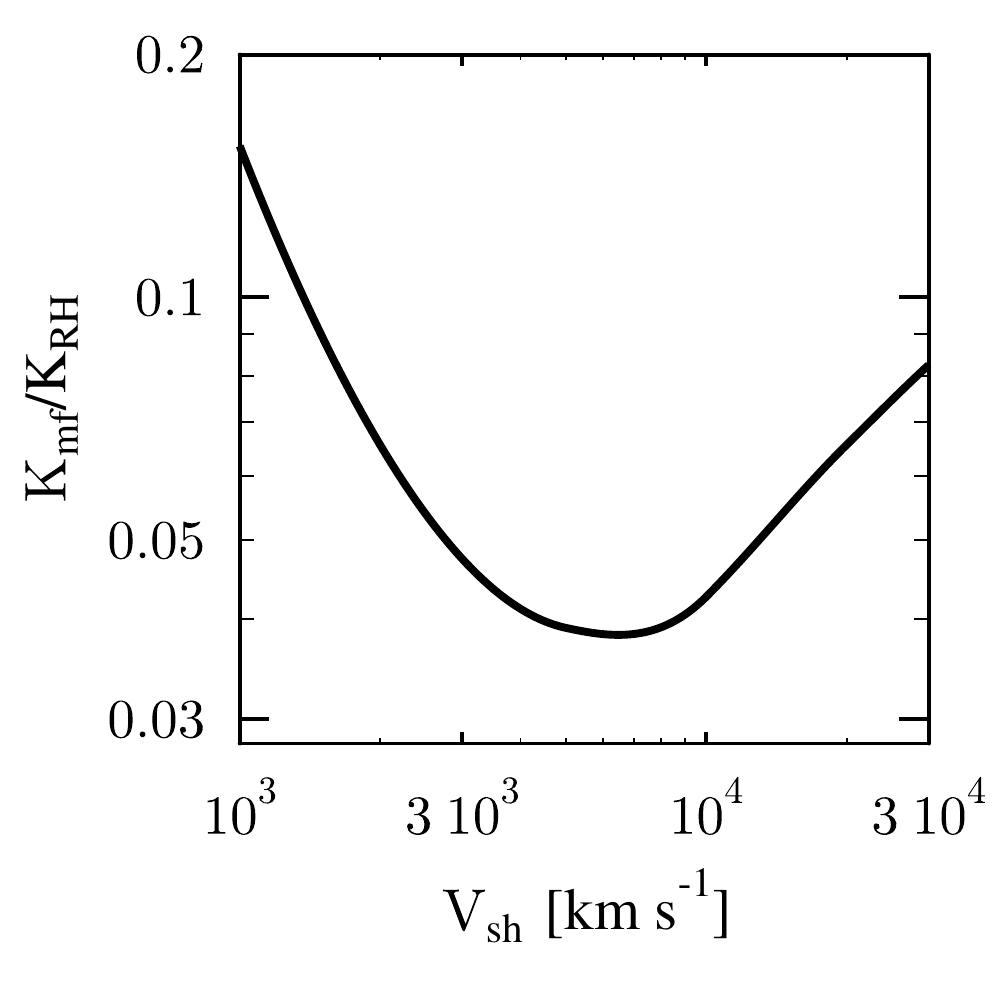}
\caption{{\it Left panel}: The post shock proton temperature as a
function of shock velocity simulated by a non-linear Monte Carlo
model with an account of the efficient particle acceleration and
magnetic field amplification. The far upstream gas temperature was
about $2\times10^4$ K corresponding to the photo-ionized intergalactic
gas accreting by a cluster. The postshock temperature (shown with
dash-dotted line) was simulated for a model with turbulent cascade
of CR-driven magnetic fluctuations \citet{vbe09}. 
 The solid curve (labeled as RH) is the standard
Rankine-Hugoniot single fluid postshock temperature (it is presented
for a comparison). {\it Right panel}: The ratio of the postshock gas
entropy (labeled as $K_{\rm mf}$ multi-fluid) to the standard single
fluid (Rankine-Hugoniot $K_{\rm RH}$) postshock gas entropy  as a
function of shock velocity simulated for cosmic ray modified
collisionless shock. } 
\label{fig:entropy_bykov}
\end{figure}

These effects may strongly affect the thermal properties of shocks
with high sonic and alfvenic Mach numbers as it is expected to be
the case in the external accreting shocks at cluster outskirts. 
In Fig.~\ref {fig:entropy_bykov} we illustrate the 
possible effect of the non-thermal components on the ion temperature (left panel) and the entropy (right panel) 
in the downstream of the multi-fluid shock simulated with non-linear Monte-Carlo model described in 
\citet{vbe09}.
Both post-shock ion temperature and gas entropy $K_{\rm mf}$ in 
the multi-fluid collisionless shock can be strongly reduced compared to that  in the standard single
fluid Rankine-Hugoniot adiabat. This is because of a substantial increase of the gas compression ratio 
in a strong collisionless shock converting a sizable part of the shock ram pressure into relativistic particles. Some 
fraction of the accelerated particles escape the system thus allowing the gas compression ratio to be much larger than 4.        
As
it was discussed in \S~\ref{sec:shocks} most of the kinetic energy
dissipation occurs at the cluster inner shocks with the modest Mach
numbers where the effects discussed above are likely much less
prominent. Indeed, the ratio of the thermal gas heating to CR
acceleration rates in weak shocks of sonic Mach number 
${\mathcal M}_{\rm s}< 2$ is proportional to 
$({\mathcal M}_{\rm s} -1)\,\beta_{\rm p}$, providing
inefficient CR acceleration by weak shocks in the case of the large
ratio of the thermal and magnetic pressures $\beta_{\rm p}\gg$ 1
expected in the inner regions of the cluster. A recent search of
$\gamma$-ray emission from stacked Fermi-LAT count maps of some dozens
of clusters of galaxies
\citep[][]{FermiLAT_cl1_13, FermiLAT_cl3_13, FermiLAT_cl2_13}
established stringent upper limits on the  average CR to thermal
pressure ratio to be below of a few percent within the radius
$r_{200}$ depending on the assumed index of the power-law CR
distribution and $\gamma$-ray background models. Detection of $\gamma$-ray
emission from extended regions around the external accretion shocks
in the vicinity of filaments and clusters is a challenging task
given its low surface brightness because of the low gas density.

\section{Concluding remarks and open problems}
\label{sec:conclusions}

In this review, we have discussed recent results on structure formation focusing our attention on the first objects in the Universe and the most massive clusters of galaxies at the present day. These extreme scenarios allow us to clearly illustrate the relevance of the physics of plasma on the formation of cosmic structures along a wide range of spatial and temporal scales.    
In the hierarchical paradigm of structure formation, the first objects are the building blocks of subsequent structure formation, leading to larger galaxies and galaxy clusters through accretion and merger events \citep[e.g.][]{Somerville12}. 
Despite the disparity of involved scales,  a number of physical processes, such as radiative cooling, turbulence and feedback, appear to be common, suggesting them to be ubiquitous in the non-linear regime of cosmic structure formation.

In the early Universe, the first objects are expected to form in halos with $10^5-10^8$~M$_\odot$ at redshift $10-30$ \citep[e.g.][]{Bromm09, Bromm11}. Here we distinguish the so-called minihalos with virial temperatures above $1000$~K from the atomic cooling halos with virial temperatures above $10^4$~K. Minihalos are the expected formation sites for the first primordial stars, with typical masses in the range of $10-100$~M$_\odot$ \citep{Abel02, Bromm02, Yoshida08, Clark11, Greif11, Hosokawa11, Turk12, Latif13a, Susa13}. Their formation is governed by the chemistry and cooling of molecular hydrogen, as well as additional processes such as turbulence \citep[e.g.][]{Turk12, Latif13a}, radiative feedback \citep[e.g.][]{Hosokawa11, Susa13} and magnetic fields \citep[e.g.][]{Tan04, Machida06, Sur10, Schober12, Sur12, Turk12}. 

\begin{figure}
\centering\includegraphics[width=10cm, angle=0]{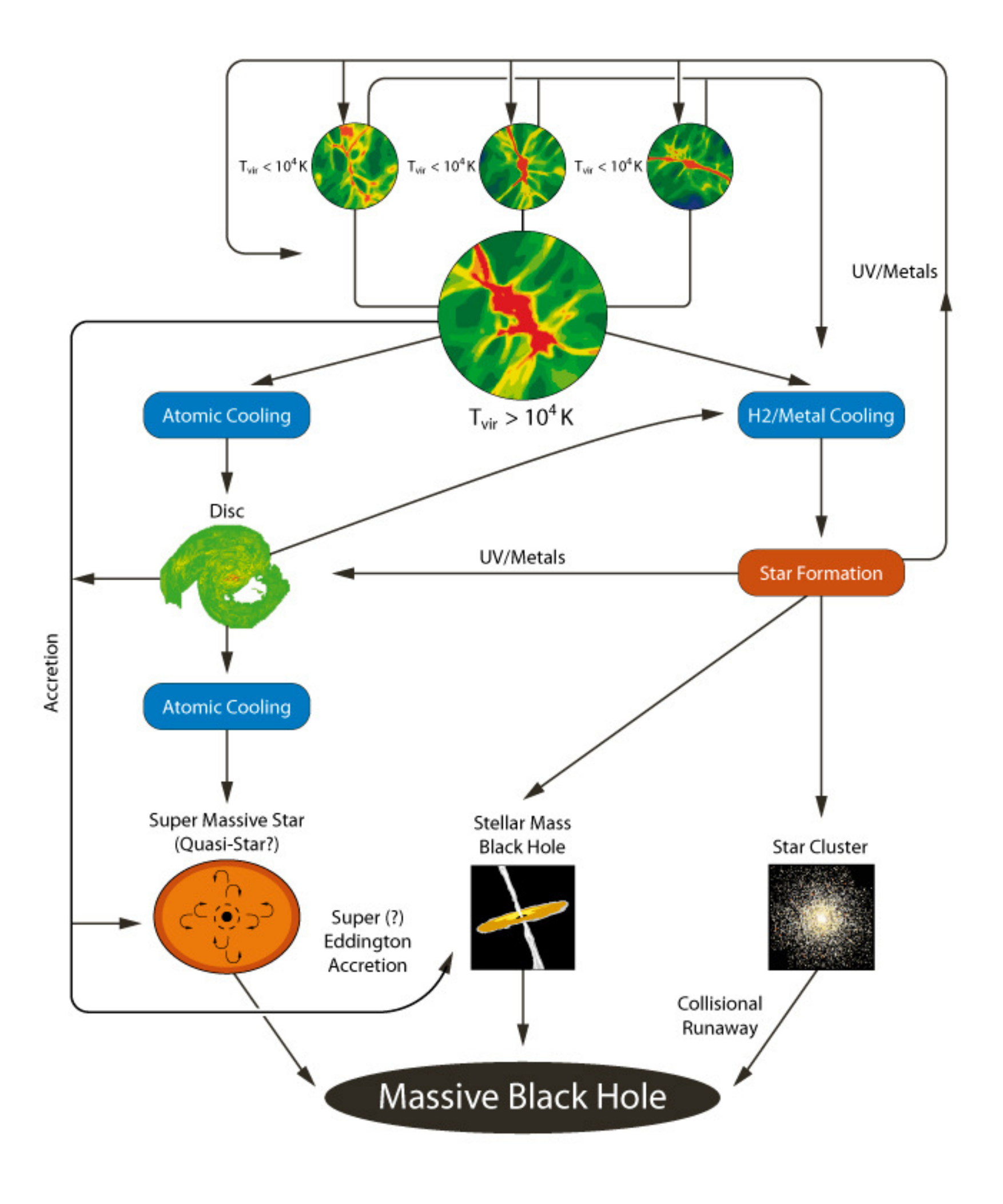}
\caption{Flowchart summarizing possible paths for the formation of the first SMBHs 
in high-redshift atomic cooling halos.  Figure from \citet{Regan09a}.}
\label{fig:SMBH_flowchart}
\end{figure}

The atomic cooling halos  show a more complex evolution depending on their local conditions, in particular regarding their metallicity and dust content. In this review, we restricted ourselves to the formation of massive black holes in primordial halos (see Fig.~\ref{fig:SMBH_flowchart} for an illustrative summary), while a more general discussion is given by \citet{Bromm11}. In the presence of strong photodissociating backgrounds, H$_2$ formation is suppressed  \citep{Omukai01, Machacek01,  Johnson08, Schleicher10, Shang10, Latif11}, leading to a close-to-isothermal collapse regulated via atomic hydrogen lines. Recent numerical simulations confirm that massive central objects can indeed form, due to the high accretion rates of more than $1$~M$_\odot$~yr$^{-1}$ \citep{Latif13b}. In the presence of such accretion rates, feedback can be expected to be weak \citep{Hosokawa12, Schleicher13} and does not impede the accretion. Indeed, even trace amounts of dust, corresponding to $10^{-5}-10^{-3}$ times the dust-to-gas ratio in the solar neighborhood, may already trigger strong cooling and fragmentation at high densities \citep{Schneider04, Omukai05}, but also stimulate the formation of molecular hydrogen at low to moderate densities \citep{Cazaux09, Latif12}. The extremely metal poor star SDSS J1029151+172927 \citep{Caffau11}  shows chemical abundances at which metal line coolant is inefficient, and where only trace amounts of dust grains were able to trigger cooling and fragmentation 
\citep{Klessen12, Schneider12}\footnote{An alternative formation scenario for the star SDSS J1029151+172927 has been recently proposed by \citet{MacDonald_2013}, who suggest that it may have been a subgiant formed with significantly higher metallicity in the vicinity of a SN-Ia.}. 
For metallicities above $10^{-2}$ solar, on the other hand, metal line cooling can be expected to be significant \citep{Bromm03, Omukai05}. The fragmentation of such metal-enriched atomic cooling halos is in fact poorly understood \citep[see e.g.][for first modeling attempts]{Safranek10} and needs to be investigated in further detail.

\begin{figure}
\centering
\includegraphics[width=11.0truecm]{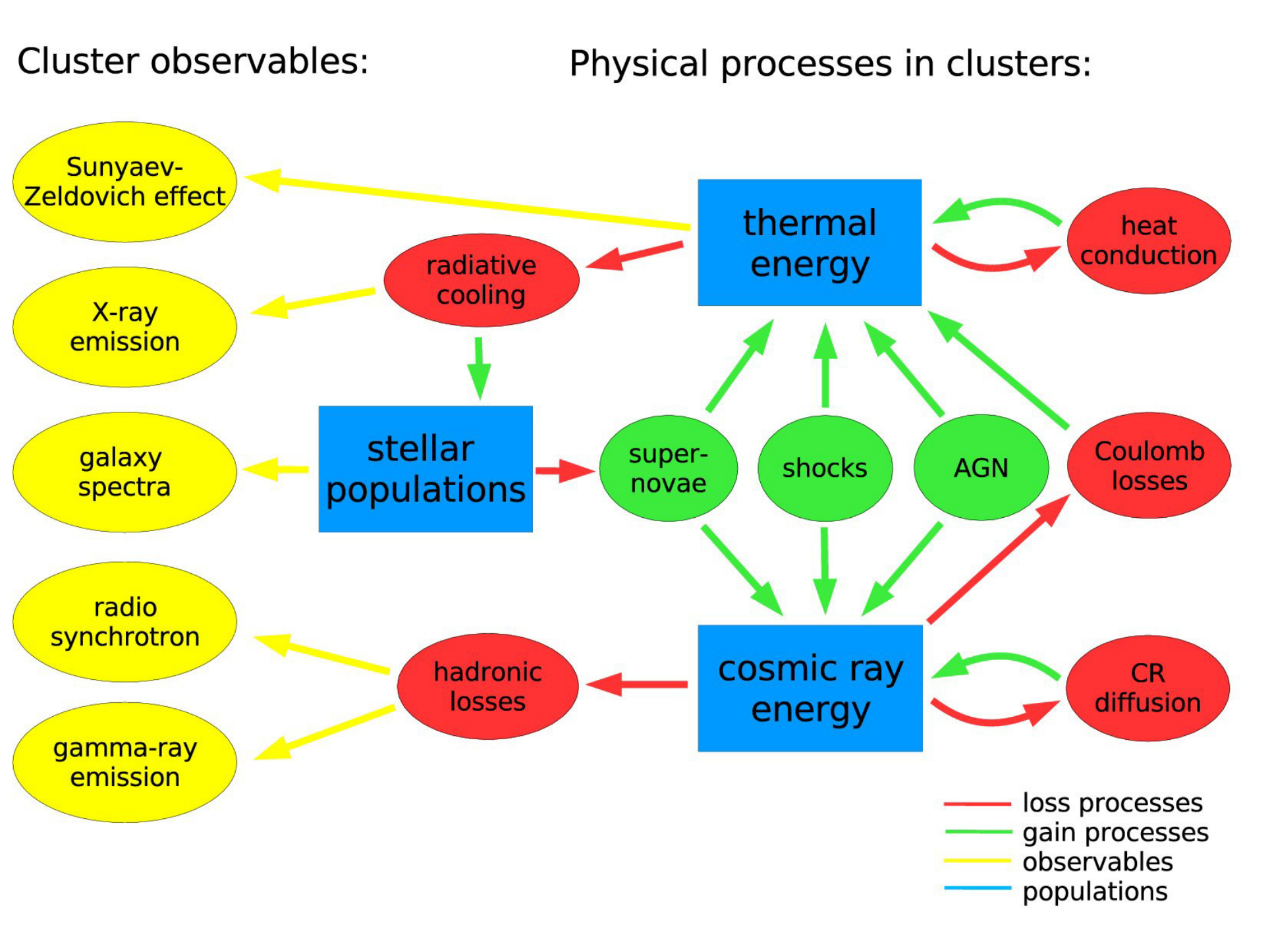}
\caption{Flowchart summarizing  the connections between the main physical processes taking place 
 in galaxy clusters together with the different observational channels through which they can be detected. 
Figure from \citet{Pfrommer_2008}.}
\label{fig:CR_flowchart}
\end{figure}

In the hierarchical paradigm of structure formation, where the first  objects are the building blocks of subsequent structure development, clusters of galaxies, with masses of up to $10^{15}M_\odot$ at $z=0$, occupy the most massive extreme of the cosmic hierarchy.
The formation and evolution of galaxy clusters is a complex and non-linear event
resulting from the intricate interaction of a  number of physical processes acting on a wide 
range of scales \citep[see][for a recent review and references therein]{kravtsov_borgani12}.
As an example, Fig.~\ref{fig:CR_flowchart}   
shows a simplified summary of some of the main processes operating  in galaxy clusters. 
On large scales, the hierarchical process of structure formation induces the development of  strong  cosmological shocks, surrounding galaxy clusters and filaments, that contribute to heat and compress the hot intra--cluster plasma. 
Within galaxy clusters, weaker internal shocks, mainly originated by subhalo mergers or accretion phenomena, change the energetic balance of the gas and allow the halos to virialize. These shocks can also generate ICM turbulence  and mixing, amplify magnetic fields, and accelerate thermal distributions of particles giving rise to a  non-thermal population of CRs.  
In dense regions within galaxies and galaxy clusters, the intra--cluster gas can cool radiatively, leading to both star formation and gas accretion onto SMBHs residing at the center of massive cluster galaxies. These processes can then provide a significant  energy contribution to the ICM in the form of SNe explosions or AGN feedback. As shown in Fig.~\ref{fig:CR_flowchart}, all these processes, which are highly interconnected between them, are manifested by means of different observational channels. 

In the last years, the new generations of
supercomputing and programming facilities have been crucial to deepen in our understanding of the complicated physical processes
taking place within the intra--cluster plasma and  shaping the observational properties of galaxy clusters.  
In order to explain the observations, cosmological hydrodynamical simulations have tried to implement the most relevant physical processes 
self--consistently with the cosmic evolution.
In particular, in addition to gravitationally--induced phenomena inherent to structure formation, the standard non--gravitational processes commonly included in these simulations are radiative cooling, star formation and SN feedback. 
In the last years, the inclusion of the effects of thermal and/or kinetic AGN feedback is also becoming a common practice, despite the fact that the particularities of the heating mechanism are still uncertain.
In spite of the relatively simplicity employed in modeling these complex processes, 
simulations have been able to  significantly reproduce most of the observational cluster properties, at least for massive systems at relatively outer cluster  regions ($0.1R_{500} \mincir r \mincir R_{500}$), where clusters are assumed to be nearly self--similar.
However, inner cluster regions and  smaller systems show  a number of significant issues that still need to be solved.
In these inner regions,   simulations still show an excess of gas cooling, which produces an excess in both the star formation  
and the metal production. In addition, simulations are still not able to solve the cooling flow problem or to reproduce the diversity of  the observed temperature and entropy radial profiles of relaxed and unrelaxed systems. 
On the other hand, cluster outskirts ($r\magcir R_{500}$) are also affected by strong
deviations from hydrostatic equilibrium caused, primarily, by sources of
non-thermal pressure support such as CRs or magnetic fields, which generally are not modeled in simulations.
These results indicate that, in addition to the processes already included, 
a number of additional physical processes, mainly related with the complex physics of plasma, such as
turbulence, viscosity  or thermal conduction, must be also properly taken into account.
Therefore, although  AGN feedback seems to be the most favored energy source to regulate cooling in clusters,  
a subtle  interplay with a number of supplementary physical phenomena
may be needed to explain the observational properties of galaxy clusters and groups, from the core regions out to the outskirts. 

In the near future, a significant numerical effort will be aimed at performing  
larger and better--resolved cosmological simulations with a more accurate modeling of the physics
of galaxy evolution. 
In addition to these technical improvements, 
forthcoming instruments, like the {\it JWST} \citep{Gardner_2006} and the new generation 
of large ground--based telescopes, are expected to detect light from the first galaxies, contributing to interpret early structure formation.
Besides, a number of large observational surveys in different wavebands, such as 
{\it eROSITA} \citep{erosita_2012}, {\it Euclid} \citep{euclid_2012}, {\it WFXT} \citep{wfxt_2011} or the {\it LSST} \citep{lsst_2009},
will provide a significantly large number of clusters.
These numerical and observational efforts, together with a more accurate treatment of the physics of plasma, will definitely shed some more light on the
nature of the physical processes governing the formation of structures in the Universe, 
from the first non-linear objects  to the present--day massive galaxy clusters.

\begin{acknowledgements}
We would like to thank the ISSI staff for their hospitality and for providing an inspiring
atmosphere at the International Space Science Institute Workshop in Bern in 2013. 
We also would like to thank the anonymous referee for
his/her constructive comments. 
SP acknowledges support by the PRIN-INAF09 project ``Towards an
Italian Network for Computational Cosmology'' and by the PRIN-MIUR09
``Tracing the growth of structures in the Universe''. 
DRGS thanks for funding from the German Science Foundation (DFG)
in the DFG priority program SPP 1573 ÒPhysics of the Interstellar 
MediumÓ under grant SCHL 1964/1-1, and via the collaborative 
research center (CRC) 963/1 ÒAstrophysical flow instabilities and 
turbulenceÓ (project A12). 
AMB was supported in part by  RAS Presidium and OFN 15 and 17 programs.
We further thank for stimulating discussions with Stefano Borgani,
Stefano Bovino, Muhammad Latif, Wolfram Schmidt, Jens Niemeyer and Barbara Sartoris.
\end{acknowledgements}

\bibliographystyle{aa}
\bibliography{issi_review}

\end{document}